\documentclass[a4paper,onecolumn, 
superscriptaddress,10pt,accepted=2023-08-20,issue=4, volume=6, shorttitle=papers]{compositionalityarticle}
 \pdfoutput=1
\usepackage[utf8]{inputenc}
\usepackage[english]{babel}
\usepackage[T1]{fontenc}
\usepackage{amsmath}
\usepackage{hyperref}

\usepackage{tikz}
\usepackage{lipsum}

\input xypic
\usepackage{amsbsy,amssymb,amscd,amsfonts,latexsym,amstext,delarray,amsmath,graphicx} 

\providecommand{\qed}{\hspace*{\fill}\nolinebreak[2]\hspace*{\fill}$\Box$}

\newenvironment{proof}%
{%
\begin{list}{\em Proof. }%
{\setlength{\labelsep}{0mm}\setlength{\leftmargin}{0mm}%
\setlength{\labelwidth}{0mm}\setlength{\listparindent}{\parindent}%
\setlength{\parsep}{\parskip}\setlength{\partopsep}{0mm}}%
\item%
}%
{%
\qed\end{list}%
}

\newtheorem{thm}{Theorem}[section]
\newtheorem{prop}[thm]{Proposition}
\newtheorem{cor}[thm]{Corollary}
\newtheorem{lem}[thm]{Lemma}

\newtheorem{defn}[thm]{Definition}

\newtheorem{rem}[thm]{Remark}

\numberwithin{equation}{section}

\def\bI{{\mathbb I}}

\def\bL{{\mathbb L}}

\def\bS{{\mathbb S}}

\def\F{{\mathbb F}}
\def\N{{\mathbb N}}

\def\R{{\mathbb R}}
\def\Z{{\mathbb Z}}

\def\cA{{\mathcal A}}
\def\cB{{\mathcal B}}
\def\cC{{\mathcal C}}
\def\cD{{\mathcal D}}
\def\cE{{\mathcal E}}
\def\cF{{\mathcal F}}
\def\cG{{\mathcal G}}

\def\cI{{\mathcal I}}
\def\cJ{{\mathcal J}}
\def\cK{{\mathcal K}}
\def\cL{{\mathcal L}}
\def\cM{{\mathcal M}}
\def\cN{{\mathcal N}}

\def\cP{{\mathcal P}}
\def\cQ{{\mathcal Q}}
\def\cR{{\mathcal R}}
\def\cS{{\mathcal S}}
\def\cT{{\mathcal T}}
\def\cU{{\mathcal U}}

\def\cW{{\mathcal W}}

\def\Hom{{\rm Hom}}
\def\id{{\rm id}}

\def\fA{{\mathfrak A}}
\def\fB{{\mathfrak B}}

\begin{document}

\title{Homotopy-theoretic and categorical models of neural information networks}
\date{}
\author{Yuri I.~Manin}
\affiliation{Max Planck Institute for Mathematics, Bonn, Germany}
\thanks{this author passed away on January 7, 2023, 
when this paper was in the final phase of revision.}

\author{Matilde Marcolli}
\email{matilde@caltech.edu}
\orcid{0000-0002-2045-2907}
\affiliation{California Institute of Technology, Department of Mathematics and Department of Computing and Mathematical Sciences, Pasadena, USA}
\thanks{Partially supported by
NSF grants DMS-1707882 and DMS-2104330, by NSERC Discovery Grant RGPIN-2018-04937 
and Accelerator Supplement grant RGPAS-2018-522593, and by FQXi grants FQXi-RFP-1 804
and FQXi-RFP-CPW-2014, SVCF grant 2020-224047.}

\maketitle

\newcommand{\authorsforheader}{Manin and Marcolli}
\newcommand{\paperdoi}{https://doi.org/10.46298/compositionality-6-4}
\newcommand{\receiveddate}{2023-04-01}
\newcommand{\accepteddate}{2023-08-20}

\begin{abstract}
In this paper we develop a novel mathematical formalism for the modeling of 
neural information networks endowed with additional structure in the form
of assignments of resources, either computational or metabolic or informational.
The starting point for this construction is the notion of summing functors and 
of Segal's Gamma-spaces in homotopy theory. 
This paper analyzes functorial assignments of different levels of structure
(resources) to networks and their subsystems. Resources are described by
categories, involving concurrent/distributed computing architectures, binary codes, 
and associated information structures and 
information cohomologies, including a cohomological version of integrated information. 
A categorical form of the Hopfield network dynamics is introduced, which recovers the 
usual Hopfield equations when applied to a suitable category of weighted codes, 
where the variables of the dynamics are these functorial assignments of
resources to a network (summing functors).  
\end{abstract}

\tableofcontents

\section{Introduction and motivation}

The main goal of this paper is the development of a new mathematical formalism
for the modeling of networks endowed with several different levels of structure.
The types of structures considered are formalized in terms of categories, whose
objects represent various kinds of resources, such as computational architectures
for concurrent/distributed computing, codes generated by spiking activity of
neurons, probabilities and information structures, and possible other categories
describing physical resources with metabolic and thermodynamical constraints.
Morphisms in these categories represent ways in which resources can
be converted and computational systems can be transformed into one another.
All these different levels of structure are in turn related via several functorial mappings.
We model a configuration space of consistent ways of assigning such resources
to a network and all its subsystems, in the form of categories of ``network summing
functors'' with invertible natural transformations as morphisms. These provide a
categorical model of a moduli space of all possible assignments of resources to
subnetworks (subject to various types of constraints), considered up to equivalence.


It is useful to consider an analogy with the usual description of physical systems,
where one first introduces a suitable {\em configuration space}. This is the {\em kinematic}
part of the model, which describes the underlying geometry (variables and constraints),
in which the dynamics takes place. One then introduces the {\em dynamics} in the form
of an equation of motion given by a dynamical system on the assigned configuration space.
We are going to proceed along the same lines here. 
The categories of summing functors play the role of the physical configuration
space, which determines the geometry and kinematics of the model. Namely, the
basic variables of our model are the summing functors. Our physical configuration space
is then given by a category of summing functors with invertible natural
transformations, which describes these functorial assignments of resources up to
equivalence. We then introduce dynamical systems on these categories of summing functors, describing
the time evolution of the assignments of resources to the network with given constraints.
The main advantage of adopting this categorical viewpoint lies in the fact that
the entire system, with all its levels of structure, transforms simultaneously and
consistently (for example, consistently over all possible subsystems as well as
over all functorial relations between different layers of structure), under dynamical
evolution, and in the course of interacting with and processing external stimuli.
More precisely, we show that a discretized form of the Hopfield network dynamics
can be formulated in this categorical setting, thus providing an
evolution equation for the entire system of the network with
all its resources and constraints, and we show that one recovers
the usual Hopfield network dynamics when specializing this to
a category of weighted codes. 


The way we incorporate these different levels of structure is based on a notion
from homotopy theory, the concept of Gamma-space introduced by Graeme
Segal in the 1970s to realize homotopy-theoretic spectra in terms of symmetric
monoidal categories. A Gamma-space functorially maps finite sets to
simplicial sets, by assigning to a set the nerve of its category of summing functors
with target a fixed category of resources. We extend this notion of Gamma-spaces to
a similar notion of Gamma networks, which assign to a network a topological
model (the nerve) of our configuration space given by the category of network
summing functors. 
We view this as a functorial construction of a topological ``configuration space''
of all possible mappings of subsystems of a finite system to resources, in a way
that is additive on independent subsystems. The categorical dynamics we
introduce at the level of the category of summing functors induces a topological
dynamical system on the nerves obtained via the associated Gamma network.
Segal's Gamma-spaces extend to endofunctors of the category of simplicial
sets. As such they can be used to construct Gamma networks that have as
input certain simplicial sets that naturally arise in an activated network
responding to a stimulus, such as clique complexes or nerves of coverings
associated to receptor fields of neural codes, as well as simplicial sets
associated to various forms of categorical information structures. The output
is a new simplicial set that combines the topology of the input with additional
topological structures coming from the category of resources, through the
associated Gamma-space and spectrum. Thus, our configuration space
also acts as an encoder that takes as input homotopy types coming from
the activity of the network and produces as output a new collection of homotopy types,
that also incorporate the topology of the configuration space itself. We show that these
homotopy types have associated measurements of integrated information
and that this encoding of homotopy types increases the integrated
information by an amount described in terms of Shannon entropy of the
Gamma-space. 
In an appendix we enrich the formalism of summing functors and Gamma-spaces with
both probabilistic and persistent structures.

\subsection{Background motivation}

In the rest of this introductory section we review some general
background motivations behind the approach developed in this paper. The content
of the paper and the main results are then summarized in \S \ref{ResultsSec}.

\subsubsection{Cognition and computation}

A main motivation of this  paper, as well as of many others,
was briefly summarized in \cite{ManMan17}: it is the heuristic value of comparative study
of {\it ``cognitive activity''} of human beings (and more generally, other biological systems)
with  {\it ``computational processing''} by engineered objects, such as computing devices.


In \cite{ManMan17} it was stressed, in particular, that such a comparison should
be {\it not restricted,} but rather {\it widened}, by the existence of wide spectra
of space and time scales relevant for understanding both of
``cognition'' and ``computation''.


In particular, in \cite{ManMan17} it was argued that  we must not {\it a priori} decide that brain should be compared to a computer,
or neuron to a chip.  We suggested, that there exist fruitful similarities between spatio-temporal
activity patterns of a {\it one brain}  and {\it the whole Web}; or between similar patterns
on the levels of {\it history of civilizations}, and several functional neuronal circuits developing 
{\it in a brain of a single human being from birth to ageing.}


It was noticed long ago  that various mathematical models of such processes 
have skeletons of  common type: {\it oriented graphs} describing paths of transmission
and/or transformation of information. Mathematical machinery
of topological nature (geometric realization of graphs by simplicial complexes,
and their topological invariants) must be  connected, in such studies,
with mathematical machinery of information theory (probability distributions, entropy, 
complexity ...):  cf.~\cite{Mar19} and \cite{Hess}. 


The primary  goal of this paper consists in the enrichment of
the domain of useful tools: paths in oriented graphs can be considered as
compositions of morphisms between objects of categories, and assignments
of resources of different types (computational, metabolic, informational) to
networks can be regarded as functors between suitable categories. 
Topological invariants of their geometric realizations might include
{\it homotopical} rather than only (co)homological invariants.
Respectively, we continue studying their
possible interaction with information-theoretic properties started e.g.~in \cite{Mar19} and \cite{Man15}.


As in classical theoretical mechanics, such  invariants embody 
{\it configuration and  phase spaces} of systems that we are studying,
{\it equations of motion, conservation laws}, etc. 
In the setting we develop here, the main {\em configuration space} is the space of
all consistent functorial mappings of a network and its subsystems to a monoidal
category of resources (computational systems, codes, information structures).
As in the case of classical mechanics, this kinematic setup describing the
configuration space is then enriched with dynamics, in the form of 
categorical Hopfield networks. 


One can view classical mechanics in categorical terms as well, by considering the
assignment of configuration spaces to classical physical systems and their
subsystems. This is a useful viewpoint, for instance, when considering the physics of
open systems, and was developed in \cite{Baez}. The assignments of configuration
spaces to systems and subsystems form a category (of spans/cospans) of Riemannian
manifolds and  surjective Riemannian submersions in the Lagrangian formulation of
classical mechanics, and of symplectic manifolds with surjective Poisson maps in the
Hamiltonian formulation, with the Legendre transform relating the Lagrangian and 
Hamiltonian formalism realized functorially. The categorical setting we consider here
is different, but it has some aspects in common with this categorical formulation of
classical mechanics, in the sense of focusing on configuration spaces for systems and
subsystems, realized by our categories of summing functors. 

\subsubsection{Homotopical representations of stimulus spaces}

One of the main motivations behind the viewpoint developed in this paper comes from
the idea that the neural code generates a representation of the stimulus space
in the form of a homotopy type. 


Indeed, it is known from \cite{Cu17}, \cite{CuIt1}, \cite{CuIt2}, \cite{Man15}, \cite{Youngs} 
that the geometry of the stimulus space can be reconstructed {\em up to homotopy} from
the binary structure of the neural code. The key observation behind this reconstruction 
result is a simple topological property: the binary code words in the neural
code represent the overlaps between the place
fields of the neurons, where the place field is the preferred region of the stimulus space 
that cause the neuron to respond with a high firing rate. The neural code determines in this
way a simplicial complex, given by the simplicial nerve of the open covering of the stimulus space.
Under the reasonable assumption that the place fields
are  convex open sets, the homotopy type of this simplicial complex is the same as the homotopy type of 
the stimulus space. Thus, the fact that the binary neural code captures the complete
information on the intersections between the place fields of the individual neurons 
is sufficient to reconstruct the stimulus space, but only up to homotopy.


The homotopy equivalence relation in topology is weaker but also more flexible
than the notion of homeomorphism. The most significant topological invariants,
such as homotopy and homology groups, are homotopy invariants. Heuristically,
homotopy describes the possibility of deforming a topological space in a
one-parameter family. In particular, a {\em homotopy type} is an equivalence
class of topological spaces up to (weak) homotopy equivalence, which roughly 
means that only the information about the space that is captured by its homotopy
groups is retained. There is a direct connection between the formulation of topology
at the level of homotopy types and ``higher categorical structures''. 
Homotopy theory and higher categorical structures have come to play an 
increasingly important role in contemporary mathematics,
including important applications to theoretical physics  
and to computer science. We will argue
here that it is reasonable to expect that they will also play a role in the mathematical
modeling of neuroscience.  This was in fact already suggested by Mikhail Gromov in \cite{Gromov}.


This suggests that a good mathematical modeling of network architectures in the brain
should also include a mechanism that generates homotopy types, through the information 
carried by the network via neural codes. One of the main goals in this paper is to show
that, indeed, a mathematical framework that models networks with additional
computational and information structure will also give rise to a mechanism that acts on
homotopy types. The Gamma-spaces associated to 
our configuration spaces of assignments of resources to
networks are functors that take as inputs homotopy types generated by the network
activities (such as clique complexes of activated subnetworks, nerve complexes of
response fields and neural codes) and encode these inputs into another class
of homotopy type (which we call a representation). The new homotopy types obtained in
this way combine the nontrivial input homotopy
types that encode information about the stimulus space with topological
information about the categories of resources, in a way that increases informational complexity
(see \S \ref{IntegInfoGammaNet} and especially Proposition~\ref{IIHEQ}).

\subsubsection{Homology and stimulus processing}

Another main motivation for the formalism developed in this paper is the
detection, in neuroscience experiments and simulations, of a
peak of non-trivial persistent homology. This arises in the clique complex of the network
of neurons activated during the processing of external stimuli. A related motivation is
given by increasing evidence of a functional role of these nontrivial topological structures.


The analysis of the simulations of neocortical microcircuitry in \cite{Hess},
as well as experiments on visual attention in rhesus monkeys \cite{Rouse},
have shown the rapid formation of a peak of non-trivial homology generators 
in response to stimulus processing. These findings are very intriguing for two reasons: 
they link topological structures in the activated neural circuitry to phenomena like 
attention, and they suggest that a sufficient amount of topological complexity serves
a functional computational purpose. 


This suggests a possible mathematical setting for modeling neural information
networks architectures in the brain. The work of \cite{Hess}
proposes the interpretation that these topological
structures are necessary for the processing of stimuli
in the brain cortex, but does not offer a theoretical explanation of why
topology is needed for stimulus processing. However, there
is a well-known context in the theory of computation
where a similar situation occurs, which may
provide the key for the correct interpretation, namely 
the theory of concurrent and distributed computing \cite{Bub12}, \cite{FaRaGou06}, \cite{Herl1}. 


In the mathematical theory of distributed computing, one considers a collection of sequential 
computing entities (processes) that cooperate to solve a problem (task).
The processes communicate by applying operations to objects
in a shared memory, and they are 
asynchronous, in the sense that they run at arbitrary varying speeds. 
Distributed algorithms and protocols decide how and when each process 
communicates and shares with others. The main questions are how to 
design distributed algorithms that are efficient in the presence of noise,
failures of communication, and delays, and how to understand when a 
distributed algorithm exists to solve a particular task. 


Protocols for distributed computing can be modeled using simplicial sets.
An initial or final state of a process is a vertex, any $d+1$ mutually compatible initial 
or final states are a $d$-dimensional simplex, and each vertex is labeled by a different 
process. The complete set of all possible initial and final states is then a simplicial set.
A decision task consists of two simplicial sets of initial and final states and a simplicial map (or
more generally correspondence) between them. The typical structure describing a 
distributed algorithm consists of an
input complex, a protocol complex, and an output complex, with a certain number of 
topology changes along the execution of the protocol, \cite{Herl1}. 


There are very interesting topological obstruction results in the theory of distributed
computing, \cite{Herl1}, \cite{HeRa95}, which show that a sufficient amount of non-trivial
homology in the protocol complex is necessary for a decision task problem to be solvable.
Thus, the theory of distributed computing shows explicitly a setting where a sufficient
amount of topological complexity (measured by non-trivial homology) is necessary
for computation.


This suggests that the mathematical modeling of network architectures in the brain
should be formulated in such a way as to incorporate additional structure keeping track of associated 
concurrent/distributed computational systems. This is indeed one of the main aspects of the formalism
described in this paper: we will show how to associate functorially to a network and its subsystems
a computational architecture in a category of transition systems, which is suitable for the modeling
of concurrent and distributed computing.  For additional discussion of topological and categorical
models of concurrent and distributed computing see for instance
\cite{Bub12}, \cite{BubWor06}, \cite{Gau03}, \cite{Gau00}, \cite{Gau08}, \cite{GoHaKr10} 
\cite{HeRa00}.

\subsubsection{Informational complexity and integrated information}

In recent years there has been some serious discussion in the neuroscience
community around the idea of possible computational models of consciousness 
based on some measure of informational complexity, in particular in the form of
the proposal of Tononi's {\em integrated information} theory (also known as the
$\Phi$ function) \cite{Tono}, see also \cite{Koch}, \cite{MasTon} for a general overview.
This proposal for a quantitative correlate of consciousness roughly measures
the least amount of effective information in a whole system
that is not accounted for by the effective information of its separate parts.
The main idea is therefore that integrated information is a measure of informational
complexity and causal interconnectedness of a system. 


This approach to a mathematical modeling of
consciousness has been criticized on the ground that it is easy to construct simple
mathematical models exhibiting high values of the $\Phi$
function. Generally, one can resort to
the setting of coding theory to generate many examples of sufficiently good codes
(for example the algebro-geometric Reed--Solomon error-correcting 
codes) that indeed exhibit precisely the typical form of high causal 
interconnectedness that leads to large values of integrated information. This
indicates that integrated information alone does not suffice to imply consciousness.
Thus, it seems that it would be preferable to interpret integrated information
as a consequence of a more fundamental model of how networks in the brain 
process and represent stimuli, leading to high informational complexity and causal
interdependence as a necessary but not in itself sufficient condition. 


One of the goals of this paper is to show that integrated information can
be incorporated as an aspect of the model of neural information network
that we develop, and that many of its properties, such as the low values
on feedforward architectures, are already built into the topological structures
that we consider. One can then interpret the homotopy types generated by the topological
model we consider as the ``representations'' of stimuli produced by the network
through the neural codes, and the space of these homotopy types as a kind
of ``qualia space'', \cite{MarTsao18}. While we
will not pursue in the present paper the development of such a model of qualia,
this motivation lies in the background of some of the results on integrated
information that we obtain in this paper, in particular our result on the gain in
integrated information caused by the encoding of homotopy types through Gamma-spaces.

\subsubsection{Perception, representation, computation}

We conclude this overview of motivational background by some broader
and more general considerations. 
At these levels  of generalization, additional challenges
arise, both for researchers and students. Namely, even when we focus
on some restricted set of observables, passage from one space/time scale
to a larger or smaller one might require a drastic change 
of languages we use for description of these levels. The typical
example is passage from classical to quantum physics. In fact,
it is only one floor of the Babel Tower of imagery that humanity uses
in order to keep, extend and transmit the vast body of knowledge,
that makes us human beings: cf.~a remarkable description of this in
\cite{HF14}.


Studying neural information, we meet this challenge,
for example, when we try to pass from one subgraph of the respective
oriented graph to the next one by adding just one oriented arrow
to each vertex. It might happen that each such step 
{\it implies a change of language}, but in fact such languages themselves
cannot be reconstructed before the whole process is relatively well
studied.


Actually, the drastic change of languages arises already in the passage between two wide communities
of readers to which this paper is addressed: that of mathematicians
and that of neuroscientists. 
Therefore, before moving to the main part of this paper, we wanted 
to make the mathematicians among our readers 
aware of this necessity of permanent change of languages.


A very useful example of successful approach to this problem is the book \cite{Sto18},
in particular its Chapter~5, ``Encoding Colour''.
Basically, this Chapter explains {\it mathematics} of color perception,
by the retina in the human eye. But for understanding its {\it neural
machinery}, the reader will have to return temporarily back each time, 
when it is necessary. Combination of both is a good lesson in 
neural information theory.
Below we will give a brief sketch of that chapter.


Physics describes light on the macroscopic level as a superposition of
electromagnetic waves of various lengths, with varying intensity.
Light perception establishes bounds for these wavelengths,
outside of which they stop to be perceived as light.
Inside  these bounds, certain bands may be perceived
as light of various ``pure'' colors: long wavelengths ({\it red}),
medium wavelengths ({\it green}), and short wavelengths ({\it blue}).


The description above refers to the ``point'' source of light. The picture
perceived by photoreceptors in the eye and transmitted to neurons in the brain,
is a space superposition of many such ``point source'' pictures,
which then is decoded by the brain as a ``landscape'', or a ``human face'',
or ``several barely distinguishable objects in darkness'', etc. 


We will focus here upon the first  stages of this encoding/decoding of an
image in the human eye
made by the retina. There are two types of photoreceptors in the retina:
cones (responsible for color perception in daylight conditions) and rods 
(providing images under night-time conditions).


Each photoreceptor (as other types of neurons) receives information 
in the form of action potentials
in its cell body, and then transmits it via its axon (a kind of ``cable'')
to the next neuron in the respective neuronal network. 
Action potentials
are physically represented by a flow of ions. Communication between two
neurons is mediated by synapses (small gaps, collecting ions from several
presynaptic neurons and transferring the resulting action potential into the cell body
of the postsynaptic neuron).


Perception of visual information by the human eye starts with light
absorption by (a part of) the retinal photoreceptors and subsequent
exchange of arising action potentials in the respective part of the neural network.
Then retinal ganglion cells, forming the optic nerves, transmit
the information from the retina to the brain.


Encoding color bands into action potentials, and subsequently encoding relative
amplitudes of respective potentials into their superpositions,
furnish the first stage of ``color vision''. Mathematical modeling of this
stage in \cite{Sto18} requires a full machinery of information theory
and of chapters of statistical physics involving entropy and its role
in efficient modeling of complex processes.


Our focus here is more abstract and general, as we deal with a formalism
for describing networks endowed with different types of resources
related by certain mutual constraints. The steps of encoding information
coming from external stimuli can be regarded as a way of assigning codes
to a network and probabilities and information measures to these resulting
codes.  Enrichment of all these models
by topology, or vice versa, enrichment of topology by information formalisms
plays an important role in our approach, as we will be discussing in the
rest of the paper. 

\subsection{Structure of the paper and main results}\label{ResultsSec}

In \S \ref{SummingSec} we introduce the general problem of modeling
networks with associated resources. We present our main configuration
spaces, parameterizing assignments of resources to networks, given by
categories of summing functors. In 
\S \ref{CatSumFunctSec} we first present the case of 
categories of summing functors from the category of
subsets of a finite set with inclusions to a category
of resources, which is a category with sums and zero object, or
more generally a symmetric monoidal category. We think
here of the finite set as representing either 
the set of vertices (nodes) or of edges of a network. We give a
simple characterization of these summing functors. 
In \S \ref{NetSumFuncSec} we extend this notion by incorporating the
network structure. Instead of considering finite sets, in \S \ref{GraphsSec}  we consider
directed finite graphs, seen as functors from a category with two
objects $V$, $E$ and two non-identity morphisms, source and target,
$s,t: E\to V$. We introduce two preliminary examples of
network summing functors, where the compatibility between
vertices and edges of the directed graph is described via 
either an equalizer or a coequalizer construction.
In \S \ref{SumNetSec}  we introduce our more general definition
of ``network summing functors'' and we show in \S \ref{EqSec}
and \S \ref{CoeqSec} that the equalizer and coequalizer examples 
determine subcategories of the category of network summing functors.
We then show that other subcategories of interest can be identified
by specifying other forms of additional constraints at vertices and edges 
that the network summing functors should satisfy. In particular, in \S \ref{GraftingSec} 
we describe the case of network summing
functors that are obtained through grafting operations, in cases
where the category of resources has an additional compositional
structure described by a properad. 
In \S \ref{ExInclSec} we describe another class of
network summing functors, which satisfy inclusion-exclusion
relations, in cases where the category of resources is either
abelian or triangulated. These cases are presented to illustrate
the fact that specific subcategories of our category of summing
functors may be suitable for different types of models, depending
on what kinds of resources on networks one is describing. 


In \S \ref{NetsResourcesSec} we analyze more closely the
notion of category of resources. We recall in \S \ref{MetabSec} various forms of resources
that are associated to neuronal networks, in particular informational
and metabolic constraints and computational resources. We then review in 
\S \ref{ResourcesSec} the mathematical theory of resources and convertibility
of resources developed in  \cite{CoFrSp16} and \cite{Fr17} using symmetric 
monoidal categories. We recall in \S \ref{ExResSec} some simple examples
of categories of resources, from  \cite{CoFrSp16} and \cite{Fr17}.
We discuss briefly in \S \ref{MeasSemigrSec} the notion of measuring
semigroups associated to categories of resources, which was also introduced in
\cite{CoFrSp16} and \cite{Fr17} to keep track of resource convertibility. We will
be using this notion of measuring
semigroup to define the threshold-dynamics of Hopfield
networks in our categorical setting in \S \ref{HopfieldSec}. 
In \S \ref{InfoLossSec} we also recall the categorical characterization of
information loss of \cite{BaFrLei11}. In \S \ref{AdjunctSec} we describe
how adjunction of functors can be viewed in this setting as optimization of
resources. This particular observation is not directly needed for our
applications, but we have included it because it provides some further
insight and intuition about the categorical formalism in discussing
resources. 


In \S \ref{GammaNetCompSec} we look more specifically into how to model
assignments of computational structures as resources attached to networks. 
We focus in \S \ref{CompResSec} on one particular categorical model of
computational resources for concurrent and distributed computing
architectures, given by the category of transition systems of \cite{WiNi95}.
While there are many categorical models of concurrent and distributed computing,
we have chosen this one as it is sufficiently flexible to accommodate various
existing computational models of individual neurons, and at the same time
it has a simple structure that makes it clear the category has the required
properties of a category of resources in the sense recalled in \S \ref{ResourcesSec}. 
In \S \ref{GammaCompArchSec} we mention briefly some of the existing
approaches to computational models of individual neurons and how they
can be made to fit in the category of transition systems, though a more
detailed account for specific neuron models will be given elsewhere, \cite{Mar-new}. 
In \S \ref{CompArchSec} we introduce a class of summing functors
obtained via grafting operations in the category of transition systems,
which provides a good configuration space in this setting. We finish
this section with some subsections aimed at illustrating interesting
possible directions of investigation related to this type of resources and
summing functors: in \S \ref{NeuromodSec} we outline the problem of
including in this setting a good computational model of neuromodulation.
In particular for this specific problem, we discuss in \S \ref{TimeDelaySec} 
how one can use a class of automata with time delays as transition systems.
We finish in \S \ref{TopQuesSec} with some questions on the possible
role of the $3$-dimensional topology of the network and of topological invariants
that depend on the $3$-dimensional embedding of graphs. 


In \S \ref{GammaCodesSec} we consider neural codes
generated by networks of neurons and associated
probabilities and information measures. We introduce
neural codes in \S \ref{CodesProbSec} and we recall
their main structure and properties. In 
\S \ref{CodesCatSec} we construct
a simple category of codes and we show that one
can think of the neural codes as determining summing
functors to this category of codes. 
In \S \ref{ProbCodesSec} we then
show that the probabilities associated
to neural codes by the firing frequencies of the neurons
fit into a functor from this category of codes
to a category of probability measures. 
However, we
show that this construction is not fully satisfactory because
it does not in general translate to a functorial assignment
of information measures (see \S \ref{InfomeasSec}). 
In \S \ref{LinNeurSec} we show
that our setting, with a category of weighted codes, recovers
a simple model of the linear neurons. (We discuss threshold-nonlinearities
in \S \ref{HopfieldSec}.) 
The problem with functorial assignments of information measures is 
solved in \S \ref{GammaNetInfoSec}, 
using the more sophisticated formalism of
cohomological information theory introduced by Baudot and Bennequin~\cite{BauBen1}
and developed by Vigneaux \cite{Vign}.
In \S \ref{CohomInfoSec} we give a very quick review of the
cohomological information setting of  \cite{Vign}, with finite information
structures, probability functors, and the Hochschild cohomology
interpretation of information functionals like Shannon and Tsallis
entropy. We start in \S \ref{NetSumInfoSec}, by considering the subcategory
of network summing functors given by the equalizer condition, 
discussed in \S \ref{EqSec}. We construct a functor from the category
of codes to the category of finite information
structures, and from there to an abelian category of modules
as in \cite{Vign}.  We obtain an associated category of summing functors
by composition. These describe assignments of informational resources
to the network. We show that these satisfy inclusion-exclusion 
properties as discussed in \S \ref{ExInclSec}. A variant of this
construction is described in \S \ref{InfoChSetsSec}, with a functor
from codes to information structures and then to the category of chain 
complexes, and a resulting category of network summing functors. 
In \S \ref{CodesNerveSec} we show that the formalism of
finite information structures and probability functors of \cite{Vign} 
incorporates as a particular case the assignment to a neural code
of the simplicial set given by the nerve of the covering associated
to the receptor fields of the neurons.
In \S \ref{TransCodesInfoSec} we further refine these functorial
relations between the different categories of resources
introduced in the previous sections by constructing a functor
from the category of transition systems to the category of codes,
describing the codes generated by the automata. We also
construct a functor from transition systems to information structures,
and we show that it agrees with the composition of the
functor to codes with the functors from codes to information
structures described in \S \ref{NetSumInfoSec}. 
We show in \S \ref{CliqueInfoSec} that we can also fit into
the formalism of finite information structures and probability functors 
the clique complexes of networks, by exhibiting a specific
choice of finite information structures and probability
functor for which the output simplicial set is the clique complex. 
These various cases are meant to show the functorial consistency
between the various categories of resources of interest to us
(neural codes, computational systems, information structures)
and how significant examples of topological structures
associated to neuronal networks, such as nerves of coverings
of receptor fields and clique complexes of activated subnetworks,
fit inside the same broader formalism. 


In all the sections of the paper up to this point we have
only dealt with a static model, in the sense that we have
focused on constructing the configuration space 
parameterizing the assignment of resources to a network
and the relations between these configuration spaces
determined by the relations between different types of
resources. In \S \ref{HopfieldSec} we make the setting
dynamical, in the sense that we introduce equations of
motion on our kinematic space. This is done by introducing
a suitable form of the Hopfield equations of networks 
which is categorical in the sense that the variables of
the equation are now summing functors. 
We start by recalling in \S \ref{ContDiscrHopfSec}
the classical Hopfield equations of networks, in both
the continuous and the discretized form. The
equations are non-linear due to the presence of
a threshold non-linearity that accounts for the
non-linear properties of neurons. In 
\S \ref{CatNonlinSec} we discuss how one
can formulate threshold non-linearity in a
categorical setting using the measuring
semigroups on categories of resources, that
we recalled in \S \ref{MeasSemigrSec}. 
We then formulate in 
\S \ref{DiscrHopfSec} the categorical
form of the Hopfield equations with
variables that are summing functors and
the dynamics determined by an
endofunctor and by the threshold non-linearity. 
We show that the resulting dynamics in the
category of endofunctors induces a topological
dynamical system on the associated nerve, which
can be used to study the dynamics through traditional
topological dynamical systems methods. 
While in the present paper we do not present a
detailed study of the properties of these equations,
which is left to future work, 
we do discuss in \S \ref{DynWCodesSec} a basic
consistency check, by showing that, in a
very special case with the category of
resources given by our category of weighted
codes of \S \ref{LinNeurSec}, one recovers the
classical Hopfield equations of networks. This
in particular shows how to extend the result of
\S \ref{LinNeurSec} from the over-simplistic
linear neuron to a more realistic non-linear case. 


In \S \ref{GammaGeneralSec} we introduce another
level of structure, focused more on simplicial sets and homotopy types.
We have already seen the role of the nerve of the category of
summing functors in discussing the Hopfield equations in \S \ref{DiscrHopfSec}, 
as an associated topological dynamical system.
We focus here more generally on functorial assignments of
simplicial sets to networks. We present these through the classical
Segal construction of Gamma-spaces, which are functorial assignments
of simplicial sets to finite sets, through the construction of the
nerve of a category of summing functors. We think of these nerves
as the geometric realizations of our categorical configuration spaces.
In \S \ref{GammaSegalSec} we review Segal's notion of Gamma-spaces
and the construction of Gamma-spaces associated to categories of
resources. We then recall in \S \ref{GammaSpectraSec} how Gamma-spaces
extend to endofunctors of the category of simplicial sets through a
coend construction. In \S \ref{HtpyTypesSec} we observe how, 
correspondingly, a Gamma-space generates a collection of
homotopy types from input simplicial sets. In 
\S \ref{SpectraHTSec} we recall the relation between Gamma-spaces
and homotopy-theoretic spectra. In \S \ref{GammaCliqueSec},
\S \ref{randomgraphsec}, and \S \ref{FeedSec1} we discuss certain
special cases that are useful in preparation for the more general discussion in 
\S \ref{GammaNetSec}. In particular, in 
\S \ref{GammaCliqueSec} we analyze the topological
properties of the output simplicial sets when the input 
of the Gamma-space endofunctor is a clique complex of a network;
in \S \ref{randomgraphsec} we present a similar discussion for the
case where the input is the (un-oriented) clique complex of an
Erd\H{o}s--R\'enyi random graph; while in \S \ref{FeedSec1} we discuss
briefly the case of feedforward networks. In 
\S \ref{GammaNetSec} we then introduce our notion of Gamma networks,
which generalizes Gamma-spaces, as functors from directed graphs to
simplicial sets, and we focus on two main classes of Gamma networks:
those obtained by composing a functorial assignments of simplicial
sets to graphs (such as clique complexes or assignments coming
from probability functors) with a classical Gamma space, and those
obtained by taking the nerve of a category of network summing
functors. Combinations (via smash product of Gamma-spaces)
of these two types cover most of the needs for our model. The
special cases discussed in \S \ref{GammaCliqueSec},
\S \ref{randomgraphsec}, and \S \ref{FeedSec1} are all examples
of the first kind. 


In  \S \ref{IntegInfoSec} we enrich our setting with a notion of
integrated information. This is a notion of informational complexity
of a system, such as a network with resources in our setting,
which is designed to capture the amount of information carried by
the system that cannot be accounted for in terms of any
partition into independent subsystems. In that sense it is a measure
of both information and causal interrelatedness between subsystems.
Integrated information has been introduced in neuroscience 
(see \cite{BaTon}, \cite{Koch}, \cite{MasTon}, \cite{Tono}) as a possible
quantitative correlate of consciousness. We are interested here in
how two aspects of our model affect integrated information: our
categorical Hopfield dynamics, and the mapping of simplicial sets
via Gamma networks. In \S \ref{InfoGeomSec} we recall the
mathematical formulation of integrated information, using the
construction of \cite{OizTsuAma}, based on information geometry.
As an example of the type of structure that integrated information
detects, we recall in \S \ref{FeedSec2} the reason why it is
trivial on feedforward network architectures. In 
\S \ref{KLintifoSec} and \S \ref{IIHSec} we present a way of formulating
integrated information in the setting of cohomological information 
theory of \cite{Vign}, by first recalling in \S \ref{KLintifoSec} how the Kullback--Leibler
divergence is formulated in that formalism, and then presenting in
\S \ref{IIHSec} our cohomological construction of integrated information.
In \S \ref{CatDynIIHSec} we show that we can assign a measurement of
integrated information to the summing functors that are solutions of our
categorical Hopfield equation, in such a way as to keep track of the
changes in integrated information along the dynamics. 
In \S \ref{IntegInfoGammaNet} we consider Gamma networks that
are obtained as composition of a probability functor from a 
category of random graphs with a classical Gamma-space, where the 
Gamma-space 
accounts for the type of resources associated to the network. We show
that there is an associated cohomological integrated information
and that this form of integrated information increases under
composition with the Gamma-space, by an amount described
in terms of Shannon entropy associated to the Gamma-space. 
This shows that the encoding of homotopy types affected by a
Gamma-space increases the amount of integrated information 
they carry. We conclude this section by formulating in 
\S \ref{HTCodesInfoSec} some questions about the possible
role of generalized cohomologies associated to the spectra
defined by Gamma-spaces in combination with the cohomological
formulation of information functionals. 


The Appendix discusses two generalizations of the summing
functors and Gamma-spaces formalism, one that incorporates probabilities
and one that incorporates persistent structures. In 
\S \ref{Simplprob}  we present a general setting for the
categorical formulation of probabilities and its specialization to
the simpler case of probabilities over finite sets. In 
\S \ref{ProbGammaSec} we show how to use this category of
probabilities to construct a probabilistic version of Gamma-spaces, following
the setting of \cite{Mar19}. As an example we describe the case of probabilistic
transition systems in \S \ref{ProbTransSec}. In 
\S \ref{persGamma} we describe how to include a notion of persistence
for Gamma-spaces and corresponding persistent spectra. In
\S \ref{probpersGamma} we show that these two generalizations can
be combined to obtain Gamma-spaces that are both probabilistic and
persistent. We discuss in \S \ref{Persist1Sec} and \S \ref{Persist2Sec} 
how these generalizations can be useful to
incorporate descriptions of constraints and of time and scale
dependence. Finally, in \S \ref{VarNerveSec} we also discuss
briefly the possible role of generalizations of the nerve construction.  

\subsubsection{Comparison with other approaches}\label{CompareOtherSec}

The idea of considering assignments of various types of data to networks,
as well as the use of topological methods, have also been considered 
in other forms, for example along the lines of constructions involving
bundle/sheaf-theoretic methods. These include, for instance, the approach
of \cite{ScoPe}, \cite{ScoPe2}, based on vector bundles, with a notion of
approximate and discrete Euclidean vector bundle and a dimensionality-reduction 
method for large data sets based on embeddability of such bundles.
Such a construction can be organized in a categorical form, and it encodes topological
information about the data sets. Another viewpoint that pursues similar ideas
is the cellular sheaves method of  \cite{HaGr}, that extends spectral graph theory
to a spectral theory (with a Hodge Laplacian) on cellular sheaves of vector spaces
over cell complexes. When considered over graphs, this allows for assignments
of data to networks, encoded by vectors, with applications to distributed
algorithms, such as consensus problems, or distributed optimization. This
sheaf-theoretic context also has a natural categorical formulation.


Some of the motivations for adopting the type of construction described
in this present paper, with summing functors and Gamma-spaces, rather
than a simple elaboration on one of the pre-existing approaches mentioned
above, are summarized by the following observations.
\begin{enumerate}
\item Not all optimization problems are reducible to real-valued (or 
vector-valued)
functions: there are more general settings where one deals with objects in more
abstract categories. A general discussion of such categorical notions of optimization
is given in \cite{Mar-Pareto}.
\item A discussed briefly in \cite{Mar21}, our process of building homotopy types from 
network Gamma spaces provides a unifying context where several different 
occurrences of simplicial sets and homotopy arising in a neuroscience-related setting
are simultaneously accounted for. For example, three different roles of topology
in neuroscience are clique complexes of subnetworks that activate in response to
stimuli, nerve complexes of neural codes that encode homotopy types of external
stimuli, and simplicial sets of probabilities in information structures. We will see that 
these are all accounted for simultaneously in the same formalism, through the construction
of simplicial sets and homotopy types through Gamma-spaces.
\item The functoriality of the construction (through categories of summing functors)
allows for the possibility of describing simultaneously several different types of
assignments to networks, such as computational architectures (automata), neural codes, 
information structures, along with (functorial) relations between them, in such a way
that the dynamics simultaneously involves all these levels of structure, compatibly with
their relations.
\item In addition to direct applications to models of neuronal networks,
the formalism considered here makes it possible also to study dynamical
systems with threshold non-linearity in other categories, of independent
interest in other mathematical setting. An example related to rational
points on arithmetic algebraic varieties and ``invisible varieties'', inspired 
by our previous work \cite{ManMar21}, will be discussed separately, in
a forthcoming paper.
\end{enumerate}


While this paper is mostly dedicated to presenting the general
construction and its properties, specific examples of the resulting
categorical Hopfield dynamics are described in detail in \cite{Mar-new},
where a very simple example of threshold  non-linear dynamics is 
presented with resources given by a category of deep neural networks (DNN).
It is shown that the simplest possible case of Hopfield dynamics with 
that category of resources reproduces, in a functorial form, the
backpropagation mechanism for the weights of the DNN based on 
gradient descent. Other explicit examples of categorical Hopfield
equations with different categories of resources will be presented
separately. 

\section{Summing functors on networks}\label{SummingSec}

In this section we introduce the main formalism we will be using
in the modeling of networks with associated resources and their
dynamical behavior. Namely we construct certain ``moduli spaces''
(described by categories) parameterizing all possible assignments
of resources of a given type (also described by categories) to
a network and its subsystems.  These categories of summing
functors provide our configuration space attached to a network.
The focus of most of this paper will be on understanding
relations between these configuration spaces for various
specific choices of categories of resources, representing
neuronal computational architectures, neural codes, and information
structures, and in introducing equations on these configuration
spaces describing the dynamical evolution of the network
and its resources.

\subsection{The category of summing functors}\label{CatSumFunctSec}

Let $\cC$ be a category with a categorical sum (coproduct) and a zero object. 
A zero object is an object $0 \in {\rm Obj}(\cC)$ that is both initial and terminal, namely
for any object $C\in {\rm Obj}(\cC)$ there is a unique morphism $0\to C$ and a unique
morphism $C\to 0$. The categorical sum $C_1\oplus C_2$ is characterized by the
following universal property. There are morphisms $\iota_i: C_i\to C_1\oplus C_2$
such that, for any object $C\in {\rm Obj}(\cC)$ and any pair of morphisms $f_i: C_i \to C$, 
there exists a unique morphism $f: C_1\oplus C_2 \to C$ such that the following
diagram commutes
$$ \xymatrix{ & C & \\ C_1 \ar[ur]^{f_1} \ar[r]_{\iota_1} & C_1\oplus C_2 \ar[u]^f & C_2\, . \ar[l]^{\iota_2} \ar[ul]_{f_2} } $$


More generally, one can consider categories $\cC$ that are unital symmetric
monoidal categories. This is especially relevant in view of interpreting $\cC$ 
as a category of resources, in the sense we will discuss in \S \ref{NetsResourcesSec}.
The main point in the paper where we will need to work with this more
general setting of unital symmetric monoidal categories, instead of restricting to
the case of categories with zero object and sum, is when we introduce the
categorical Hopfield dynamics in \S \ref{HopfieldSec}. 
In the setting of \cite{Tho95}, which we will refer to in \S \ref{GammaGeneralSec}, morphisms in the
category of small symmetric monoidal categories are taken to be 
lax symmetric monoidal functors, that is, functors $F: \cC \to \cC'$ together with
a natural transformation $f: F(A)\oplus F(B)\to F(A\oplus B)$ with commutativity
of the diagrams determining the compositions $F(\alpha) \circ f \circ (1\oplus f) =f\circ (f\oplus 1) \circ \alpha$,
with $\alpha: F(A)\oplus (F(B)\oplus F(C))  \to  (F(A)\oplus F(B))\oplus F(C)$ the associativity natural isomorphism,
and $F(\gamma)\circ f=f\circ \gamma$, with $\gamma: F(A)\oplus F(B)\to F(B)\oplus F(A)$ the
commutativity natural isomorphism. In our setting it is preferable to 
work with strict symmetric monoidal functors, where the natural transformation $f$ is the identity.

In the following, we will refer to symmetric monoidal categories, without making explicit
the unital condition, except where it is explicitly used, as in the setting of categories of resources
mentioned above.


Most of the cases we will be discussing in the following sections fit into the
stronger case of a category $\cC$ with sums and zero object. These include the 
category of computational systems as in \S \ref{CompResSec}, a category of neural 
codes as in \S \ref{CodesCatSec}, or categories of information structures 
as discussed in \S \ref{GammaNetInfoSec}. Thus, we will assume throughout our
discussion that $\cC$ has sum and zero object, except where we need to adopt
the more general setting of unital symmetric monoidal categories, as in \S \ref{HopfieldSec}. 


Let $X$ be a finite pointed set, with $*$ denoting the base point. For most of this section
we do not need to work with pointed sets, but the presence of base points will become
relevant for the homotopy-theoretic constructions used in \S \ref{GammaGeneralSec} and
\S \ref{SpectraHTSec}. Adding a base point should simply be regarded as a computational artifact 
(introduced for the purpose of homotopy theory),
while the ``relevant set'' is just the complement $X\smallsetminus \{ * \}$. 


The notion of summing functors was introduced in \cite{Segal} (see also \cite{Carlsson})
in the construction of Gamma-spaces, which we will discuss in \S \ref{GammaGeneralSec}.

\begin{defn}\label{SumFunctDef}
Let $P(X)$ denote the category whose objects are pointed
subsets $A\subset X$ with morphisms given by inclusions.
A summing functor $\Phi_X: P(X) \to \cC$ is a functor with
the property that the object $\{ * \}$ of $P(X)$ has image
$\Phi_X(\{ * \})=0$, the zero object of $\cC$, and for any
$A,A'\in {\rm Obj}(P(X))$ with $A\cap A'=\{ * \}$ one has
\begin{equation}\label{summing}
\Phi_X(A\cup A')=  \Phi_X(A)\oplus \Phi_X(A')\, . 
\end{equation}
\end{defn}


In the following, we will interpret the complement $X\smallsetminus \{ * \}$
as describing a certain system of neurons, with $A\subset X$ ranging over
all possible choices of subsystems $A\smallsetminus \{ * \}$. 
A summing functor $\Phi_X: P(X) \to \cC$ describes a way of assigning 
to every subsystem $A$ a corresponding object $\Phi_X(A)$ in the category $\cC$. 
The target category $\cC$ represents a certain type of resources, either computational 
architectures, describing resources of concurrent or distributed computing in the form 
of the category of transition systems  described in \S \ref{CompResSec}, or other forms 
of resources associated to the neurons. 
The summing-functor property \eqref{summing}, that a union of two disjoint sets 
(which after adding a basepoint means $A\cap A'=\{ * \}$)
is mapped to the coproduct $\Phi_X(A)\oplus \Phi_X(A')$, describes
the requirement that this assignment of resources is 
{\em additive on independent subsystems}.


Summing functors are themselves organized into a category, which is a subcategory of
the category of functors ${\rm Func}(P(X),\cC)$.


\begin{defn}\label{SumFunctCatDef}
Let $\cC$ be a category with sums and zero object.
The category $\Sigma_\cC(X)$ of summing functors has objects the 
summing functors $\Phi_X: P(X) \to \cC$ as in Definition~\ref{SumFunctDef}
and morphisms given by the {\rm invertible} natural transformations.
\end{defn}


Note that if we allow all natural transformations as morphisms rather than
restricting to only the invertible ones, the resulting category would not be
interesting in a topological sense, since the nerve would be contractible,
given that the category $\cC$ has a zero object so the category of summing functors
has an initial object. Restricting to only invertible natural transformations as
morphisms precisely avoids having an initial or terminal object in the category
of summing functors, hence allowing for non-contractible topologies:
with this restriction to invertible natural transformations,
the nerve of the category $\Sigma_\cC(X)$ of summing functors becomes topologically
very non-trivial, as we will recall more in
detail in \S \ref{GammaGeneralSec} and \S \ref{SpectraHTSec},  Indeed, it 
was shown in \cite{Tho95} that, for $\cC$ ranging over
symmetric monoidal categories, the nerves of the corresponding categories
of summing functors generate (in a sense we will make more precise in
\S \ref{GammaGeneralSec}) all connective spectra. In our
perspective it is a feature of the model to be able to generate a large supply of sufficiently complex homotopy types
(this will be further discussed in \S \ref{SpectraHTSec} and in following work, see also \cite{Mar21}, \cite{MarTsao18}). 


Another reason why it is desirable, in our setting, to restrict morphisms between
summing functors to be invertible natural transformations is that we want to
interpret summing functors as consistent assignments of resources to a system.
Invertible natural transformations identify which of such assignments should be
regarded as {\em equivalent} to each other. So we can interpret $\Sigma_\cC(X)$
as a categorical ``moduli space'' of all possible such assignments up to
equivalence. Note, however, that classical geometric intuition here may 
be misleading, as one does {\em not} take a quotient by equivalence:
one simply maintains all the equivalences explicitly as morphisms in the category. 
A better intuition is provided by the notion of ``action groupoid'':
given a space $\Omega$ with a group action by a group $G$, instead of considering the
quotient $\Omega/G$ where points in the same orbit are identified, one
considers the action groupoid (sometimes denoted by $\Omega / / G$),
which is a category with objects the points $\omega\in \Omega$ and
morphisms the elements $(g,\omega)\in G \times \Omega$ with source
 $s(g,\omega)=\omega$ and target $t(g,\omega)=g\cdot \omega$.
This construction ``resolves'' the quotient $\Omega/G$ in the sense that
the identifications of points in $\Omega/G$ are replaced by (invertible) 
morphisms in the category $\Omega / / G$. It is well known that the
action groupoid $\Omega / / G$ is a better behaved notion of quotient
than $\Omega/G$ in the case of non-free actions \cite{BrownGrpds}. 
Thus, one should view here the category $\Sigma_\cC(X)$ of summing functors
as playing a similar role as the action groupoids, in describing assignments of 
resources to subsystems of $X$ and keeping track of their equivalence structure.


Note that the summing condition \eqref{summing} gives an equivalent and
very simple description of summing functors, stated as follows.


\begin{lem}\label{PhiPtsLem} Let $\cC$ denote a category with sums and zero object.
\begin{enumerate}
\item A summing functor $\Phi_X: P(X) \to \cC$ as in Definition~\ref{SumFunctDef}
is completely determined by its values $\Phi_X(x):=\Phi_X(A_x)$ 
on the sets $A_x=\{ x, * \}$ for $x\in X\smallsetminus \{ * \}$.
\item Let $\hat\cC$ denote the category with the same objects as $\cC$ and
with morphisms the invertible morphisms of $\cC$. For $X$ a finite
pointed set with $\# X =n+1$, the category $\Sigma_\cC(X)$ of
summing functors is equivalent to $\hat\cC^n$, the $n$-fold product 
with objects the $n$-tuples of objects in $\hat\cC$
and morphisms the $n$-tuples of arrows (invertible morphisms) with component-wise composition.
\end{enumerate}
\end{lem}

\begin{proof}
  The first statement is obtained directly from Definition~\ref{SumFunctDef}.
For the second statement, an invertible natural transformation $\eta: \Phi \to \Psi$ of
summing functors $\Phi, \Psi\in \Sigma_\cC(X)$ consists of a family of
isomorphisms $\eta_A : \Phi(A) \to \Psi(A)$ in the category $\cC$ that are
compatible with the morphisms of $P(X)$, given by the inclusions of
pointed subsets $j: A\hookrightarrow A'$. This compatibility with inclusions
shows that, in fact, the isomorphisms $\eta_A$ must be of the form
$\eta_A=\oplus_{x\in A\smallsetminus \{ * \}} \eta_x$, with the isomorphisms
$\eta_x: \Phi(x)\to \Psi(x)$, as can be seen inductively on the number of
elements of $A$. 
\end{proof}


Note that in the proof of Lemma~\ref{PhiPtsLem} we are explicitly using the fact that
$\oplus$ is a coproduct, with the pointed inclusions $j: A \hookrightarrow A'$ in $P(X)$
inducing the canonical morphisms $\Phi_X(A) \to \Phi_X(A)\oplus \Phi_X(A'\smallsetminus A \cup \{ * \})=\Phi_X(A')$
defined by the universal property of the coproduct. 


In the following we will also consider cases where the category $\cC$ is, more
generally, a unital symmetric monoidal category. For this case we write here the monoidal product
and the unit as $(\oplus, 0)$ rather than in the more usual form $(\otimes, {\mathbb I})$,
for consistency of notation with Lemma~\ref{PhiPtsLem}. Note, however, that here
$\oplus$ is not a coproduct and $0$ is not a zero object. 


In this more general setting Lemma~\ref{PhiPtsLem} no longer holds as stated.
Indeed, first observe that using the same definition of summing functor implies the existence
of morphisms $0\to \Phi_X(A)$ for all $A\in P(X)$, coming from the inclusions
$j: \{ * \} \hookrightarrow A$. Since $0$ is no longer required to be an initial object 
of $\cC$, morphisms $0\to C$ for $C\in {\rm Obj}(\cC)$ need not always exist. 
This then imposes a constraint on the possible range of the summing functors,
namely summing functors $\Phi_X: P(X) \to \cC$ have range in the subcategory
of $\cC$ of the ``no-cost resources'', namely the subcategory of $\cC$ with objects
those $C \in {\rm Obj}(\cC)$ with ${\rm Mor}_\cC(0,C)\neq \emptyset$
(see \S \ref{ResourcesSec} for an explanation of the ``no-cost'' terminology).


The summing-functor property $\Phi_X(A\cup A')=\Phi_X(A)\oplus \Phi_X(A')$ for $A,A'\in P(X)$
with $A\cap A'=\{ * \}$ gives an identification 
\begin{equation}\label{PhiXAmonoid}
 \Phi_X(A)\simeq \bigoplus_{x\in A, x\neq *} \Phi_X(x)\, ,
\end{equation}
up to the associators and braiding isomorphisms of the symmetric
monoidal structure, that relate the different bracketing and reordering of 
terms in the right-hand side of 
\eqref{PhiXAmonoid}. Indeed, the coherence theorem for unital symmetric monoidal categories ensures
that all these different choices of bracketing and reordering differ by a canonical isomorphism. We still have, as in
Lemma~\ref{PhiPtsLem}, that the values (up to isomorphism) of a summing functor 
$\Phi_X: P(X)\to \cC$ on objects $A\in P(X)$ are 
completely determined by the collection of objects $\{ \Phi_X(x) \}_{x\in A}$. 


In the special case where the unital symmetric monoidal category $\cC$ is a commutative monoidal
category, \eqref{PhiXAmonoid} is an identification, and ordering and 
bracketing of the right-hand side 
is irrelevant. Indeed, a commutative monoidal
category is a permutative category (strictly associative and unital) that is also strictly commutative,
so that the natural transformations that give the associators, braiding, and unitors of the monoidal
category are all identities. Examples of commutative monoidal categories include Petri nets  
and categories of line bundles and invertible sheaves \cite{BaezMa}.


Inclusions $j: A\hookrightarrow A'$ correspond to some morphisms 
$\Phi_X(j): \Phi_X(A) \to \Phi_X(A')=\Phi_X(A)\oplus \Phi_X(A'\smallsetminus A \cup \{ * \})$ that are no longer 
canonically determined by the universal property of a coproduct. 
Thus, invertible natural transformation $\eta: \Phi_X\to \Psi_X$ between
summing functors are now determined by the invertible morphisms $\eta_x: \Phi_X(x) \to \Psi_X(x)$, together with
this additional datum of the morphisms 
$\Phi_X(j)$ and $\Psi_X(j)$ for inclusions $j: \{ x, * \} \hookrightarrow A$, with $\Psi_X(j)\circ \eta_x=\eta_A \circ \Phi_X(j)$. 
In the special case of a commutative monoidal category, composition of morphisms and the monoidal product $\oplus$
satisfy the interchange relation
$(\varphi \circ \psi)\oplus (\varphi'\circ \psi') = (\varphi \oplus \varphi') \circ (\psi \oplus \psi')$. 
This implies that the morphisms $\Phi_X(j)$ and $\Psi_X(j)$ are completely determined by the morphisms
$\varphi_{x,y}:=\Phi_X(j:\{x,*\}\hookrightarrow \{ *,x,y \})$ with
$\varphi_{x,y}: \Phi_X(x)\to \Phi_X(x)\oplus \Phi_X(y)$ (and similarly for the $\Psi_X(j)$), and these in turn
are determined by the morphisms $\varphi_x : 0 \to \Phi_X(x)$ determined by the inclusions $\{ * \} \hookrightarrow \{ *, x \}$.
(As observed above, summing functors necessarily have range in the subcategory of ``no-cost'' resources so that these
morphisms can exist.)


Thus, in the case of commutative monoidal categories, we have obtained 
the following simple modification of Lemma~\ref{PhiPtsLem}.


\begin{cor}\label{PhiPtsCor}
Let $(\cC,\oplus,0)$ be a commutative monoidal category. Let 
$\cC^{\text{\rm no-cost}}$ be the full subcategory
with objects those $C\in {\rm Obj}(\cC)$ with ${\rm Mor}_\cC(0,C)\neq \emptyset$.
\begin{itemize}
\item A summing functor $\Phi_X: P(X)\to \cC$, defined as in Definition~\ref{SumFunctDef}, takes values in the
subcategory $\cC^{\text{\rm no-cost}}$.
\item Such a summing functor is completely determined by a collection of objects $$\{ \Phi_X(x) \}_{x\in X\smallsetminus *}
\in {\rm Obj}(\cC^{\text{\rm no-cost}}) $$ and a collection of morphisms $$\{ \varphi_x : 0 \to \Phi_X(x) \}_{x\in X\smallsetminus *}.$$
\item Invertible natural transformations $\eta: \Phi_X\to \Psi_X$ of two summing functors are completely determined by
the isomorphisms $\{ \eta_x : \Phi_X(x)\to \Psi_X(x) \}$.
\end{itemize}
\end{cor}


It is desirable in general to work with arbitrary unital symmetric monoidal categories, not just with the
more restrictive class of commutative monoidal categories.  In \cite{Tho95}, Thomason extended Segal's 
construction of \cite{Segal} to the case where $\cC$ is an arbitrary unital symmetric monoidal category, see also
 \cite{Tho78}, \cite{Tho82}. 
 

In this general setting of arbitrary unital symmetric monoidal categories (see the Appendix of \cite{Tho82}), 
one proceeds by modifying the notion of summing functor of Definition~\ref{SumFunctDef}, and replacing
its characterization in terms of the collection of objects $\{ \Phi_X(x) \}_{x\in X\smallsetminus *}$ of Lemma~\ref{PhiPtsLem} 
into a definition. For our purposes, we take the definition of the category of summing functors for 
unital symmetric monoidal categories to be of the following form. 


\begin{defn}\label{SumPhiSymmMon}
Let $(\cC,\oplus,0)$ be a symmetric monoidal category. For a finite pointed set $X$, the
category $\Sigma_\cC(X)$ has objects 
$$ \Phi_X:= \{ \Phi_X(x) \}_{x\in X\smallsetminus *} $$
given by objects in the Cartesian product $\hat\cC^n$, with $\# X = n+1$, and
morphisms given by morphisms in $\hat\cC^n$.
\end{defn}


Note that here, because the category $\cC$ does not have, in general, an initial or a terminal object,
one does not have to restrict to invertible natural transformations of summing functors in order to ensure
a non-trivial topology of the resulting category of summing functors. Thus, in \cite{Tho82} one just
considers the category $\cC^n$ instead of $\hat\cC^n$. However, in our setting we are interested
in maintaining this constraint, as we want to interpret the category of summing functors as assignments
of resources {\em up to equivalence}, hence we modified the setting of \cite{Tho82}, \cite{Tho95}, to
include the requirement that summing functors take values in $\hat\cC$. One may worry here whether 
this restriction might affect the main result of \cite{Tho95}, that Gamma-spaces obtained from this
construction realize {\em all} connective spectra. However, this is still the case. Indeed, our setting
includes in particular the case where $\cC$ is a unital symmetric monoidal {\em groupoid}, in which
case $\cC=\hat\cC$, and it is known by Theorem~5.3 of \cite{Fue-Keu} that Gamma-spaces associated
to unital symmetric monoidal groupoids already suffice to realize all connective spectra. Thus, this restriction
does not affect the main homotopy-theoretic properties we will be discussing in \S \ref{GammaGeneralSec}.
This leaves an ambiguity of two slightly different possible definitions of summing functors in the case of 
commutative monoidal categories, so one will need to specify, in those cases, which notion 
of $\Sigma_\cC(X)$ is used.
In the following, we will mostly discuss summing functors based on Definition~\ref{SumFunctDef},
without specifying explicitly how to incorporate the case of Definition~\ref{SumPhiSymmMon}, 
except where directly needed, as the latter case usually follows by a simple modification. 


The reason why it is useful to consider both of these slightly different notions
of category of summing functors will be discussed more in detail in \S \ref{DiscrHopfSec}
below, when we introduce categorical Hopfield equations with threshold non-linearities.
We will see that, while the case of categories of resources with a zero-object reduces to a linear 
dynamics, the more general case of symmetric monoidal categories gives rise to
genuinely non-linear models. (As shown above, summing functors necessarily take values
in ``no-cost resources'', that is, in the subcategory of objects $C$ with a morphism $0\to C$,
while as we will see in \S \ref{DiscrHopfSec} the threshold dynamics is designed to detect 
the opposite convertibility $C\to 0$.)
The case of the symmetric monoidal category of deep neural
networks introduced in \cite{Mar-new} provides such an example with non-linear dynamics.

\subsection{Networks and summing functors} \label{NetSumFuncSec}

Our goal is to assign resources to networks of neurons. So far, we have only
described a notion of consistent assignments of $\cC$-type resources to subsets of a given finite set.
We next describe how to introduce the network structure. 
  The setting we described in the previous subsections can
 be modified  by regarding the data of
 neurons and connections as a directed graph and incorporating
 it in the construction.
 

A categorical description of networks and flows on networks was introduced in \cite{Harper}. 
In that generality, one considers networks to be directed graphs, where a priori no restriction on
edges is imposed (that is, one allows multiple edges and looping edges). In more specific
cases (for example when considering cliques), one only allows graphs without these types of edges. 
The standard description of directed graphs in
 categorical terms is as follows (see e.g.~\cite{Harper}).
 
 
 \begin{defn}\label{DirGrDef}
 Let ${\bf 2}$ denote the category
 with two objects $E,V$ and as only non-identity
 morphisms two parallel morphisms $s,t: E\to V$ (called source and
 target morphisms).
 A directed finite graph is a functor $G: {\bf 2}\to \cF$
 where $\cF$ is the category of finite sets. 
 \end{defn}
 
 
 In the following we will refer to a functor $G\in {\rm Func}({\bf 2}, \cF)$
 as a directed graph, or equivalently as a (directed) network, to the
 set $V_G=G(V)$ as either the set of {\em vertices} or, equivalently,  
 as the set of {\rm nodes} of $G$, and to
 $E_G=G(E)$ as the set of edges. 
 
 
 Note that some variants of this categorical notion of directed graphs
 are possible, and useful to consider in some cases. For example, with
 the notion given in Definition~\ref{DirGrDef}, morphisms of directed
 graphs do not include contraction of edges (mapping an edge to a vertex).
 If one wants to work with directed graphs where it is important to
 also consider such transformations, then a simple modification of
 the category ${\bf 2}$ achieves this purpose. We refer the reader to
 \S 2.1.1 of \cite{Mar-new} where different categorical formulations of
 directed graphs are compared. A specific example where contractions
 of edges are need is also discussed in \cite{Mar-new}.
 
 
 Because of the need to work with pointed sets for homotopy-theory purposes,
 we can alter slightly this standard definition with the addition of base-point data.
 Again, these base-point data should be regarded only as an artificial computational
 device introduced here for later use (see \S \ref{GammaGeneralSec} and \S \ref{SpectraHTSec}).
 For the purpose of what we discuss here, the reader can easily ignore this extension to
 the pointed case and just rephrase everything in terms of the original Definition~\ref{DirGrDef}.  
 
 
 \begin{defn}\label{PtDirGrDef}
 A pointed directed finite graph is a functor $G: {\bf 2}\to \cF_*$ to the category of pointed finite sets.
 \end{defn}
 
 
 Note that this definition differs from other notions of flow graphs,
 since we do not require the distinguished root vertex to
 be a source or a sink, nor do we require the existence of 
 directed paths from the root to all other vertices. 
 Moreover, since the source and target maps are mapped by the functor
 $G$ to morphisms of pointed sets, these graphs have 
 a distinguished looping edge with both source and target equal to 
 the root vertex. This root vertex and its looping edge do not play a
 direct role in our model and are only an artificial device to
 introduce base points for homotopy-theoretic purposes.
 
 
 For all the practical aspects of the model, we can assume
 that we work with directed graphs $G: {\bf 2}\to \cF$ in the
 usual sense. Indeed the pointed directed graphs we will be 
 considering are obtained from an ordinary directed graph
 in the following way. 
 
 \begin{lem}\label{GGstar}
 Given a functor $G: {\bf 2}\to \cF$,
 we associate to it a pointed directed graph $G^*: {\bf 2} \to \cF_*$
 defined by $E_{G^*}=E_G \sqcup \{ e_* \}$ and $V_{G^*}=V_G \sqcup \{ v_* \}$
 with $s,t: E_{G^*} \to V_{G^*}$ given by the source and target maps
 $s,t: E_G \to V_G$ for all edges $e\in E_G$ and as $s,t: e_* \mapsto v_*$.
 \end{lem}
 
 Thus the pointed graphs $G^*$ we will consider here are just ordinary directed graphs $G$ together 
 with a disjoint base-point vertex with a single looping edge attached to it. In the following,
 in cases where we consider the case without looping edges, we mean that the underlying
 $G$ has no looping edges. 
 
 
 \begin{lem}\label{summingEG}
 The source and target maps $s,t: E\to V$ determine functors between
 the categories of summing functors (still denoted $s,t$),
 $$ s,t: \Sigma_{\cC}(E_{G^*})\to \Sigma_\cC(V_{G^*}) \, .$$
 \end{lem}
 
\begin{proof}
  The source and target maps $s,t: E\to V$ transform summing functors
 $\Phi_{E_{G^*}}: P(E_{G^*})\to \cC$ into functors $\Phi_{V_{G^*}}^{s,t}: P(V_{G^*})\to \cC$
 given by
 $$ \Phi_{V_{G^*}}^s (A) := \Phi_{E_{G^*}}(s^{-1}(A) ) = \oplus_{e\in E_{G^*}\,:\, s(e)\in A} \Phi_{E_{G^*}}(e), $$
 for $A\in P(V_{G^*})$ where $\Phi_{E_{G^*}}(e)$ means the functor $\Phi_{E_{G^*}}$
 evaluated on the pointed set $\{ e, * \}\in P(E_{G^*})$. Because of the way the
 pointed directed graph $G^*$ is constructed from the directed graph $G$, we see that
 the functor $ \Phi_{V_{G^*}}^s$ obtained in this way is by construction still a summing functor.
 Indeed, for $A\cap A' = \{ v_* \}$ in $P(V_{G^*})$, we have
 $$ \Phi_{V_{G^*}}^s (A\cup A') = \bigoplus_{e\in E_{G^*}\,:\, s(e)\in A\smallsetminus \{ v_* \}} \Phi_{E_{G^*}}(e)
 \oplus \bigoplus_{e\in E_{G^*}\,:\, s(e)\in A' \smallsetminus \{ v_* \}} \Phi_{E_{G^*}}(e) \oplus \Phi_{E_{G^*}}(e_*), $$
 where $\Phi_{E_{G^*}}(e_*)$ is the zero object in $\cC$, so the above equals
 $\Phi_{V_{G^*}}^s (A) \oplus \Phi_{V_{G^*}}^s (A')$. The case of $\Phi_{V_{G^*}}^t$ is similar.
\end{proof}
 
 
 We can interpret this explicitly in terms of our model in the following way. The
 directed graph $G$ represents a network of neurons (the nodes $V_G$) and
 connections between them (the directed edges $E_G$). The introduction of the
 artificial base vertex $v_*$ with its single looping edge $e_*$ is merely a
 computational artifact that does not affect the structure of the network. 
 The category $\Sigma_\cC(V_{G^*})$ parameterizes all the possible
 consistent assignments of resources of type $\cC$ over subsets of 
 nodes in $V_G$ (in fact at the individual nodes of $G$, by Lemma~\ref{PhiPtsLem}).
 In a similar way $\Sigma_{\cC}(E_{G^*})$ describes 
 assignments of resources of type $\cC$ to the edges of the network. The induced source and
 target maps can be used to express possible compatibility requirements between the assignments
 at nodes and at edges. The images of the source and target maps describe
 assignments of $\cC$-resources at sets $A$ of nodes
 of the network that come from an assignment at either the outgoing or the  incoming edges
at those nodes. We will describe in \S \ref{EqSec} and \S \ref{CoeqSec} below some specific 
examples of possible ways of imposing constraints 
relating assignments of resources at vertices and edges.

 \subsubsection{Conservation laws at vertices}\label{EqSec}
 
 The first and simplest example of compatibility condition between assignments of resources to vertices and edges
 consists of imposing the standard physical conservation law at vertices. This is a typical feature, for example, of 
 electrical networks with flows of electric currents, where conservation at vertices holds because of Kirchhoff's current law.
 For biological neuronal networks, this very simple conservation law is not always adequate, but we present it here as the
 first case  because of its very simple description. In categorical terms, this kind of conservation law is literally 
 expressed by the {\rm equalizer} construction. 
 
 
 \begin{prop}\label{equalizer}
 The equalizer $\Sigma_\cC^{\operatorname{eq}}(G)$ of the two functors 
 $$s,t: \Sigma_\cC(E_{G^*}) \rightrightarrows 
 \Sigma_\cC(V_{G^*})$$ is a category consisting of the summing
 functors $\Phi_E\in \Sigma_\cC(E_{G^*})$ that satisfy the Kirchhoff 
 conservation law at vertices
 \begin{equation}\label{Kirchhoff}
  \bigoplus_{e\,:\, s(e)=v} \Phi_E(e) = \bigoplus_{e\,:\, t(e)=v} \Phi_E(e) \, . 
 \end{equation} 
 \end{prop}
 
\begin{proof}
 Consider the two functors $s,t: \Sigma_\cC(E_{G^*}) \rightrightarrows 
 \Sigma_\cC(V_{G^*})$ as above, between the small categories of summing functors, 
 induced by the source and target morphisms of the directed graph
 $G: {\bf 2}\to \cF$. The equalizer of this diagram is the small category $\Sigma_\cC^{\operatorname{eq}}(G)$ with functor
 $\iota: \Sigma_\cC^{\operatorname{eq}}(G) \to \Sigma_\cC(E_{G^*})$ such that $s\circ \iota = t\circ \iota$ satisfying the
 universal property expressed  for any $\cA$ and $q$ with $s\circ q =t\circ q$ by the commutative diagram
 $$ \xymatrix{ \Sigma_\cC^{\operatorname{eq}}(G) \ar[r]^{\iota} & \Sigma_\cC(E_{G^*})  \ar@<-.2ex>[r]^s \ar@<.2ex>[r]_t &  \Sigma_\cC(V_{G^*}) \\ \cA \ar[u]^{\exists u} \ar[ur]^q & & } $$
 This can be realized as summing functors 
$\Phi_E: P(E_{G^*}) \to \cC$ in $\Sigma_\cC(E_{G^*})$ such that, for all $A\in P(V_{G^*})$
\begin{equation}\label{Vconserve}
\Phi_E(s^{-1}(A))=\Phi_E(t^{-1}(A)). 
\end{equation}
The relation \eqref{Vconserve} is exactly expressing the Kirchhoff conservation law at vertices since by Lemma~\ref{PhiPtsLem} 
it can be reduced to the case where $A$ has a single (non base-point)
vertex where it reduces to \eqref{Kirchhoff}. 
\end{proof}

\subsubsection{Vertex constraints by coequalizer} \label{CoeqSec}

Another way of imposing a Kirchhoff-type conservation at vertices is provided by the dual coequalizer construction.
While the equalizer construction selects those summing functors on edges, with the given target category $\cC$, 
that satisfy conservation at vertices, the coequalizer construction modifies the target category to a suitable quotient
where the conservation laws hold. 


The coequalizer construction is more subtle for various reasons: the nerve functor (that we will
 be using in \S \ref{GammaGeneralSec}) only preserves directed colimits, and in general coequalizers 
 in the category of small categories are more subtle to construct. However, we can still consider the following
 construction at the level of summing functors. 


 Following \cite{BeBoPa99}, coequalizers in the category of small categories can be described
 in terms of a quotient construction based on the notion of generalized congruences. For a small
 category $\cC$, let ${\rm Mor}^+(\cC)$ denote the set of $n$-tuples of 
 (not necessarily composable)
 morphisms of $\cC$ for arbitrary $n$. For $\phi \in {\rm Mor}^+(\cC)$ one denotes by ${\rm dom}(\phi)$
 and ${\rm codom}(\phi)$, respectively, the objects of $\cC$ given by the domain of the first morphism 
 in the tuple and the codomain of the last morphism in the tuple.
 
 \begin{defn}\label{gencongr}{\rm \cite{BeBoPa99}}
 A generalized congruence on $\cC$ consists of an equivalence relation on the set of
 objects ${\rm Obj}(\cC)$ and a partial equivalence relation on the tuples of morphisms of $\cC$
 with the properties 
 \begin{enumerate}
 \item if $A\sim B$ for $A,B\in {\rm Obj}(\cC)$ then ${\rm id}_A\sim {\rm id}_B$;
 \item if $\phi\sim \psi$ for $\phi,\psi\in {\rm Mor}^+(\cC)$ then ${\rm dom}(\phi)\sim {\rm dom}(\psi)$
 and  ${\rm codom}(\phi)\sim  {\rm codom}(\psi)$;
 \item if $\phi_1 \phi_2\sim \psi$ with $\phi_i,\psi\in {\rm Mor}^+(\cC)$ then 
 ${\rm dom}(\phi_2)\sim {\rm codom}(\phi_1)$;
 \item if $\phi \sim \psi$ and $\chi \sim \xi$ for $\phi,\psi,\chi,\xi \in {\rm Mor}^+(\cC)$ with
 ${\rm codom}(\phi)\sim {\rm dom}(\chi)$ then $\phi \chi \sim \psi \xi$;
 \item if ${\rm codom}(\phi) ={\rm dom}(\psi)$ for single morphisms $\phi,\psi\in {\rm Mor}(\cC)$
 then the chain $\phi \psi$ is composable and $\phi \psi \sim \psi \circ \phi$ in ${\rm Mor}^+(\cC)$.
 \end{enumerate}
  \end{defn}
  
  
 The quotient $\cC/_\sim$ of $\cC$ by a generalized congruence is a small category with objects
 the equivalence classes of objects ${\rm Obj}(\cC/_\sim)={\rm Obj}(\cC)/_\sim$ and morphisms 
 given by equivalence classes of
 tuples $\phi_1\cdots \phi_n$ in ${\rm Mor}^+(\cC)$ with ${\rm codom}(\phi_i)\sim {\rm dom}(\phi_{i+1})$ 
 (that is, chains that become composable in the quotient), 
 with the composition determined by concatenation of tuples of paths. There is a quotient functor
 $Q: \cC \to \cC/_\sim$. 
 A generalized congruence is principal if it is generated by a relation on single morphisms. 
 
 
 It is shown in \cite{BeBoPa99} that, given two functors $F,G : \cA \to \cC$ in the category of
 small categories ${\rm Cat}$, the coequalizer ${\rm coeq}(F,G)$ with functor $Q: \cC \to {\rm coeq}(F,G)$
 is the quotient category $\cC/_\sim$ with quotient functor $Q: \cC \to \cC/_\sim$ with respect to the principal
 generalized congruence generated by $F(A)\sim G(A)$ in ${\rm Obj}(\cC)$ and $F(\phi)\sim G(\phi)$
 for $\phi\in {\rm Mor}(\cC)$. 

 
 For a fixed network specified by a directed graph $G\in {\rm Func}({\bf 2},\cF)$,
 let $G^*$ be the pointed directed graph obtained as above. As above, 
 given a summing functor $\Phi_E: P(E_{G^*}) \to \cC$, we consider the two
 functors $\Phi_V^s$ and $\Phi_V^t$ from $P(V_{G^*})$ to $\cC$ given by
 $\Phi^s_V(A)=\Phi_E(s^{-1}(A))$ and $\Phi^t_V(A)=\Phi_E(t^{-1}(A))$
 for all pointed subsets $A\in P(V_{G^*})$ and $s,t$ the source and target
 maps of $G$. 
 
 \begin{prop}\label{CoeqProp}
 The coequalizer $\rho_G: \cC \to \cC^{\operatorname{coeq}}_G(\Phi_E)$ of the functors $\Phi_V^s, \Phi_V^t$ 
 gives a category $\cC^{\operatorname{coeq}}_G(\Phi_E)$ of resources that is optimal with respect to the
 property that resources associated to the systems $\Phi_E(s^{-1}(A))$ and
 $\Phi_E(t^{-1}(A))$ satisfy the conservation law at vertices
 \begin{equation}\label{coEqresconserve}
  \rho_G(\Phi_E(s^{-1}(A)))= \rho_G(\Phi_E(t^{-1}(A))), \ \ \  \forall A\in P(V_{G^*}). 
 \end{equation} 
 The multiple coequalizer $\rho_G: \cC \to \cC^{\operatorname{coeq}}_G$ over the family $\{ (\Phi_V^s, \Phi_V^t) \,|\, \Phi_E \in \Sigma_{\cC}(E_{G^*}) \}$
 gives a category $\cC^{\operatorname{coeq}}_G$ such that the conservation law \eqref{coEqresconserve} holds for all $\Phi_E \in \Sigma_{\cC}(E_{G^*})$.
 \end{prop}

\begin{proof}
 Consider the coequalizer $\cC^{\operatorname{coeq}}_G(\Phi_E):={\rm coeq}(\Phi_V^s, \Phi_V^t)$ of the functors 
 $$\Phi_V^s, \Phi_V^t: P(V_{G^*}) \rightrightarrows \cC,$$
 with the functor $\rho_G: \cC \to \cC^{\operatorname{coeq}}_G$ satisfying $\rho_G \circ \Psi_V^s = \rho_G \circ \Psi_V^t$. This
 is characterized by the universal property given by the commutativity of the diagrams
 $$ \xymatrix{ P(V_{G^*})   \ar@<-.2ex>[r]^s \ar@<.2ex>[r]_t &  \cC \ar[r]^{\rho_G} \ar[rd]^{\rho} & 
 \cC^{\operatorname{coeq}}_G(\Phi_E) \ar[d]^{\exists g} \\ 
& & \cR } $$
 for all small categories $\cR$ and functors $\rho: \cC \to \cR$ such that $\rho\circ \Phi_V^s = \rho \circ \Phi_V^t$,
 and a functor $g: \cC^{\operatorname{coeq}}_G(\Phi_E) \to \cR$ with $g\circ \rho_G =\rho$.
 
 By the result of \cite{BeBoPa99} recalled above, we can describe the coequalizer
 $\rho_G: \cC \to \cC^{\operatorname{coeq}}_G$ as the quotient functor to $\cC^{\operatorname{coeq}}_G(\Phi_E)=\cC/_{\sim_{G,\Phi_E}}$ where
 $\sim_{G,\Phi_E}$ is the principal generalized congruence on $\cC$ generated by the
 relations $\Phi_E(s^{-1}(A))\sim \Phi_E(t^{-1}(A))$ for all $A\in P(V_{G^*})$ and
 the same equivalence on morphisms corresponding to pointed inclusions of sets in $P(V_{G^*})$. 
 
 The universal property of the coequalizer shows that the category $\cC/_{\sim_{G,\Phi_E}}$ is the
 optimal choice of a category $\cR$ of resources with a functor $\rho: \cC \to \cR$ from
 systems to resources that implements the conservation laws \eqref{coEqresconserve} 
 of resources at vertices for the summing functor $\Phi_E$. 
\end{proof}
  
 \begin{defn}\label{CoeqSum}
 If the category $\cC^{\operatorname{coeq}}_G$ obtained as the multiple coequalizer $\rho_G: \cC \to \cC^{\operatorname{coeq}}_G$ in Proposition~\ref{CoeqProp}
 admits a symmetric monoidal structure then we can consider the category of summing functors
 $$ \Sigma^{\operatorname{coeq}}_\cC(G):= \Sigma_{\cC^{\operatorname{coeq}}_G}(E_{G^*}). $$
 \end{defn} 
 
 This category describes the imposition of constraints \eqref{coEqresconserve} at vertices.
 
 
 An advantage of the coequalizer construction is that, instead of selecting a smaller subcategory of summing
 functors with fixed target category, it imposes the conservation law at vertices by suitably altering only the target
 category.  
  
 \subsection{Categories of summing functors on networks} \label{SumNetSec} 
 
 The examples of constructions of categories of summing functors on networks described in \S \ref{EqSec} and \S \ref{CoeqSec}
 via equalizers and coequalizers are special cases (realized by subcategories) of a more general setting that we introduce
 here. The subcategories obtained via equalizers and coequalizers correspond  to choosing only those summing functors 
 that are determined by certain specific types of constraints at vertices. 
 We will then show in \S \ref{GraftingSec} another example of a construction of a category of summing functors on networks
 that also fits into the general framework discussed here, but which arises from different types of constraints coming from
 grafting operations. 
 
 
 We assume that $\cC$ is either a category with zero object and sum, or more generally a symmetric monoidal category. 
 
 
 Given a directed graph $G: {\bf 2} \to \cF$, a subgraph is another functor $G' : {\bf 2} \to \cF$ with a natural transformation
 $\alpha: G' \hookrightarrow G$, meaning that $\alpha_V: V_{G'}\hookrightarrow V_G$ and $\alpha_E: E_{G'} \hookrightarrow E_G$ are inclusions.
 The case of pointed directed graphs is analogous with $\alpha_V$, $\alpha_E$ inclusions of pointed sets. 
 
 
 \begin{defn}\label{SummingNetDef}
 Given $G: {\bf 2} \to \cF$, let $P(G)$ be the category whose objects are the subgraphs $G'\hookrightarrow G$ with morphisms
 given by the inclusions. A network summing functor is a functor $\Phi: P(G) \to \cC$ that maps the empty subgraph to the zero object and such that
 $$ \Phi(G' \sqcup G'') =\Phi(G')\oplus \Phi(G'') $$
 for non-intersecting subgraphs. The category $\Sigma_\cC(G)$ consists of network summing functors with invertible natural transformations.
 \end{defn}
 
 \begin{rem}\label{pointSumNet} {\rm 
 For $G^*:{\bf 2} \to \cF_*$ a pointed graph with base
 vertex $v_*$ with looping edge $e_*$, the category  $\Sigma_\cC(G^*)$ consists of functors 
 $\Phi: P(G^*)\to \cC$ that map the pointed component $\Phi(\{ v_*,e_*\})=0$ to the zero object of $\cC$ and satisfy $\Phi(G'\cup G'')=\Phi(G')\oplus \Phi(G'')$
 for $G',G''\in P(G^*)$ with $G'\cap G''=\{ v_*, e_* \}$. For the graph $G^*$ obtained by adding to a non-based graph $G$ a separate component $\{ v_*, e_*\}$,
 the categories $\Sigma_\cC(G)$ and $\Sigma_\cC(G^*)$ are equivalent, so we will use the same notation $\Sigma_\cC(G)$.}
 \end{rem}
 
 
 The categories of summing functors $\Sigma^{\operatorname{eq}}_\cC(G)$ and $\Sigma^{\operatorname{coeq}}_{\cC}(G)$ considered in \S \ref{EqSec} and \S \ref{CoeqSec} are
 (sub)categories of network summing functors. Indeed, we can view a $\Phi\in \Sigma^{\operatorname{eq}}_\cC(G)$ as an object in $\Sigma_\cC(G)$
 by precomposition with the functor $P(G^*)\to P(E_{G^*})$, hence $\Sigma^{\operatorname{eq}}_\cC(G)\subset \Sigma_\cC(G)$. 
 In the same way a functor $\Phi \in \Sigma^{\operatorname{coeq}}_{\cC}(G)$ can be seen as an
 object in the category $\Sigma_{\cC^{\operatorname{coeq}}_G}(G)$. One can see from these examples that, in more concrete problems, one will want to restrict summing
 functors to some suitable subcategory of $\Sigma_\cC(G)$ that corresponds to specific types of constraints one wants to impose
 dictated by the structure of the network (such as conservation laws at vertices in these examples). 
 
\subsubsection{Graphs in terms of vertices and flags}\label{GraphsSec}

There are other variants of the standard categorical description of directed graphs of Definition~\ref{DirGrDef} that
can also be useful in our setting, especially for the formulation of \S \ref{GraftingSec} below. 
If one does not need the directed structure, but would like graphs to have
some ``external edges'' (external ports, which in the non-directed case serve simultaneously as inputs and outputs), 
then the physics description of graphs in terms of vertices
and half-edges (flags) instead of vertices and edges would be more suitable.

\begin{defn}\label{GraphsFlagsDef}
Let ${\bf 2}_F$ be the category with two objects $V,F$ and non-identity morphisms $\partial: F \to V$ and
$\iota: F \to F$ with $\iota^2=1_F$, as well as $\iota\circ \partial$. 
A finite graph is a functor $G: {\bf 2}_F \to \cF$ to the category of finite sets.
\end{defn}

Here $V_G:=G(V)$ is the set of vertices and $F_G:=G(F)$ is the set of half-edges. The morphism $\partial$ assigns
to each half-edge the vertex it is attached to, and the involution $\iota$ glues together the loose ends of the half-edges. Here we do not
assume that $\iota$ is fixed-point free: the fixed points of $\iota$ are the external edges of the graph, while
the pairs of flags $f\neq f'$ with $f'=\iota(f)$ are the half-edges glued together to form an (internal) edge of $G$.
The resulting graphs can have multiple and looping edges. 


A pointed version can be obtained as in the previous case, by replacing the target category $\cF$ with finite
pointed sets $\cF_*$. Since the induced morphisms determined by $\partial$ and $\iota$ have to be maps
of pointed sets, we obtain that the base vertex $v_*$ has a base external edge $f_*=\iota(f_*)$ attached to it.
Given a graph  $G: {\bf 2}_F \to \cF$ the associated based $G^* : {\bf 2}_F \to \cF_*$ simply has an added
component consisting of $v_*$ with the external edge $f_*$. Note that,
in the case of the category of pointed graphs $G^* : {\bf 2}_F \to \cF_*$, one can use the base
vertex with external edge as a way to incorporate data of an assigned external input to the
network. 


As in the case of the description of graphs of Definition~\ref{DirGrDef}, one can then consider
categories of summing functors $\Sigma_\cC(V_{G^*})$ and $\Sigma_\cC(F_{G^*})$. 


In the case of directed graphs, one can also accommodate external edges in two possible ways.
One is simply to consider any univalent vertices as ``external'' vertices and the corresponding
edges as ``external edges'', the other is to adapt the flag definition of graphs of Definition~\ref{GraphsFlagsDef}
to the directed case in the following way.

\begin{defn}\label{GraphsFlagsOrDef}
Consider the category ${\bf 2}^{i/o}$ with objects $\{ V, E, F_i ,F_o \}$ and 
morphisms freely generated by
\begin{equation}\label{VEIO}
 E \stackrel{f_i}{\rightarrow} F_i \stackrel{t}{\rightarrow} V \stackrel{s}{\leftarrow} F_o \stackrel{f_o}{\leftarrow} E \, .
\end{equation} 
A (finite) directed graph with external edges (also called an ``open-ended'' graph) is a functor $G: {\bf 2}^{i/o}\to \cF$, with $\cF$ the
category of finite sets, where the morphisms $f_i,f_o$ are mapped to injective maps.
\end{defn}

We interpret here the sets $E(G):=G(E)$ and $V(G):=G(V)$ as directed (internal) edges and vertices, and we
interpret the sets $F_i(G):=G(F_i)$ and $F_o(G):=G(F_o)$ as the incoming/outgoing flags (oriented
to/from the vertex). The morphisms $t: F_i(G)\to V(G)$ and $s:  F_o(G)\to V(G)$ are the
boundary morphisms that associate to a flag the corresponding vertex (target or source depending
on orientation) and the morphisms $f_i: E(G)\to F_i(G)$ and $f_o: E(G)\to F_o(G)$ assign to an
edge its two flags (half-edges), respectively attached to source and target vertex. The set $E_{ext}(G)$
of external edges of $G$ is then given by the set 
$$ E_{ext}(G)=(F_i(G)\smallsetminus f_i(E)) \sqcup (F_o(G)\smallsetminus f_o(E)) \, . $$
The case of the category ${\bf 2}$ and directed graphs $\cG={\rm Func}({\bf 2},\cF)$ without external edges 
corresponds to the case where the outer arrows of \eqref{VEIO} are identity maps. In this case edges attached to valence-one
vertices are not considered external. 


Definition~\ref{GraphsFlagsOrDef} allows for directed cycles (for example, pairs of
vertices with a directed edge between them in both directions). In the following,
in general we will be restricting to acyclic graphs, for compatibility with the properad
composition, see Lemma~\ref{ProperadSum}. 
 
 \subsubsection{Constraints through grafting operations}\label{GraftingSec}
 
We now describe another construction of an interesting subcategory of summing functors, where instead of simple conservation conditions
at vertices one uses more interesting grafting operations, in a case where additional compositionality structure is present on the target category $\cC$.
This type of construction will be useful in the case where we consider resources given by certain classes of computational architectures (see \S \ref{CompResSec}
and \S \ref{GammaCompArchSec}). 


The compositionality structures referred to above can be expressed in terms of the notion of {\em properad} \cite{Val}  (see also \cite{Kock}).


\begin{defn}\label{Properad}
Let ${\rm Cat}$ denote the category of small categories. A {\rm properad} in ${\rm Cat}$ is a collection $\cP=\{ \cP(m,n) \}_{m,n\in \N}$
of small categories with composition functors (grafting operations)
\begin{equation}\label{ProperadComp}
\circ^{i_1,\ldots, i_\ell}_{j_1,\ldots, j_\ell} : \cP(m,k) \times \cP(n,r) \to \cP(m+n-\ell, k+r-\ell)\, ,
\end{equation}
for non-empty $\{ i_1,\ldots, i_\ell \}\subset \{ 1,\ldots, k \}$ and $\{ j_1,\ldots, j_\ell \}\subset \{ 1,\ldots, n \}$, $i_s< i_{s+1}$ and $j_s < j_{s+1}$
for $s=1,\ldots,\ell-1$. These composition operations satisfy associativity and unity conditions.  
A {\rm symmetric properad} also has symmetric group actions of $\Sigma_m \times \Sigma_n$ on $\cP(m,n)$ with respect to which the
compositions are bi-equivariant. We will assume properads to be symmetric.
\end{defn}


The unit ${\bf 1}\in \cP(1,1)$ of the properad satisfies 
${\bf 1} \circ^1_{j} P=P$ for all $P\in \cP(n,r)$
and $P' \circ^{i}_1 {\bf 1}=P'$
for all $P'\in \cP(m,k)$, for all $j\in \{ 1,\ldots, n \}$ and all $i\in \{ 1,\ldots, k \}$.
We will not write out here explicitly the associativity condition for 
the properad composition laws \eqref{ProperadComp}, but it can be
deduced directly from the definition of the composition law.

It is in general assumed that properads are symmetric, especially in the 
context of graphs, which would otherwise require 
additional data of planar structures compatible with composition. In the
symmetric case an abstract set rather than an ordered set suffices for
indexing.


For a more detailed discussion of the properties of properads and the
compatibility between the properad composition and the monoidal structure
in the case where $\cC$ is unital symmetric monoidal, see \S 1.1.1 of \cite{Mar-new}. 
An explicit example of properad in ${\rm Cat}$ and its properad composition is described
in \cite{Mar-new} in the form of a category of deep neural network architectures. 


We consider here open-ended subgraphs $G'\in P(G)$ of the open-ended graph $G$ that are {\em full},
in the sense that if a vertex is in the subgraph then all its incident half-edges are also in the subgraph,
and if two vertices are in the subgraph then all internal edges between them are also in the subgraph.
For an acyclic graph $G$, we also require the subgraphs to be convex, in the sense that
if two vertices are in the subgraph, so are all the intermediate vertices along directed
paths connecting them. 


Given a directed graph $G$ and two subgraphs $G',G''\in P(G)$ as above with $V_{G'}\cap V_{G''}=\emptyset$, let 
$E(G',G'')\subset E_G$ denote the set of edges with one endpoint in $V_{G'}$ and the other in $V_{G''}$,
and let $G'\star G''$ denote the subgraph of $G$ with $V_{G'\star G''}=V_{G'} \cup V_{G''}$ and $E_{G'\star G''}=E_{G'} \cup E_{G''} \cup E(G',G'')$.
For the purpose of the following construction we assume that the graph $G$ has a certain number $\deg^{\operatorname{in}}(G)\geq 1$ of incoming external legs 
and a number $\deg^{\operatorname{out}}(G)\geq 1$ of outgoing external legs. Similarly for a subgraph $G'\subset G$. Let $E(G',G\smallsetminus G')$ denote the
set of edges in $G$ with one end in $V_{G'}$ and the other end in $V_G \smallsetminus V_{G'}$.


We write $\deg^{\operatorname{in}}(G')$
(respectively, $\deg^{\operatorname{out}}(G')$) for the number of edges in $E(G',G\smallsetminus G')$ with target vertex (respectively, source vertex) in $G'$, plus the number of external (half)edges of $G$ with target
(respectively, source) vertex in $G'$.
Then the following is a direct consequence of the definitions.


Recall that, for a vertex $v\in V_G$ the corolla $C(v)$ consisting of $v$ together with 
all the attached (half)edges, with $\deg^{\operatorname{in}}(v)$ incoming and $\deg^{\operatorname{out}}(v)$ outgoing
(half)edges.


\begin{lem}\label{ProperadSum}
Let $\cC$ be a symmetric monoidal category such that there is a family of full subcategories $\cC(n,m)$ for $n,m\in \N$
with the properties:
\begin{itemize}
\item ${\rm Obj}(\cC)=\cup_{n,m\in \N} {\rm Obj}(\cC(n,m))$;
\item the monoidal structure $(\otimes, {\mathbb I})$ satisfies
$$ \otimes: \cC(m,k) \times \cC(n,r) \to \cC(m+n, k+r)\, ; $$
\item the family $\{ \cC(n,m) \}_{n,m\in \N}$ is a properad in ${\rm Cat}$.
\end{itemize}
Let $G$ be a directed acyclic graph. For 
two subgraphs $G',G''\in P(G)$ as above with $V_{G'}\cap V_{G''}=\emptyset$, 
we say that $G' < G''$ if there are no directed paths from vertices of 
$G''$ to vertices of $G'$. 
Then there is a full subcategory $\Sigma_\cC^{\operatorname{prop}}(G)\subset \Sigma_\cC(G)$ given by 
the summing functors $\Phi: P(G)\to \cC$  with the following properties:
\begin{enumerate}
\item for all full convex open-ended subgraphs $G'\in P(G)$,  
$$ \Phi(G') \in {\rm Obj}(\cC(\deg^{\operatorname{in}}(G'), \deg^{\operatorname{out}}(G'))\, , $$
\item for any vertex, $\Phi(\{ v \})=\Phi(C(v))$ where $C(v)$ is the corolla of the vertex $v$ in $G$,
\item for any $G' < G''\in P(G)$ with $V_{G'}\cap V_{G''}=\emptyset$, 
\begin{equation}\label{graftingPhi}
\Phi(G'\star G'') = \Phi(G') \circ_{E(G',G'')} \Phi(G''), 
\end{equation}
where $E(G',G'')\subset E_G$ is the set of edges with source endpoint in $V_{G'}$ 
and target in $V_{G''}$ and
$\circ_{E(G',G'')}$ is the properad composition 
$$ \circ_{E(G',G'')}: \cC(\deg^{\operatorname{in}}(G'), \deg^{\operatorname{out}}(G')) \times \cC(\deg^{\operatorname{in}}(G''), \deg^{\operatorname{out}}(G'')) $$ $$ \to \cC(\deg^{\operatorname{in}}(G'\star G''), \deg^{\operatorname{out}}(G'\star G''))\, . $$
\end{enumerate}
\end{lem}


Note that in (3) of Lemma~\ref{ProperadSum} the properad composition
requires $E(G',G'')\neq \emptyset$. In the case with $E(G',G'')=\emptyset$
one can replace the properad composition with the monoidal operation.
This would correspond to generalizing properads to {\em props}, 
where composition along an empty overlap of outputs and
inputs is also allowed.


In our formulation of Lemma~\ref{ProperadSum}, the requirement that the $\cC(n,m)$
are full subcategories is motivated by the case of subcategories of
a category of computational systems (automata) where one fixes the number of inputs
and outputs. This is in contrast with the usual example of the category of 
vector spaces, with $\cC(n,m)$ given by spaces of linear maps from the $n$-th to the $m$-th 
powers, which would not be full subcategories. 


\begin{cor}\label{ProperadSumCor}
Let $G$ be a directed acyclic graph.
A network summing functor $\Phi \in \Sigma_\cC^{\operatorname{prop}}(G)$ is completely determined by its value on corollas.
\end{cor}

\begin{proof}
  At each vertex $v\in V_G$ consider the corolla $C(v)$. The functor $\Phi$ assigns values $\Phi_v:=\Phi(C(v))\in \cC(\deg^{\operatorname{in}}(v),\deg^{\operatorname{out}}(v))$. 
Consider a first vertex $v \in V_{G}$ and the associated value $\Phi_v$. 
Choose then a second vertex $w\in V_{G}$ with value $\Phi_w$. If $v\leq w$,
the subgraph $C(v)\star C(w)$ will have value
$\Phi(C(v)\star C(w))=\Phi_v \circ_{E(v,w)} \Phi_w$, with $E(v,w)$ the set of directed edges of $G$ 
connecting $v$ to $w$. Inductively, if $\Phi(G')$ has been
constructed for all subgraphs $G'\subset G$ with up to $n$ vertices that are lowersets 
for the partial order of the directed graph $G$, 
and $\# V_{G} >n$, then choose another vertex $u$ of $G$ not in $G'$. If $u\geq v$ for some $v\in G'$,
the subgraph $G'\star \{ u \}$
has $\Phi(G'\star \{ u \})=\Phi(G')\circ_{E(G',u)} \Phi_u$, and this determines the value on all lowerset subgraphs with $n+1$ vertices. The order of choice of the new
vertices does not matter because of the associativity condition of the 
properad operations, and the presence of external edges does not change the result
because of the unity condition of the properad, since external
edges are compositions with the properad unit. 
\end{proof}


We will see a more concrete instance of this type of construction in \S \ref{CompResSec}
and \S \ref{GammaCompArchSec}. 
External edges and the properad unit are also further discussed 
in \S 2.1.2 and \S 2.1.3 of \cite{Mar-new}.

\subsubsection{Inclusion-exclusion properties}\label{ExInclSec}

In the case where the category $\cC$ is an abelian category or a triangulated category, one can also make 
requirements on the dependence of the summing functors on subnetworks
through imposing inclusion-exclusion behavior. 

\begin{itemize}
\item If $\cC$ is an {\em abelian category}, one can in particular consider those
summing functors in $\Sigma_\cC(G)$ that satisfy an inclusion-exclusion relation, in the form of
exact sequences, namely summing functors such that, for all $G', G''\in P(G)$,
there is an exact sequence in $\cC$
$$ 0 \to \Phi_G(G'\cap G'') \to \Phi_G(G')\oplus \Phi_G(G'') \to \Phi_G (G'\cup G'')\to 0. $$
\item If $\cC$ is a {\em triangulated category}, one can consider those summing functors in $\Sigma_\cC(G)$
such that, for all $G', G''\in P(G)$, one has a Mayer--Vietoris type distinguished triangle
$$ \Phi_G(G'\cap G'') \to \Phi_G(G')\oplus \Phi_G(G'') \to \Phi_G (G'\cup G'')\to \Phi_G(G'\cap G'')[1]. $$
\end{itemize}
This choice determines a subcategory $\Sigma_\cC^{\text{incl/excl}}(G)\subset \Sigma_\cC(G)$ of summing
functors that satisfy a form of inclusion-exclusion.


The category of computational systems described in \S \ref{CompResSec}
does not have the structure needed to formulate this 
kind of inclusion-exclusion properties, although it is suitable for the grafting
conditions described in \S \ref{GraftingSec}, but the category of information
systems that we will discuss in \S \ref{GammaNetInfoSec} is an abelian category, 
so this type of summing functors will be relevant in that context.

\section{Neural information networks and resources}\label{NetsResourcesSec}  

In the previous section we have been referring to
a category $\cC$ which has zero object and sum or is a symmetric monoidal category as
a ``category of resources'', with the category of summing functors representing a
configuration space parameterizing all the possible assignments of resources to subsets of a set
or to subnetworks of a network. In this section we explain more precisely what we
mean by ``resources''.   Our discussion here is based primarily on the 
``mathematical theory of resources'' developed in \cite{CoFrSp16} and \cite{Fr17}.
This section serves as a general introduction to our understanding of resources,
while in the following sections, \S \ref{GammaNetCompSec} and \S \ref{GammaCodesSec},
we provide some more explicit and directly relevant examples of such categories of resources.


In modeling of networks of neurons, one can consider three different but closely
related aspects: the transmission of information with related questions of
coding and optimality, the sharing of resources and related issues of metabolic
efficiency, and the computational aspects. The third of these characteristics has
led historically to the development of the theory of neural networks, starting with
the McCulloch--Pitts model of the artificial neuron \cite{McCPit43} in the early
days of cybernetics research, all the way to the contemporary very successful theory
of deep learning \cite{GoBeCou16}. For the first two aspects mentioned above, a good discussion of the
computational neuroscience background can be found, for instance, in \cite{Sto18}.
One of our goals is to present ways of modeling the assignment to a network of resouces
describing its computational capacity, in terms of concurrent and distributed computing architectures,
consistently with informational and metabolic constraints. 

\subsection{Networks with informational and metabolic constraints}\label{MetabSec}

We consider here a kind of neuronal architecture consisting of
populations of neurons exchanging information via synaptic
connections and action potentials, subject to a tension of two
different kinds of constraints: metabolic efficiency and 
coding efficiency for information transmission. As discussed in \S 4 of \cite{Sto18},
metabolic efficiency and information rate are inversely related. The
problem of optimizing both simultaneously is reminiscent of another
similar problem of coding theory: the problem of simultaneous optimization,
in the theory of error-correcting codes, between efficient encoding (code rate)
and efficient decoding (relative minimum distance).  For a discussion
of error-correcting codes in the context of neural networks, see \cite{Man18}.
In order to model the
optimization of resources as well as of information transmission, we
rely on a categorical framework for a general mathematical theory
of resources, developed in \cite{CoFrSp16} and \cite{Fr17}, and on a
categorical formulation of information loss \cite{BaFrLei11}, \cite{BaFr14}, \cite{Mar19}. 
Before discussing the relevant categorical framework, we give a very quick overview of the main
aspects of the neural information setting, for which we refer the readers to \cite{Sto18} for
a more detailed presentation. 

\subsubsection{Types of neural codes}

There are different kinds of neural codes. There are 
binary codes that account only for the on/off information of which neurons in a
given population/network are firing. In these binary codes,
each code word is a binary string of some length $N$, which represents the
total number of time intervals $\Delta t$ considered. There is one code word
for each neuron in the given neuron population, with the $i$-th entry equal to $0$ or $1$
depending on whether that neuron has been firing during the $i$-th time interval.
Thus, we can view the code words as a binary (and coarse-grained by the choice of $\Delta t$)
representation of the spike train of the individual neurons. 
Comparing the $i$-th entry of all the code words shows which neurons in the
population considered have been simultaneously firing during that time interval.
This type of code allows for an
interesting connection to homotopy theory through a reconstruction of the homotopy
type of the stimulus space from the code, see \cite{Cu17}, \cite{Man15}. Different types
of coding are given by rate codes, where the input information is encoded in
the firing rate of a neuron, by spike timing codes, where the precise timing of spikes
carries information, and by correlation codes that use both the probability of a spike
and the probability of a specific time interval from the previous spike.

\subsubsection{Spikes, coding capacity, and firing rate}

Using a Poisson process to model spike generation, spikes are regarded
as mutually independent, given a firing rate of $y$ spikes per second. All long spike
trains generated at that firing rate are equiprobable. The information contained
in a spike train is computed by the logarithm of the number of different ways of
rearranging the number $n$ of spikes in the total number $N$ of basic time intervals 
considered. The neural coding capacity (the maximum coding 
rate $R$ for a given firing rate $y$) is given by the output entropy $H$ divided
by the basic time interval $\Delta t$. This can be approximated (\S 3.4 of \cite{Sto18})
by $R_{\max}=- y \log(y \Delta t)$. 

\subsubsection{Metabolic efficiency and information rate}

One defines the metabolic efficiency of a transmission channel as
the ratio $\epsilon=I(X,Y)/E$ of the mutual information 
$I(X,Y)$ of output $Y$ and input $X$ to the energy cost $E$ 
per unit of time, where the energy cost
is a sum of the energy required to maintain the channel
and the signal power. The latter represents the power required to 
generate spikes at a given firing rate. The energy cost of a spike
depends on whether the neuron axon is myelinated or not, and in
the latter case on the diameter of the axon. A discussion of optimal
distribution of axon diameters is given in \S 4.7 of \cite{Sto18}.


This description of metabolic efficiency shows in particular that an assignment
of informational resources (in the form of mutual information measurements)
to a network also governs the assignment of metabolic resources, once the
data about the channels that determine the energy costs $E$ are assumed
as known. This provides an example of interdependence between different
types of resources, which we will be discussing more extensively in \S \ref{GammaNetCompSec} 
and \S \ref{GammaCodesSec}.

\subsubsection{Connection weights and mutual information}

Over a fixed time interval $T$ subdivided into $N$ discrete steps $\Delta t$, and
a population of $K$ neurons that respond to a stimulus,
the output can be encoded as a $K\times N$ matrix $X=(x_{k,n})$, where the
$x_{k,n}$ entry records the output of the $k$-th neuron during the $n$-th time interval
in response to the stimulus. When this
output is transmitted to a next layer of $R$ cells (for example, in the visual system,
the output of a set of cones transmitted to a set of ganglion cells) an $R\times K$ 
weight matrix $W=(w_{r,k})$ assigns weights $w_{r,k}$ to each connection so
that the next input is computed by $y_{r,n}=\sum_{k=1}^K w_{r,k} x_{k,n}$. Noise
on the transmission channel is modeled by an additional term, $\eta=(\eta_{r,n})$
given by a random variable so that $Y= W X + \eta$. The optimization with respect
to information transmission is formulated as the weights $W$ that maximize
the mutual information $I(X,Y)$ of output and input. 


We see here another example of the interdependence between different
types of resources assigned to a network, where informational resources
depend on underlying resources of weighted codes,
as we will discuss more in detail in \S \ref{GammaCodesSec}.

\subsubsection{Resources and constraints}

In all the examples described above, one can see certain kinds of {\em resources}
associated to a network (energy and metabolic resources, neural codes, information)
subject to {\em constraints}, which are either intrinsic to a certain kind of resurce
or that involve the relation between different kinds of resources (such as the relation
between metabolic efficiency and information rate). What we want to argue in the
rest of this section is the fact that a categorical framework is especially suitable for
describing resources and assignments of resources to networks, in the form of
symmetric monoidal categories of resources and summing functors that describe
the assignments to networks. The categorical language also provides a setting
for describing constraints and relations between resources, in the form of
functors between categories of resources and universal properties, which are
a way of categorically describing optimality constraints. 

\subsection{The mathematical theory of resources}\label{ResourcesSec}

A general mathematical setting for a theory of resources was developed in \cite{CoFrSp16}
and \cite{Fr17}. We recall here the main setting and the relevant examples we need for the
context of neural information.


A theory of resources, as presented in \cite{CoFrSp16}, is a symmetric monoidal category
$(\cR,\circ,\otimes,{\mathbb I})$, where the objects $A\in {\rm Obj}(\cR)$ represent resources.
The product $A\otimes B$ represents the combination of resources $A$ and $B$, 
with the unit object ${\mathbb I}$ representing the empty resource.
The morphisms $f: A\to B$ in ${\rm Mor}_\cR(A,B)$ represent possible 
conversions of resource $A$ into resource $B$. In particular, no-cost resources
are objects $A\in {\rm Obj}(\cR)$ such that ${\rm Mor}_\cR({\mathbb I},A)\neq \emptyset$
and freely disposable resources are those objects for which ${\rm Mor}_\cR(A,{\mathbb I})\neq \emptyset$.
The composition of morphisms $\circ: {\rm Mor}_\cR(A,B)\times {\rm Mor}_\cR(B,C)\to {\rm Mor}_\cR(A,C)$
represents the sequential conversion of resources. 

\subsubsection{Examples of resources}\label{ExResSec}

Among the cases relevant to us are the two examples based on classical information mentioned in
\cite{CoFrSp16}, and another example of \cite{CoFrSp16} more closely related to the setting of \cite{Mar19}.
\begin{itemize}
\item {\em Resources of randomness}: the category $\cR={\rm FinProb}$ has objects the pairs $(X,P)$
of a finite set $X$ with a probability measure $P=(P_x)_{x\in X}$ with $P_x\geq 0$ and $\sum_{x\in X}P_x=1$,
and with morphisms ${\rm Mor}_\cR((X,P),(Y,Q))$ the maps $f: X\to Y$ 
satisfying the measure-preserving
property $Q_y=\sum_{x\in f^{-1}(y)} P_x$, and with product $(X,P)\otimes (Y,Q)=(X\times Y, P\times Q)$
with unit $(\{ * \}, 1_*)$ a point set with measure $1$. 
\item {\em Random processes}: the category $\cR={\rm FinStoch}$ with objects the finite sets $X$
and maps given by stochastic matrices $S=(S_{yx})_{x\in X, y\in Y}$ with $S_{yx}\geq 0$ for all $x\in X$ and $y\in Y$
and $\sum_{y\in Y} S_{yx}=1$ for all $x\in X$.
\item {\em Partitioned process theory}: the category considered in this case is the coslice category
${\mathbb I}/\cR$ of objects of $\cR$ under the unit object. This has objects given by the 
morphisms $f:{\mathbb I}\to A$, for $A\in {\rm Obj}(\cR)$, and morphisms 
$$ {\rm Mor}_{{\mathbb I}/\cR}((f:{\mathbb I}\to A), (g:{\mathbb I}\to B))=\{ (\xi: A\to B) \in {\rm Mor}_\cR(A,B)\, |\, \xi\circ f =g \}. $$
The category $\cF\cP$ of \cite{Mar19} has objects $(X,P)$ the pairs of a finite set with a probability
distribution $P=(P_x)_{x\in X}$ and morphisms ${\rm Mor}_{\cF\cP}((X,P),(Y,Q))$ given by the 
stochastic maps $S=(S_{y,x})_{x\in X, y\in Y}$ such that $Q= SP$. It is the coslice category $\cF\cP={\mathbb I}/{\rm FinStoch}$ with ${\rm FinStoch}$ the category of stochastic processes as in the previous example. 
\end{itemize}

In this last example, partitioned processes in \cite{CoFrSp16}  describe a theory of processes (resources and their
conversions, described by a symmetric monoidal category $\cC$) together with a subtheory of ``free processes''. No-cost resources
are precisely those objects of $\cC$ that have a morphism from the unit object, and ``states'' for this subtheory are described by processes with
input the unit object. 

\subsubsection{Convertibility of resources}\label{MeasSemigrSec}
The question of convertibility of a resource $A$ to a resource $B$ is formulated as the question of
whether the set ${\rm Mor}_\cR(A,B)\neq \emptyset$. Thus, to the symmetric monoidal category
$(\cR,\circ,\otimes,{\mathbb I})$ of resources, one can associate a preordered abelian monoid
$(R,+,\succeq, 0)$ on the set $R$ of isomorphism classes of ${\rm Obj}(\cR)$, 
with $[A]+[B]$ the class of 
$A\otimes B$ with unit $0$ given by the class of the unit object ${\mathbb I}$ and with $[A] \succeq [B]$ iff 
${\rm Mor}_\cR(A,B)\neq \emptyset$. The partial ordering is compatible with the monoid
operation: if $[A]\succeq [B]$ and $[C]\succeq [D]$ then $[A]+[C]\succeq [B]+[D]$.


The maximal conversion rate $\rho_{A\to B}$ between resources $A,B\in {\rm Obj}(\cR)$ is given by
\begin{equation}\label{maxconvrate}
 \rho_{A\to B}:= \sup \left\{ \frac{m}{n} \,\bigg|\, n\cdot [A] \succeq m \cdot [B], \, \, m,n\in \N \right\},
\end{equation} 
where $n \cdot [A]\in R$ is the class of $A^{\otimes n}$. 
It measures the optimal (maximal) fraction of number of copies of resource $B$ that can be produced by $A$.


Given an abelian monoid with partial ordering 
$(S,*, \geq, 1_S)$, an $S$-valued measuring of $\cR$-resources is a
monoid homomorphism $M: (R,+,0)\to (S,*,1_S)$ such that 
$M(A)\geq M(B)$ in $S$ whenever $[A]\succeq [B]$ in $R$.
(Here and below we write $M(A)$ as shorthand for $M([A])$.)


For $(S,*)=(\R,+)$ and $M: (R,+)\to (\R,+)$ a measuring monoid homomorphism, we have (Theorem~5.6 of \cite{CoFrSp16})
$$ \rho_{A\to B} \cdot M(B) \leq M(A), $$
that is, the optimal fraction of copies of resource $B$ that one can obtain using resource $A$
is not bigger than the value of $A$ relative to the value of $B$. 

\subsubsection{Information loss}\label{InfoLossSec}

A characterization of information loss is given in \cite{BaFrLei11} as a map $F: {\rm Mor}_{{\rm FinProb}} \to \R$
satisfying 
\begin{enumerate}
\item additivity under composition $F(f\circ g)=F(f)+F(g)$; 
\item convex linearity $F(\lambda f \oplus (1-\lambda)g)
= \lambda F(f)+(1-\lambda) F(g)$ for $0\leq \lambda \leq 1$ and for $\lambda f \oplus (1-\lambda)g: (X\sqcup Y, \lambda P\oplus (1-\lambda) Q) \to (X'\sqcup Y', \lambda P'\oplus (1-\lambda) Q')$ the convex combination of
morphisms $f:(X,P)\to (X',P')$ and $g:(Y,Q)\to (Y',Q')$ in  ${\rm FinProb}$;
\item continuity of $F$ over ${\rm Mor}_{{\rm FinProb}}$.
\end{enumerate}
The Khinchin axioms for the Shannon entropy can then be used to show that 
an information-loss
functional satisfying these properties is necessarily of the form $F(f)=C \cdot (H(P)-H(Q))$ for some $C>0$
and for $H(P)=-\sum_{x\in X} P_x \log P_x$ the Shannon entropy.  When working with the category $\cF\cP={\mathbb I}/{\rm FinStoch}$, a similar characterization of information loss using the Khinchin axioms for the Shannon entropy
is given in \S 3 of \cite{Mar19}.  

\subsection{Adjunction and optimality of resources}\label{AdjunctSec}

The discussion in this subsection is not directly needed for our main goal in this paper, but
it is included here because it provides a better intuition on how to think of optimization 
processes in categorical terms. 


Suppose then that we have a category $\cC$ as above that models distributed/concurrent
computational architecture (a category of transition systems or of higher dimensional
automata, see \S \ref{GammaNetCompSec} below). 
We also assume that we have a category $\cR$ describing metabolic or
informational resources. The description of the resource constraints associated to a
given automaton is encoded in a strict symmetric monoidal functor $\rho: \cC \to \cR$. 
The property of being strict symmetric monoidal here encodes the requirement that
independent systems combine with combined resources.  


A stronger property would be to require that the functor $\rho: \cC \to \cR$ that assigns resources to computational systems has a left adjoint,
a functor $\beta: \cR\to \cC$ such that for all objects $C\in {\rm Obj}(\cC)$ and $A\in {\rm Obj}(\cR)$
there is a bijection 
\begin{equation}\label{adjoinCR}
 {\rm Mor}_\cC( \beta(A), C) \simeq {\rm Mor}_\cR (A,\rho(C)) .
 \end{equation} 
 The meaning of the left-adjoint functor and the adjunction formula \eqref{adjoinCR} can
 be understood as follows. In general an adjoint functor is a solution to an optimization problem.
 In this case the assignment $A \mapsto \beta(A)$ via the functor $\beta: \cR\to \cC$ is an
 optimal way of assigning a computational system $\beta(A)$ in the category $\cC$ to 
 given constraints on the available resources, encoded by the object $A\in {\rm Obj}(\cR)$.
 The optimization is expressed through the adjunction \eqref{adjoinCR}, which states that
 any possible conversion of resources from $A$ to the resources $\rho(C)$ associated
 to a system $C\in {\rm Obj}(\cC)$ determines in a unique way a corresponding modification 
 of the system $\beta(A)$ into the system $C$. Note, moreover, that the system $\beta(A)$
 is constructed from the assigned resources $A\in {\rm Obj}(\cR)$, and since some of the
 resources encoded in $A$ are used for the manufacturing of $\beta(A)$ one expects that
 there will be a conversion from $A$ to the remaining resources available to the system $\beta(A)$,
 namely $\rho(\beta(A))$. 
 The existence of the left-adjoint $\beta: \cR\to \cC$ (hence the possibility of solving this
 optimization problem) is equivalent to the fact that the conversion of resources
 $A \to \rho(\beta(A))$ is the initial object in the category $A \downarrow \rho$.
 Here, for an object $A\in {\rm Obj}(\cR)$ the comma category $A\downarrow \rho$ of objects $\rho$-under $A$
 has objects the pairs $(u,C)$ with $C\in {\rm Obj}(\cC)$ and $u: A \to \rho(C)$ a morphism in $\cR$ and
 morphisms $\phi: (u_1,C_1) \to (u_2, C_2)$ given by morphisms $\phi\in {\rm Mor}_\cC (C_1,C_2)$ such
 that one has the commutative diagram
 $$ \xymatrix{ & A \drto^{u_2} \dlto_{u_1} & \\ \rho(C_1) \rrto^{\rho(\phi)} & & \rho(C_2) \, .} $$

 
 Freyd's adjoint functor theorem gives a condition for the existence of a left-adjoint
 functor for a continuous functor $\rho: \cC \to \cR$, in the form of a completeness 
 condition on the category $\cC$ and the existence of a {\em solution set} in the 
 comma category $A\downarrow \rho$. We briefly discuss what this result means
 in our setting. 
 
 
 The functor $\rho: \cC \to \cR$ is continuous if it commutes with limits. This is
 a reasonable assumption to make regarding the functor that assigns to a
 computational system $C$ in the category $\cC$ its resources in the category $\cR$.
 As discussed in \S 3 of \cite{Per19}, categorical limits are solutions to constrained
 optimization problems that generalize to the categorical setting the usual notion
 of infimum (indeed the categorical limit agrees with the notion of greatest lower 
 bound in the case of a category given by a poset). Requiring that the functor
 that assigns resources to systems is continuous means requiring that it preserves
 the optimization properties encoded in categorical limits. 
 
 
 The completeness of the category $\cC$ depends on which models of
 concurrent and distributed computing we are considering in the category $\cC$.
 We will be working broadly with the framework of a
 category $\cC$ of transition systems introduced in \cite{WiNi95} as a model
 for computational architectures, see \S \ref{GammaNetCompSec}. 
   However, one can focus on more specific categorical models of
 concurrency. For example, among the categories considered in \cite{WiNi95}, 
 the category of synchronization trees has infinite products and
 pullbacks, hence it is also complete.
 
 
 If our category $\cC$ is complete, as in the cases mentioned above, and
 the functor $\rho: \cC \to \cR$ preserves infinite products and equalizers, then 
 the comma category $A\downarrow \rho$ is also complete for all objects $A\in {\rm Obj}(\cR)$.
 In this case 
 Freyd's adjoint functor theorem then shows that the existence of an initial object in the
 category $A \downarrow \rho$ (hence the existence of a left-adjoint functor 
 $\beta: \cR\to \cC$ for $\rho: \cC \to \cR$) follows from the existence of a {\em solution set}, 
 that is, a set $\{ T_j=(u_j,C_j) \}_{j\in J}$ of objects of $A \downarrow \rho$ such that every object $T=(u,C) \in {\rm Obj}(A \downarrow \rho)$  admits a morphism $f_j: T_j \to T$ for some $j\in J$.  
 
 
 The existence of a solution set can be interpreted in the following way. 
 If we fix the resources by choosing an object $A\in {\rm Obj}(\cR)$, 
 there is a set $\{ C_j \}_{j\in J}$ of systems in $\cC$ together with
 conversion of resources $u_j: A \to \rho(C_j)$ with the property that,
 for any system $C\in {\rm Obj}(\cC)$ for which there is a possible
 conversion of resources $u: A \to \rho(C)$ in ${\rm Mor}_\cR(A,\rho(C))$,
 there is one of the systems $C_j$ and a modification of systems
 $\phi: C_j \to C$ in ${\rm Mor}_\cC(C_j,C)$ such that the conversion
 of resources $u: A \to \rho(C)$ factors through the system $C_j$, namely
 $u=\rho(\phi)\circ u_j$. One can therefore think of the solution set
 $\{ (u_j,C_j) \}_{j\in J}$ as being those systems in $\cC$ that are optimal
 with respect to the resources $A$, from which any other system 
 that uses less resources than $A$ can be obtained via modifications.

\section{Networks with computational structures}\label{GammaNetCompSec}

In this section we focus on assignments of computational resources to a network,
which we think of as computational models of individual nodes (neurons) of the
network, together with prescriptions for their wiring together according to
the network structure. As in the previous section, we aim at constructing a
configuration space of all such possible assignments within which one can
choose an initial assignment and prescribe a dynamical evolution. We will
deal with the dynamical aspect in \S \ref{HopfieldSec}.  Here we
introduce a suitable category of computational resources, aimed at
accommodating a sufficiently broad and flexible range of models of
concurrent and distributed computing, in the form of automata describing
transition systems. We then investigate the compositional structure that
gives the compatibility of these assignments over the network. We discuss
some related questions, including how to incorporate some
computational models of neuromodulation based on a subcategory of the
category of transition systems given by time-delay automata.

\subsection{Transition systems: a category of computational resources}\label{CompResSec}

We consider here, as a special case of categories of resources, in the sense of  \cite{CoFrSp16}
and \cite{Fr17} recalled above, a category of ``reactive systems'' in the sense of \cite{WiNi95}. 
These describe models of computational architectures that involve parallel and distributed 
processing, including interleaving models 
such as synchronization trees and concurrency models based on causal independence. 
Such computational systems can be described in categorical terms, formulated as a category
of transition systems \cite{WiNi95}. 
The products in this category of transition systems represent parallel compositions where all possible
synchronizations are allowed. More general parallel compositions are then obtained as combinations of 
products, restrictions and relabeling. The coproducts in the category of transition systems represent
(non-deterministic) sums that produce a single process with the same computational 
capability of two or more separate processes. 


In the most general setting, a category $\cC$ of transition systems has objects given by
data of the form $\tau=(S,\iota, \cL,\cT)$ where $S$ is the set of possible states of the system,
$\iota$ is the initial state, $\cL$ is a set of labels, and $\cT$ is the set of possible transition
relations of the system, $\cT \subseteq S \times \cL \times S$ (specified 
by pre state, label of the transition, and post state). A transition system $\tau=(S,\iota, \cL,\cT)$ 
also has a set $S_F\subset S$ of final states.
Such a system can be represented in graphical notation
as a directed graph with vertex set $S$ and with set of labeled directed edges $\cT$.
Morphisms ${\rm Mor}_{\cC}(\tau,\tau')$ in the category $\cC$ of transition systems are given by 
pairs $(\sigma,\lambda)$ consisting of a function $\sigma: S \to S'$  with $\sigma(\iota)=\iota'$
and $\sigma(S_F)\subset S'_F$, and
a (partially defined) function $\lambda: \cL \to \cL'$ of the labeling sets such that, for any
transition $s_{in} \stackrel{\ell}{\to} s_{out}$ in $\cT$, if $\lambda(\ell)$ is defined, 
then $\sigma(s_{in}) \stackrel{\lambda(\ell)}{\to} \sigma(s_{out})$ is a transition in $\cT'$.

\begin{figure}[t]
\begin{center}
 \includegraphics[scale=0.45]{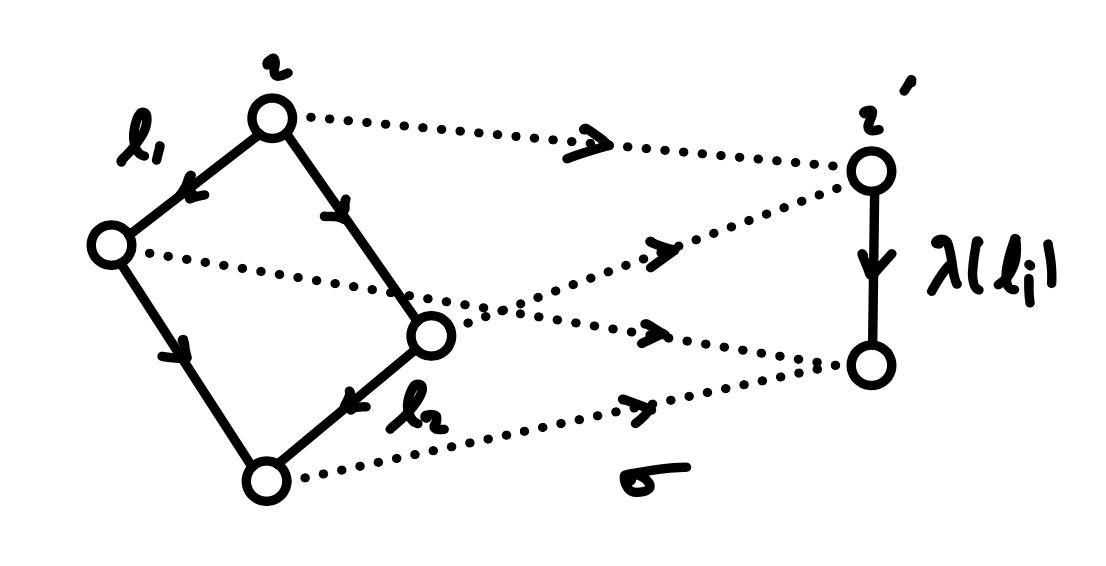}  
\caption{Example of a morphism of transition systems.  \label{fig:figure1}}
 \end{center} 
\end{figure}

Heuristically, a morphism $(\sigma,\lambda)\in {\rm Mor}_{\cC}(\tau,\tau')$ describes the
fact that the system $\tau'$ can partially simulate the system $\tau$, where ``partially''
is determined according to $\lambda$, see \cite{Nielsen}. A simple explicit example of a 
morphism of transition systems is given graphically in Figure~\ref{fig:figure1} (see \cite{Nielsen}).

As shown in \cite{WiNi95}, the category $\cC$ has a coproduct given by
\begin{equation}\label{TransCoprod}
 (S,\iota,\cL,\cT)\sqcup (S',\iota',\cL', \cT') = (S\times \{ \iota' \} \cup \{ \iota \} \times S', (\iota, \iota'), \cL \cup \cL', \cT\sqcup \cT' ) 
 \end{equation}
$$ \cT \sqcup \cT':= \{ (s_{in}, \ell, s_{out})\in \cT \}\cup \{ (s'_{in}, 
\ell', s'_{out})\in \cT'\} , $$
where both sets are seen as subsets of 
$$ (S\times \{ \iota' \} \cup \{ \iota \} \times S')\times (\cL\cup \cL') \times (S\times \{ \iota' \} \cup \{ \iota \} \times S'). $$
This coproduct $(S,\iota,\cL,\cT)\sqcup (S',\iota',\cL', \cT')$ satisfies the universal property of a categorical sum. 
The zero object is given by the stationary single-state system $S=\{ \iota \}$ with empty labels and transitions. 
There is also a product structure on $\cC$ given by
$$ (S\times S', (\iota,\iota'), \cL\times \cL', \Pi) , $$
where the product transition relations are determined by $\Pi=\pi^{-1}(\cT)\cap {\pi'}^{-1}(\cT')$, for the projections
$\pi: S\times S' \to S$ and $\pi: \cL \times \cL'\to \cL$ and $\pi': S\times S'\to S'$ and $\pi': \cL\times \cL'\to \cL'$.

\begin{figure}[t]
\begin{center}
 \includegraphics[scale=0.45]{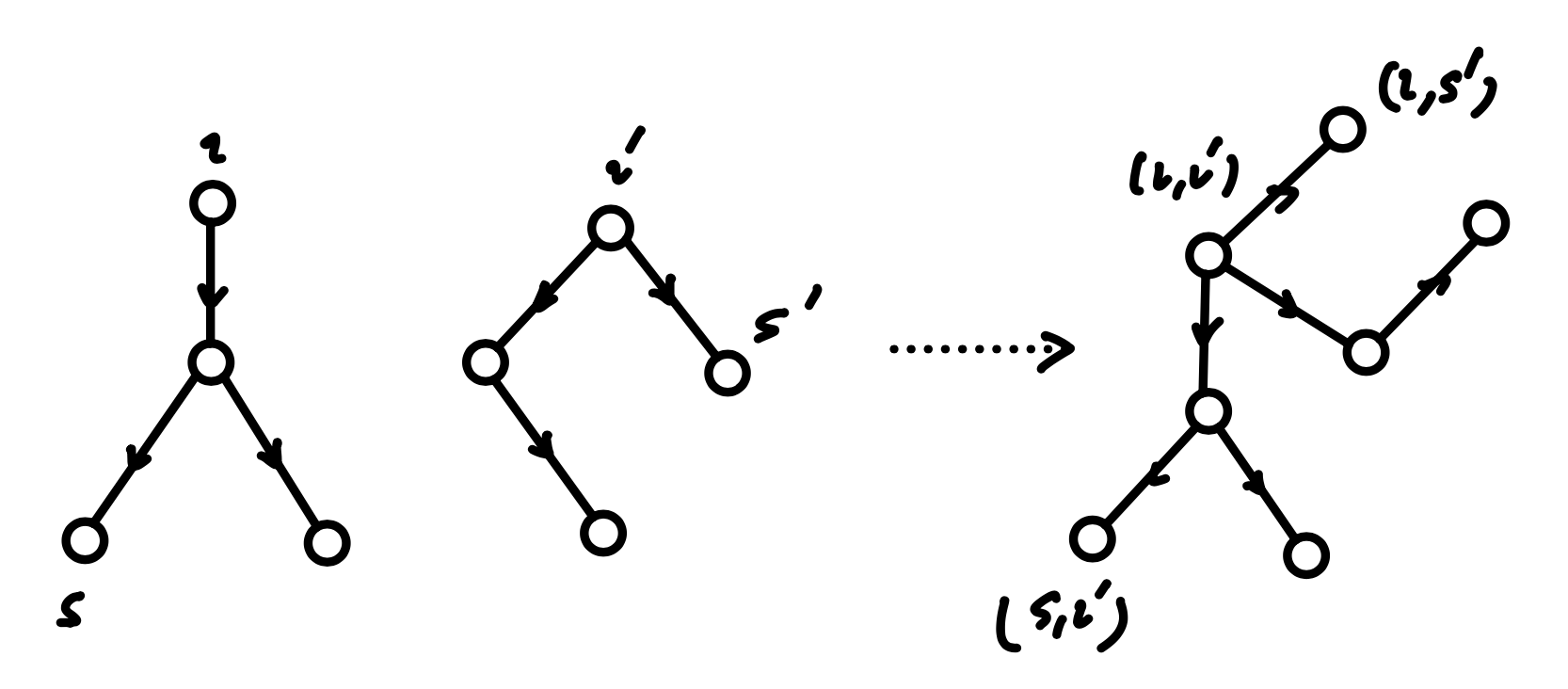}  
 \caption{A simple example of coproduct of two transition systems.\label{fig:figure2}}
 \end{center} 
\end{figure}

The coproduct of two transition systems is illustrated graphically in a simple example in Figure~\ref{fig:figure2}.
As observed in \S 2.2.5 of \cite{WiNi95}, this categorical sum in the category $\cC$ of transition systems 
represents a system that can behave as any one of its summands. 

Note that \eqref{TransCoprod} is a categorical coproduct only in the case of labeled transition systems 
with marked initial state. In the case where there is also a marked final state, this is no longer the case,
but one can still define a monoidal structure. 


A version of probabilistic transition systems is discussed in the Appendix, in \S \ref{ProbTransSec}.

\subsection{Computational architectures in neuronal networks}\label{GammaCompArchSec}

We first review here some ideas about
computational models for single neurons and how they can be made to fit with the very broad description of
computational architectures provided by the category of transition systems.


In this context we can treat a computational model for a single neuron in terms of
a sequence of simplifying steps. These follow the discussion in the introduction of 
\cite{Ibarz}.
\begin{itemize}
\item {\em Discretization in space} makes it possible to subdivide a neuron into separate ``modules'',
and replace a model of the relevant quantities such as membrane voltage in terms of
a set of PDEs into a model in terms of ODEs. This is a classical simplification of the problem,
which leads to the well-known Hodgkin--Huxley model~\cite{HoHu52}.
\item {\em Discretization in time} further replaces the continuum-time ODE with a discrete dynamical system.
We will discuss again this kind of step in relation to our categorical Hopfield network dynamics in \S \ref{HopfieldSec}.
\item {\em Discretization in field values} then makes it possible to model the discrete dynamical
system in terms of a finite state automaton.
\end{itemize}
If we follow this outline as in \cite{Ibarz}, then we would be assigning to single neurons (vertices $v\in V=V_G$
in the network) corresponding finite state automata. These are particular cases of the more general
objects in the category of transition systems of \cite{WiNi95} described in \S \ref{CompResSec}.


Another model of the computational structure of a single neuron is developed in \cite{BeSeLo}.
In this model the input-output mapping complexity of neurons is investigated by identifying
deep neural networks that can be trained to faithfully replicate the input-output function of various 
types of cortical neurons at millisecond spiking resolution. So for example a layer-$5$ cortical 
pyramidal cell requires a  convolutional deep neural network with five to eight layers, while a
minimal deep neural network with a single hidden layer suffices for the simple integrate-and-fire
neuron model. In this case, the computational structures associated to (different types of) neurons
are deep neural networks. Thus, in order to cast this model into our framework, one needs to
formulate the right categorical structure describing compositional roles of neural networks and
a relation to the category of transition systems described above. This will appear in a separate
paper~\cite{Mar-new}, so we will not include the discussion here, but we can direct the reader to 
\cite{FioCam}, \cite{FoSpiTu}, \cite{Ganchev} for some of the relevant categorical setting for
deep neural networks.


There is also another possible approach to assigning a computational system to the
individual neurons, as suggested in \cite{Bjerk}, by considering the system of
ion-gated channels in the membrane as a concurrent computing system where
synaptic inputs interact to modulate activity with shared resources 
(represented by different ion densities and thresholds), regarded as a system of 
interacting synaptic ``programs''. We do not develop this model in the present 
paper, but this would be a very natural approach in view of representing the
entire computational architecture of the network in terms of concurrent/distributed
computing. Such models would also fit within the category of transition systems
described above, and with dynamical models of interacting neuron populations
such as \cite{KnMaSi96}.

\subsection{Computational architectures and network summing functors}\label{NeuronCompSec}

We now look more closely at categories of network summing functors, as discussed in \S \ref{SummingSec},
where the target category is the category of transition systems of \cite{WiNi95} that we recalled in \S \ref{CompResSec} above.
In particular we will discuss what specific conditions on network summing functors it is reasonable to require in such
a model, or equivalenty what subcategory of $\Sigma_\cC(G)$ one wants to focus on,  with additional structure that takes 
into account local and larger-scale connections in the network. In particular, we show that a model of network
summing functors based on grafting operations, similar to what we discussed more abstractly in \S \ref{GraftingSec} is
especially suitable for assignments of computational resources to networks in the form of  transition systems.  
A model of assignment of resources more directly built on the properad grafting operations of \S \ref{GraftingSec} will
be discussed in a separate paper~\cite{Mar-new}, in relation to the deep neural networks model of computational resources of
individual neurons of \cite{BeSeLo}. 

\subsubsection{Transition systems and network summing functor}\label{CompArchSec}

Let $\cC$ be the category of transition systems of \cite{WiNi95} described
in \S \ref{CompResSec} Let $\cG:={\rm Func}({\bf 2}, \cF)$ be the category
of finite directed graphs. As before, for $G\in {\rm Obj}(\cG)$ we denote 
by
$G_*$ the associated pointed graph. For simplicity we write
$\Sigma_\cC(V_G)$ instead of $\Sigma_\cC(V_{G_*})$ with the pointed
structure implicitly understood.


\begin{defn}\label{fibersumautomata}
For $i=1,2$ let $\tau_i=(S_i,\iota_i,\cL_i,\cT_i)$ be objects in the category $\cC$ of transition systems.
Given a choice of two states $s\in S_1$ and $s'\in S_2$, the grafting of $\tau_1$ and $\tau_2$ is
the object $\tau_{s,s'}=(S,\iota,\cL,\cT)$ in $\cC$ with $S=S_1 \sqcup S_2$, $\iota=\iota_1$, 
$\cL=\cL_1\sqcup \cL_2\sqcup \{ e \}$ and $\cT=\cT_1\sqcup \cT_2 \sqcup \{ (s,e,s') \}$. Let $\cC'\subset \cC$
be the subcategory of transition systems $\tau$ that have a single final state $S_F=\{ q\} \subset S$. For 
$\tau_i\in {\rm Obj}(\cC')$, the grafting $\tau_1\star \tau_2$ is simply defined as the grafting 
$\tau_{q_1,\iota_2}$ with the final state of $\tau_1$ grafted to the initial state of $\tau_2$.
\end{defn}


A topological ordering $\omega$ of the vertices of a directed acyclic graph $G$
is a linear ordering of the set of vertices such that, whenever there is a directed edge $e$
with $s(e)=v$ and $t(e)=v'$ then $v\leq v'$ in the ordering, that is, 
a monotone map from the underlying poset of the vertices to a linear order.


\begin{lem}\label{acyclicgraft}
Let $G$ be a finite acyclic directed graph with vertex set $V=V_G$. Let $\omega$ be a 
topological ordering of the vertex set $V$. 
Suppose given a collection $\{ \tau_v \}_{v\in V}$
of objects in the subcategory $\cC'$ of $\cC$. There is a well-defined grafting $\tau_{G,\omega}$
of the $\tau_v$ that is also an object in $\cC'$. 
\end{lem}

\begin{proof}
  For $v\in V$, we have $\tau_v=(S_v,\iota_v, \cL_v, \cT_v)$. Since $\tau_v$ is in $\cC'$, the set
$S_v$ contains a unique final state $q_v$. Let $v_{in}$ denote the first vertex and $v_{out}$ 
the last vertex in the topological ordering $\omega$. 
The object $\tau_{G,\omega}=(S,\iota,\cL,\cT)$ has $S=\cup_{v\in V} S_v$ with initial state
$\iota=\iota_{v_{in}}$ and final state $q=q_{v_{out}}$. The set of labels is given by 
$\cL=\cup_{v\in V} \cL_v\cup E$ with $E=E_G$ the set of edges of $G$ and transitions
$\cT=\cup_{v\in V} \cT_v \cup \{ (q_{s(e)}, e, \iota_{t(e)}) \,|\, e\in E \}$ with $s(e),t(e)$ the
source and target vertices of $e$.
\end{proof}

\begin{figure}[t]
\begin{center}
 \includegraphics[scale=0.45]{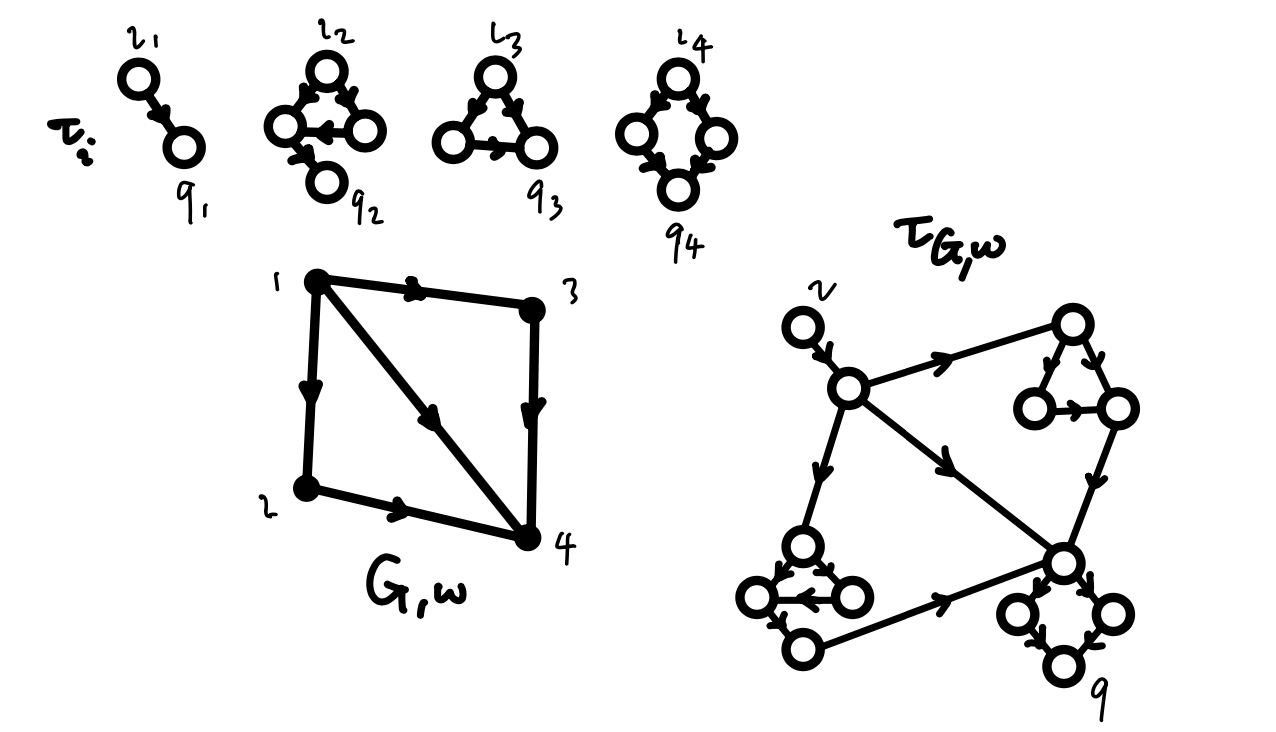}  
 \caption{A simple example of the grafting operation of Lemma~\ref{acyclicgraft}. \label{fig:figure3}}
 \end{center} 
\end{figure}

The grafting operation of Lemma~\ref{acyclicgraft} is illustrated in a
simple example in Figure~\ref{fig:figure3}.

For an arbitrary finite directed graph $G$, a strongly connected component is a subset
$V'$ of the vertex set $V_G$ such that each of the vertices in $V'$ is reachable through
an oriented path in $G$ from any other vertex in $V'$, and which is 
maximal with respect to this property. The strongly connected components
determine a partition of $V_G$. The condensation graph $\bar G$ is a directed acyclic
graph that is obtained from $G$ by contracting each  strongly 
connected component (consisting of the vertices of the component and all the 
edges between them) to a single vertex. Given two strongly connected
components $X\neq Y$, there is an edge $e_{X,Y}$ connecting the corresponding vertices 
in the condensation graph $\bar G$ if there is an edge $e_{v,w}$ in $G$ for some
$v\in X$ and $w\in Y$.


There are algorithms that construct a topological ordering on a directed acyclic
graph in linear time, such as the Kahn algorithm~\cite{Kahn}. For a given
directed graph $G$ we write $\bar\omega$ for the topological ordering of
its condensation graph $\bar G$ obtained through the application of a 
given such algorithm. 


\begin{defn}\label{Gstrongconn}
Let $G$ be a strongly connected graph and let $\{ \tau_v \}_{v\in V_G}$ be
a collection of objects $\tau_v=(S_v,\iota_v, \cL_v, \cT_v)$ in $\cC'$ with
$q_v$ the respective final states. For a given pair $(v_{in}, v_{out})$ in $V_G\times V_G$
let $\tau_{G,v_{in}, v_{out}}=(S,\iota,\cL,\cT)$ be the object in $\cC'$ with 
$S=\cup_{v\in V_G} S_v$, $\cL=\cup_{v\in V_G} \cL_v \cup E_G$, and
$\cT=\cup_{v\in V_G} \cT_v \cup \{ (q_{s(e)},e,\iota_{t(e)} ) \}_{e\in E_G}$ and with
initial and final state $\iota=\iota_{v_{in}}$ and $q=q_{v_{out}}$. Then set 
$\tau_G:=\oplus_{(v_{in}, v_{out})\in V_G\times V_G} \tau_{G,v_{in}, v_{out}}$.
\end{defn}


Notice that this definition represents correctly what one heuristically expects to
be the grafting for a strongly connected graph. In a transition system a state
is reachable if there is a directed path of transitions from the initial state $\iota$
to that state. In particular a final state is assumed to be reachable. 
A transition system is reachable if every state is reachable. 
Since in the strongly connected case any vertex can be reached via a directed path 
from any other, then any of the initial states $\iota_v$ of the systems $\tau_v$
can be taken to be the initial state of the grafting, and any final state $q_v$
can be taken as the final state of the grafting.
The grafting $\tau_G$ for a strongly connected graph $G$ represents 
a transition system that can behave as the grafting of the $\tau_v$ with
any possible pair $(\iota_v, q_{v})$ as the initial and final state. 

\begin{figure}[t]
\begin{center}
 \includegraphics[scale=0.37]{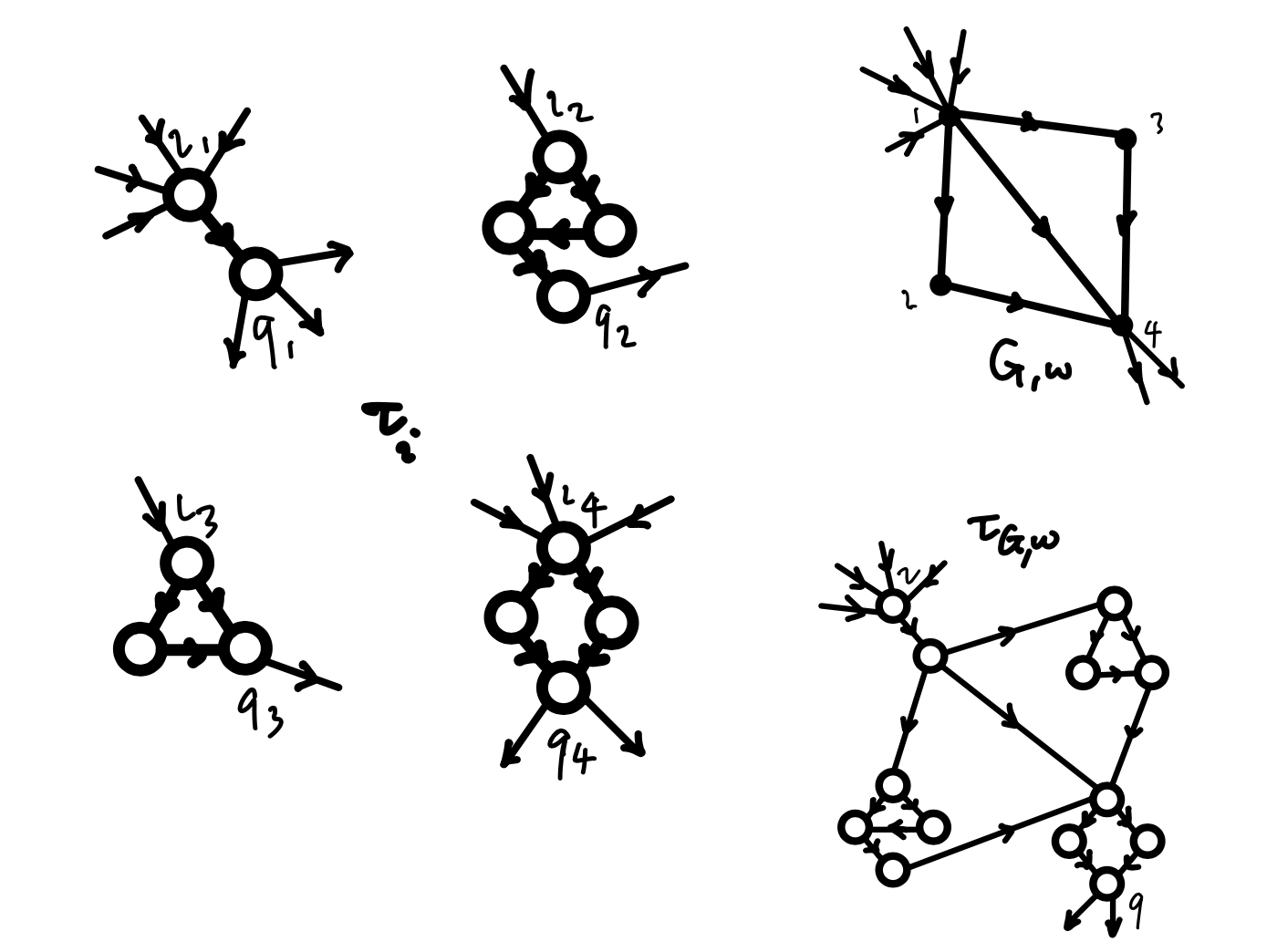}  
 \caption{Grafting operation with matching external edges. \label{fig:figure3b}}
 \end{center} 
\end{figure}

In Lemma~\ref{acyclicgraft} 
(see also the example in Figure~\ref{fig:figure3}) we have described
simple grafting operations at vertices. More generally, and more realistically, the grafting 
should also involve a matching of external (half)edges and can be formulated following 
the setting of \S \ref{GraphsSec} and \S \ref{GraftingSec}.
The corresponding modifications of Lemma~\ref{acyclicgraft} is
are straightforward. An example illustrating this form of grafting is given in
Figure~\ref{fig:figure3b}. An explicit example where the grafting is directly modeled on
Lemma~\ref{ProperadSum}, with a category of deep neural networks, is discussed in
\cite{Mar-new}.

In this case, we assign to the initial state $\iota$ and the final state $q$ an in-degree
and an out-degree, respectively. The meaning of these in/out degrees and the attached
half-edges is that the output computed at the final state $q$ is made available as pre state 
on all the outgoing external half-edges, and similarly, the initial state $\iota$ is made available 
as post state on each of the incoming external half-edges. 
When endowed with these additional data, we can organize the objects of the 
category $\cC'$ into subsets $\cC'(n,m)$ consisting of those transition systems $\tau$
with $n=\deg^{\rm in}(\iota)$ and $m=\deg^{\rm out} (q)$. The category $\cC'$ then has
a properad composition that matches outputs to inputs. 
We now construct an associated category of network summing functors that
satisfy grafting conditions, as discussed in \S \ref{GraftingSec}. 


\begin{prop}\label{compfunctorprop}
Given a network $G$, there is a faithful functor $\Upsilon: \Sigma_{\cC'}(V_G) \to \Sigma^{\operatorname{prop}}_{\cC'}(G)$,
with $\cC'$ the subcategory of transition systems of Definition~\ref{fibersumautomata}, 
with the target category as introduced
in \S \ref{GraftingSec}, with the $\cC'(n,m)\subset \cC'$ and the properad composition as described here above.
\end{prop}

\begin{proof}
  By Lemma~\ref{PhiPtsLem}, a summing functor $\Phi\in \Sigma_{\cC'}(V_G)$ is completely determined
by the assignment of the objects $\Phi(v) \in \cC'$. The morphisms are invertible natural transformations that
are in turn determined by isomorphisms of these objects. Given $\Phi\in \Sigma_{\cC'}(V_G)$ we construct an
associated summing functor, $\Upsilon(\Phi)$ in $\Sigma^{\operatorname{prop}}_{\cC'}(G)$, where the
composition operations on the target category $\cC'$ are the grafting operations described above 
in Lemma~\ref{acyclicgraft} and Definition~\ref{Gstrongconn}, in the modified form that accounts for matching of
external edges at the grafting of final and initial state, as discussed above. For $G'\subseteq G$ we set 
$\Upsilon(\Phi)(G')$ to be equal to the object in $\cC'$ obtained as the grafting $\tau_{\bar G',\bar\omega}$
as in Lemma~\ref{acyclicgraft} of the objects $\tau_{G'_i}$, with $G_i'$ the strongly connected components
of $G'$, with $\tau_{G'_i}$ given by the grafting of Definition~\ref{Gstrongconn} of the $\Phi(v)$ associated to 
the vertices of $G'_i$, once matching of external edges is included (as in Figure~\ref{fig:figure3b}). 
Since morphisms in $\Sigma_{\cC'}(V_G)$ are given by isomorphisms of the $\Phi(v)$,
these induce isomorphisms of the grafted objects, hence invertible natural transformations of the obtained
summing functors $\Upsilon(\Phi)$.
\end{proof}


\subsection{Larger-scale structures and distributed computing}\label{NeuromodSec}

We have shown in Proposition~\ref{compfunctorprop} how to obtain
a functorial assignment of a computational structure in the category 
of transition systems of \cite{WiNi95} to a network of neurons related
by synaptic connections, assuming a computational model for individual neurons is given. 
This construction is based on a given model of local automata that implement the
computational properties of individual neurons with their pre-synaptic and 
post-synaptic activity (for example the map-based model of \cite{Ibarz} or
the deep network model of \cite{BeSeLo}) and on the grafting of these automata 
into a  larger computational structure where their inputs and outputs
are connected according to the connectivity of the network. 


As discussed in \cite{Potj}, there are larger-scale structures involved in
the computational structure of neuronal arrangements beyond what is
generated by the pre-synaptic and post-synaptic activity. In particular,
non-local neuromodulation bridges between the microscopic and the
larger-scale structures and plays a role in synaptic plasticity and learning.
These are not captured by the construction of Proposition~\ref{compfunctorprop}.
Thus, such phenomena provide a reason why a suitable subcategory of the
category $\Sigma_\cC(G)$ of network summing functors may have to be
larger than that accounting for summing functor built from some type of
grafting operations (which reflect only the local connectivity of the network).


Neuromodulators are typically generated in neurons in the brainstem 
and in the basal forebrain and transmitted to several different 
brain regions via long-range connections. As shown in \cite{Potj},
this kind of larger-scale structure of neuromodulated plasticity, where
the neuromodulatory signal is generated within the network, 
is better accounted for by a distributed computing model. The focus
in \cite{Potj} is on efficient simulation, in a distributed environment,
of a neuromodulated network activity. Here we have a somewhat
different viewpoint as we are interested in a computational
architecture that can be realized by the network with its local and
large-scale structure. Nonetheless, the model developed in \cite{Potj}
can be useful in identifying how to go beyond the local structure
encoded in the construction given in Proposition~\ref{compfunctorprop}.

\subsubsection{Distributed computing model of neuromodulation}\label{machinesSec}

The distributed computing model considered in \cite{Potj} can be
summarized as follows:
\begin{itemize}
\item The network of neurons and synaptic connections is described 
by a finite directed graph $G$.
\item The set of vertices $V=V_G$ is partitioned into $N$ subsets $V_i$,
the different machines ${\bf m}_i$ of the distributed computing system.
\item The set of edges $E=E_G$ is partitioned into the machines ${\bf m}_i$
by the rule that an edge $e$ belongs to ${\bf m}_i$ iff the target vertex  
$t(e)$ belongs to $V_i$.
\item One additional vertex $v_{0,i}$ is added into each machine ${\bf m}_i$,
which accounts for the neuromodulator transmission.
\item There is a set $E_{0,i}$ of additional edges connected to the 
vertices $v_{0,i}$ in ${\bf m}_i$: the incoming edges $e\in E_{0,i}$ with
$t(e)=v_{0,i}$ can have source vertex $s(e)$ anywhere in the graph $G$, not
necessarily inside ${\bf m}_i$, while the outgoing edges $e\in E_{0,i}$ with $s(e)=v_{0,i}$
have their target vertex in the same machine, $t(e)\in V_i$. 
\item We obtain in this way a new directed graph $G_0$ obtained from $G$ by adding the vertices
$v_{0,i}$ and the edges in the sets $E_{0,i}$. 
\item The vertices that are sources of edges in $E_{0,i}$ with target $v_{0,i}$ are
the neurons that release the neuromodulator, while the edges in $E_{0,i}$ outgoing from
$v_{0,i}$ represent the synaptic connections that are neuromodulated. 
The nodes $v_{0,i}$ collect {\em globally} the spikes from the neuromodulator 
releasing neurons and transmits them {\em locally} to neuromodulated synapses.
\item Each edge $e$ in the sets $E_{0,i}$ carries a time delay information $d_e$ 
(in multiples of the fixed time interval $\Delta t$ of the discretized dynamics of the system).
\end{itemize}


If more than one type of neuromodulator is present at the same time, then
each neuromodulator determines a (different) partition of $G$ into
machines ${\bf m}_i$, and a corresponding set of vertices $v_{0,i}$ 
and edges $E_{0,i}$. Thus one obtains a graph $G_0$ by adding all of
these new vertices and edges for each neuromodulator present in the model. 
For simplicity we restrict to considering the case of a single modulator. 

\subsubsection{Network summing functors and automata with time delays}\label{TimeDelaySec}

In models of distributed computing one considers in particular a
generalization of finite state automata given by {\em timed automata}, see \cite{Alur}.
In general, these are described as finite state machines with a finite set of real-valued clocks, 
which can be independently reset with the transitions of the automaton. Transitions can
take place only if the current values of the clocks satisfy certain specified constraints. 


In order to model the time delays introduced in the neuromodulator model of 
\cite{Potj} one does not need this very general form of timed automata. Indeed,
it is better in this case to work with the class of {\em automata
with time delay blocks}, developed in \cite{Chatt}. These automata generate 
a class of formal languages that strictly contains the regular languages and that
is incomparable to the context-free languages (as it includes some non-context-free
languages while it cannot represent some context-free ones). 


In a finite state automaton with time delay blocks, the transitions are labeled by
the usual label symbols of the underlying finite state machine, and by an additional
symbol given by a non-negative integer number $n\in \Z_+$ which represents the
time delay block of that transition. Thus, given a directed path in the directed
graph of the finite state automaton starting at the initial state $\iota$, given by
a string $(a_1,n_1)(a_2,n_2)\cdots (a_m,n_m)$, the time-zero transition consists
of the substring of $a_i$ such that $n_i=0$, the time-one transition consists of
the substring of $a_i$ with $n_i=1$, and so on.  Thus, at time zero the
automaton carries out the computation corresponding to the string made
by the $a_i$ with $n_i=0$ (which must be in the regular language of the underlying
finite state automaton), and so on for the successive times. The sequence of
integer times is usually assumed to be non-decreasing. 

For example, an automaton with three states $s_0,s_1,s_2$, with initial state $s_0$ 
and transitions $a$ between $s_0$ and $s_1$, $b$ between $s_1$ and $s_2$ and
$c$ between $s_2$ and $s_0$ would produce the $\{ (abc)^n \,:\, n\in \N \}$ language. However, if one
introduces time delays, using timed transitions
$(a,0)$ between $s_0$ and $s_1$, $(b,1)$ between $s_1$ and $s_2$ and
$(c,2)$ between $s_2$ and $s_0$, then only the symbol with delay $n=0$ is
deposited in the output until time resets to $1$, then only the symbol with time $1$,
until the automaton returns to the state $s_0$ and time if reset,  
so that this timed automaton produces a timed language $\{ (a,0)^n (b,1)^n (c,2)^n \,:\, n\in \N\}$
and the associated untimed language (forgetting the time markings) is now $\{ a^n b^n c^n\, : \, n\in \N \}$.


We can then modify the construction of Proposition~\ref{compfunctorprop} to
accommodate this kind of model of neuromodulated networks. 

\begin{defn}\label{timeautocat}
Let $\cC^t\subset \cC'\subset \cC$ denote the time-delay subcategory of the category $\cC$ of transition
systems of \S \ref{CompResSec}, with objects $\tau=(S,\iota,\cL,\cT)$ that have
a unique final state $q$ and whose label set is of the form $\cL=\cL'\times \Z_+$, where
$\cL'$ is a label set and $n\in \Z_+$ is a time delay block as 
above. When a time delay is not explicitly written in a transition in $\cT$ it is assumed
to mean that $n=0$. These correspond to the usual transition with labeling set $\cL'$.
As in the case of the subcategory $\cC'\subset \cC$, we can consider a version
of the category $\cC^t$ where the objects $\tau$ are also endowed with incoming half-edges
at the initial state and outgoing half-edges at the final state, with subcategories $\cC^t(n,m)$
where $\deg^{\rm in}(\iota) =n$ and $\deg^{\rm out}(q)=m$.
\end{defn}

As described in \S \ref{machinesSec}, we define a distributed structure on a directed graph $G$ as
follows.

\begin{defn}\label{distrGm}
A distributed structure ${\bf m}$ on a finite directed graph $G$ is given by:
\begin{enumerate}
\item a partition into $N$ machines ${\bf m}_i$ as described in \S \ref{machinesSec},
\item two subsets of vertices $V_{s,i}, V_{t,i}$ 
inside the vertex set $V_i$ of each machine ${\bf m}_i$ 
\item a directed graph $G_0$ with $G_0 \supset G$, obtained by adding 
\begin{itemize}
\item for all $i=1,\ldots, N$, a new vertex $v_{0,i}$ to each vertex set $V_i$, with $V_{{\bf m}_i}=V_i \cup \{ v_{0,i} \}$
\item for all $i=1,\ldots, N$ and for each vertex $v\in V_{t,i}$ a new edge with source $v_{0,i}$ and target $v$,
\item for all $i,j=1,\ldots, N$ and for each vertex $v\in V_{s,j}$ a new edge with source $v$ and target $v_{0,i}$, 
\item a non-negative integer $n_e \in \Z_+$ assigned to each edge $e \in E_{G_0}$, with $n_e=0$ if
$e\in E_G$.
\end{itemize}
\end{enumerate}
Given a pair $(G,{\bf m})$ of a directed graph with a distributed structure, we denote by 
$\bar G_0({\bf m})$ the condensation graph obtained by contracting each of the
subgraphs $G_i$ given by the vertices $V_i$ and the edges between them to a single vertex. 
(Note that the condensation graph $\bar G_0({\bf m})$ is acyclic.)
\end{defn}

\begin{defn}\label{catDeltaGt}
Let $\cG^{\rm dist}$ be the category with objects $(G, {\bf m})$ given by 
a finite directed graph with a distributed structure as in Definition~\ref{distrGm}, with the
properties that 
the induced subgraphs $G_i$ of $G_0$ with vertex set $V_{{\bf m}_i}$ are strongly connected.

Morphisms $\alpha \in
{\rm Mor}_{\cG^{\rm dist}}(G,{\bf m}), (G',{\bf m}'))$ 
are given by morphisms
$\alpha: G \to G'$ of directed graphs that are compatible with the distributed structure, in the
sense that the induced morphisms $\alpha_i=\alpha |_{G_i}: G_i \to G_{j(i)}'$ map the subgraphs $G_i$ of the
distributed structure of $G$ to the subgraphs $G'_j$ of the distributed structure of $G'$.
\end{defn}


Note that we use here, as morphisms of directed graphs the natural transformation of
functors in ${\rm Func}({\bf 2}, \cF)$ (see Definition~\ref{DirGrDef}). These morphisms
allow for identifications of edges, but not for contractions of edges to vertices. A slight
variant of the category ${\bf 2}$ that also allows for edge contractions is discussed in
\S 2.1.1 of \cite{Mar-new}. 


We then have the suitable modification of the functorial construction of 
Proposition~\ref{compfunctorprop} adapted to this setting, where we consider
the category $\cC^t$ with subcategories $\cC^t(n,m)$ as in Definition~\ref{timeautocat}
and the properad structure as in the case of $\cC'$.

\begin{prop}\label{compfunctorprop2}
Given an object $(G, {\bf m})$ of $\cG^{\rm dist}$, let $P(G, {\bf m})$ be the category of subgraphs
with compatible distributed structure. Given a summing functor $\Phi \in \Sigma_{\cC^t}(V_G)$ with
values in the time-delay subcategory $\cC^t$, consider the following procedure:
\begin{itemize}
\item consider the objects $\Phi(v)$ for $v\in V_{{\bf m}_i}$, the vertex set of the subgraph $G_i$, as in Definition~\ref{catDeltaGt};
\item these determine the objects $\tau_{G_i}$ obtained by grafting as in Definition~\ref{Gstrongconn};
\item for $\bar G_0({\bf m})$ the condensation graph as in Definition~\ref{distrGm}, perform the
grafting $\tau_{\bar G_0({\bf m}),\bar\omega}$, as in Lemma~\ref{acyclicgraft}, 
of the objects $\tau_{G_i}$.
\end{itemize}
This procedure determines,  as in Proposition~\ref{compfunctorprop}, 
a summing functor $$\Upsilon(\Phi)\in \Sigma_{\cC^t}^{\operatorname{prop}}(G, {\bf m}),$$ which 
assigns to an object $(G'{\bf m}')$ in $P(G, {\bf m})$ the object in $\cC^t$ given by 
the grafting $\tau_{\bar G_0({\bf m}),\bar\omega}$.
\end{prop}

\subsubsection{Topological questions}\label{TopQuesSec}

An interesting mathematical question is then to describe the topological
structure, in terms of protocol simplicial complexes, of the distributed computing
algorithm implementing a neuromodulated network, and to investigate
how the topology of the resulting protocol simplicial complexes are related to
other topological structures we have been considering in this paper.
We leave this question to future work.


There is a further interesting aspect to the larger-scale structures of the
network and its computational properties. 
As pointed out in \cite{Potj}, the usual analysis of networks in neuroscience 
is based on the abstract connectivity properties of the network as a 
directed graph without any information on its embedding in $3$-dimensional space.
Topologically it is well known that embedded graphs are at least as 
interesting as knots and links and capture subtle topological properties
of the ambient space that are not encoded in the structure of the
graph itself, but in the embedding. We will not be developing this
aspect in the present paper, but it is an interesting mathematical
question to identify to what extent invariants of {\em embedded graphs},
such as the fundamental group of the complement (as in the case of knots
and links), can carry relevant information about the informational and
computational structure of the network beyond the local connectivity
structure. 

\section{Codes, probabilities, and information} \label{GammaCodesSec}

In this section we show a toy model construction, where we use the
setting of categories of network summing functors described in \S \ref{SummingSec} 
to describe functorial assignments of codes to neurons in a network and of associated
probabilities and information measures. This shows a possible way of
describing informational constraints in a network of neurons.

\subsection{Introducing neural codes}

There is an additional part of the modeling of a neural  information network which we
have not introduced in our construction yet. 
Neurons transmit information by generating a spike train, with a certain firing rate. As 
discussed in \cite{Sto18}, the spiking activity can be described in terms of a binary code,
in the following way. Let $T>0$ be a certain interval of observation time, during which
one records the spiking activities. We assume it is subdivided into multiples of some unit of time $\Delta t$,
with $n=T/\Delta t$ the number of basic time intervals considered. Assuming that $\Delta t$ is
sufficiently small, so that one does not expect a time interval of length $\Delta t$ to contain more
than one spike, one can assign a digital word of length $n$ to an observation by recording a digit $1$
for each time interval $\Delta t$ that contained a spike and a $0$ otherwise. When $k$ observations
are repeated, one obtains $k$ binary words of length $n$, that is, a binary code $C\subset \F_2^n$.
We assume that the neurons generate spikes at a given rate $y$ of spikes per second. This
rate is computed from observations as the number $m$ of spikes observed per observation time, $y=m/T$.

\subsubsection{Firing rates of codes}
For sufficiently large $T$ (hence for large $n$), the empirical estimate $y \Delta t =m/n$ of observing a spike
in a time interval $\Delta t$ will approximate a probability $0< p < 1$. Thus, for large $n$ the digits 
of the code words of $C$ are drawn randomly from the distribution $P$ on $\{0,1\}$ that gives
probability $p$ to $1$ and $1-p$ to $0$. This means that the relevant probability space to
consider here is the following.


Shift spaces and subshifts of finite type are a class of symbolic dynamical systems
used to model various types of dynamics, see \cite{Kitch}. In particular, given an
alphabet $A$ with $\# A=q$, the shift space $\Sigma_q^+=A^\N$ is the space of all sequences
$a_0 a_1 a_2 \ldots a_n \ldots$ with $a_i\in A$, endowed with the one-sided
shift map $\sigma: \Sigma_q^+ \to \Sigma_q^+$ that maps 
$\sigma(a_0 a_1 a_2 \ldots)=a_1 a_2 a_3 \ldots$. The set $\Sigma_q^+$ can be
topologized (as a Cantor set) with a basis for the topology given by the cylinder sets
$\Sigma_q^+(w)$, with $w=w_0\ldots w_m$ for some $m\geq 1$ a word in the
alphabet $A$, where $\Sigma_q^+(w)=\{ w a_{m+1} a_{m+2} \ldots a_n \ldots \, , \, a_i\in A \}$
is the set of infinite words starting with the word $w$.


Let $(\Sigma_2^+, \mu_P)$ denote the probability space with $\Sigma_2^+=\{0,1\}^\N$ the shift space 
given by all the infinite sequences of zeroes and ones, and with $\mu_P$ the Bernoulli measure
that assigns to the cylinder set $\Sigma_2^+(w_1,\ldots, w_n)$ of sequences starting with the
word $w_1\cdots w_n$ the measure $\mu_P(\Sigma_2^+(w_1,\ldots, w_n))=p^{a_n(w)} (1-p)^{b_n(w)}$,
where $a_n(w)$ is the number of $1$'s in $w_1\cdots w_n$ and $b_n(w)=n-a_n(w)$ is the number of zeros.


The observation that code words of $C$ are drawn randomly from the distribution $P=(p,1-p)$,
corresponds to saying that $C$ is in the Shannon Random Code Ensemble (SRCE), which is the set of
codes with this property. Note that one commonly works with the SRCE for the uniform distribution
with $p=1/2$, but one can equally consider SRCEs for a given  $P=(p,1-p)$ specified by the problem.


\begin{lem}\label{shiftP}
For large $n$ the neural code $C$ is a code in the Shannon Random Code Ensemble, generated by
the probability space $(\Sigma_2^+, \mu_P)$. Moreover, with $\mu_P$-probability one, codes obtained 
in this way represent neural codes with firing rate $y=p/\Delta t$.
\end{lem}

\begin{proof}
  As above, the probability $0< p < 1$ is the probability of observing a spike
in a time interval, for sufficiently large $n$, with the code words of $C$ drawn according to 
the distribution $P(1)=p$, $P(0)=1-p$. 
This means that the code words can be identified as parts of a sequence in $\Sigma_2^+$
generated with the stochastic process given by the Bernoulli measure $\mu_P$. As observed
above, this is the property that the code $C$ belongs to 
the Shannon Random Code Ensemble with this Bernoulli measure. 
Consider sequences $w=w_1,w_2,\ldots, w_n, \ldots$ in the shift space $\Sigma_2^+$ with 
the Bernoulli measure $\mu_P$ determined by $P$. Let $a_n(w)$ be the number
of $1$'s in the first $n$ digits of a sequence $w\in \Sigma_2^+$. Then $\mu_P$-almost everywhere
one has the limit
$$ \lim_{n\to \infty} \frac{a_n(w)}{n} \stackrel{a.e.}{=} p. $$
This means that for random codes generated from sequences in $\Sigma_2^+$ drawn according
to the Bernoulli measure, the ratio $m/n=a_n(w)/n$, which described the firing rate of the
neural code, approaches $p$ for sufficiently large $n$, hence with probability one, codes $C$
obtained in this way can be regarded as possible neural codes associated to a 
neuron with firing rate $y=p/\Delta t$, obtained from $k=\# C$ observations of $n=T/\Delta t$
time intervals, where $T$ and $\Delta t$ are given. 
\end{proof}


Thus, in order to incorporate the information of the possible neural codes produced
by the nodes of the network, given the firing rates of the neurons at the nodes, we
can assign to each node a probability space $(\Sigma_2^+, \mu_P)$ from which 
the neural codes are generated. This is entirely determined by assigning the finite
probability $P=(p,1-p)$ at the node. 

\subsubsection{Codes and finite probabilities}\label{CodesProbSec}

Models of neural codes typically assume that only the firing rates and the timing of spikes
encode information, while other characteristics such as spike amplitudes do not contribute
to encoding of the stimulus. The use of binary codes as discussed above is adequate for
this type of models, as it records only the digital $0/1$ information on whether a spike is
detected in a given time interval $\Delta t$ or not. However, it has been suggested that
other kinds of information may be present in the neural codes that are not captured by the
binary code detecting the presence and timing of spikes. This is the case, for example,
with the proposal that ``spike directivity'' contributes to the neural 
encoding~\cite{AurConJog05}.
In order to allow for the possibility of additional data in the neural code, besides the $0/1$ record 
of whether a spike is present or not in a given time interval, one can consider non-binary codes. 
Since the codes we are considering are unstructured rather than
linear, we do not need to require that the number of letters $q$ of the code alphabet is a prime
power and that the ambient space the code sits in is a vector space over a finite field.
Thus we can simply assume that a discretization of the additional data being recorded
(such as spike directivity) is chosen with a set of $q$ values, for some $q\in \N$, $q\geq 2$.
The code is now constructed with words that record, for each of the $n$
time intervals $\Delta t$, whether a spike is absent (a digit $0$) or whether it is present
(a non-zero digit) and what is the registered value of the other parameters, discretized 
over the chosen range of $q$ possible values (including $0$). In this more general setting,
we then assign to each node of our neural information network a code $C$ of length $n$
on an alphabet of $q$ letters (where in principle $q$ may vary with the node, depending on
different types of neurons present). It is not obvious in this more general setting that
Bernoulli processes on shift spaces will be adequate to model these more general codes,
but in first approximation we can assume the same model and consider these
neural codes as codes in the Shannon Random Code Ensemble generated by a Bernoulli process on the
space of sequences $\Sigma_q^+$ determined by a finite probability measure $\mu_P$ with
$P=(p_1,\ldots, p_q)$. We consider this setting in the following. It is easy to
restrict to the original case by just restricting to $q=2$ for all the codes. The firing
rate of the neuron is still related to the probability distribution $P$. Indeed, for $w\in \Sigma_q^+$,
let $a^0_n(w)$ be the number of zeroes in the first $n$ digits of the sequence $w$ and 
let $b_n(w)=n-a^0_n(w)$ be the number of the non-zero digits. The firing rate can now
be seen as the ratio $b_n(w)/n$ which has a $\mu_P$-almost everywhere limit
$$ \lim_{n\to \infty} \frac{b_n(w)}{n} \stackrel{a.e.}{=} \sum_{i=1}^q p_i = 1-p_0 \, .$$


To make more precise the construction of the probability distribution associated to
a code $C$, we focus for simplicity on the case of binary codes, though the
following discussion can be easily generalized to $q$-ary codes.


Recall that an $[n,k,d]_2$-code is a binary code $C\subset \F_2^n$ of length $n$, with cardinality $\# C =2^k$,
and with minimum distance $d=\min\{ d_H(c,c')\,|, c\neq c' \in C \}$, the minimal Hamming distance
$d_H(c,c')=\# \{ i\in \{1, \ldots, n\} \,|\, c_i\neq c_i' \}$ between code words of $C$.


\begin{defn}\label{Pbincode}
Let $C$ be an $[n,k,d]_2$-code. For every code word $c\in C$ let $b(c)$ denote the 
number of digits of $c$ that
are equal to one. The probability distribution $P_C=(p,1-p)$ of the code $C$ is given by
\begin{equation}\label{PCprob}
p= \sum_{c\in C} p(c), \ \ \ \text{ with } \ \ \ p(c)=\frac{b(c)}{n \cdot \#C}.
\end{equation}
\end{defn}


If $a(c)$ denotes the number of letters in the code word $c\in C$ that are equal to zero,
then clearly we also have $1-p=\sum_{c\in C} a(c)/(n\cdot \# C)$. 


\begin{lem}
Let $C,C'$ be binary codes of equal length $n$, both containing the word with all digits equal to zero.
Let $f: C \to C'$ be a surjective map that sends the zero word to itself, such that for all code 
words $c,c'\in C$ the Hamming distance satisfies
$d(f(c),f(c')) \leq d(c,c')$. Then the probability $P_{C'}$ is related to $P_C$ by
$p(f(c))=\lambda(c) p(c)$, where $\lambda(c)\leq 1$ is given by the ratio $\lambda(c)=b(f(c))/b(c)$.
\end{lem}


\begin{proof}
Since the Hamming distance is decreasing under the map $f$ we have
$b(f(c))=d(f(c),0) \leq d(c,0)=b(c)$. It is then clear that $p(c')=\sum_{c\in f^{-1}(c')} \lambda(c) p(c)$
is the probability distribution associated to the code $C'$.
\end{proof}

\subsubsection{Categories of codes}\label{CodesCatSec}

We discuss here two possible constructions of a category of codes. The first one is
modeled on the notion of decomposable and indecomposable codes variously
considered in the coding theory literature (see for instance \cite{Sle60}). The second
one is more directly suitable for modeling neural codes associated to populations of
neurons and their firing activities. 


Let $C$ be an $[n,k,d]_q$-code over an alphabet $\fA$ with $\# \fA=q$, so that $C\subset \fA^n$
with $\# C = q^k$. Given two such codes, $C$ an $[n,k,d]_q$-code over the alphabet $\fA$
and $C'$ an $[n',k',d']_{q'}$-code over the alphabet $\fB$, a morphism is a function $\phi: C \to C'$,
such that the image $\phi(C)\subset C'$ satisfies $d_{\fB^{n'}}(\phi(c_1),\phi(c_2))\leq d_{\fA^n}(c_1,c_2)$ 
for all code words $c_1,c_2\in C$, in the respective Hamming distances.
Note that here we do not define the morphisms $\phi: C \to C'$ as maps $\phi: \fA^n \to \fB^{n'}$ 
of the ambient spaces that map $C$ inside $C'$. 
If we restrict to only considering codes over a fixed alphabet $\fA$, then there is a sum operation
given by 
\begin{equation}\label{sumcodes1}
C\oplus C' :=\{ (c,c')\in \fA^{n+n'}\,|\, c\in C, \, c'\in C' \}. 
\end{equation}
A code $C$ is decomposable if
it can be written as $C=C'\oplus C''$ for codes $C',C''$ and 
indecomposable otherwise~\cite{Sle60}.
If $C$ is an $[n,k,d]_q$-code and $C'$ is an $[n',k',d']_q$-code then $C\oplus C'$ is an
$[n+n', k+k', \min\{ d,d' \}]_q$-code. With this choice of objects and morphisms the resulting
category does not have a zero object. If we identify the alphabet $\fA$ with a set of $q$ digits 
$\fA=\{ 0,\ldots, q-1\}$ we can consider, for each $n\in \N$, only those codes $C\subset \fA^n$ 
that contain the zero word $(0,\ldots, 0)$ as one of the code words. We can interpret these
codes as being the result of a number $\# C$ of observations of the spiking neuron, with
each observation consisting of $n$ time intervals $\Delta t$, where the observations stop
when no more spiking activity is detected in the $T=n \Delta t$ observation time, that is,
when the response to the stimulus has terminated, so the last code word is the zero word. 

\begin{lem}\label{CodesCatLem}
Let ${\rm Codes}$ be the category with objects the codes containing the zero word and
morphisms $\phi: C \to C'$ as above.
This category has a zero object given by the code $C=\{ 0 \}\subset \fA$ consisting 
of the zero word of length one and a coproduct of the form \eqref{sumcodes1}.
\end{lem} 


Our previous construction of the measure associated to a binary code in
Definition~\ref{Pbincode} satisfies the following property with respect to
the sum of codes. 

\begin{lem}\label{Psum}
The probability associated to the sum $C\oplus C'$ of \eqref{sumcodes1} is given by
$P_{C\oplus C'} =\lambda P_C + (1-\lambda) P_{C'}$ with $\lambda =n/(n+n')$.
\end{lem}

\begin{proof}
  The probability associated to the code C is given by 
\begin{equation}\label{PCp}
P_C=(p,1-p) \ \ \text{ with } \ \ 
p=\sum_{c\in C} p(c) \ \ \text{ and } \ \  p(c)=b(c)/(n\cdot \# C)
\end{equation} 
with $b(c)$ the number
of letters equal to one in the code word $c$. Similarly for the probability $P(C')$.
For code words $(c,c')$ in $C\oplus C'$ with $c\in C$ and $c'\in C'$, we have
length $n+n'$, cardinality $\# C \cdot \# C'$, and number of letters equal to one
given by $b(c,c')=b(c)+b(c')$. Thus the probability $P_{C\oplus C'}$ has 
$p=\sum_{(c,c')} p(c,c')$ with 
$$ p(c,c')=\frac{b(c)+b(c')}{(n+n') \cdot \# C \cdot \# C'} =
\frac{b(c) \cdot n}{n \cdot (n+n') \cdot \# C \cdot \# C'} + \frac{b(c')\cdot n'}{n'\cdot (n+n') \cdot \# C \cdot \# C'} $$ 
$$  =p(c) \cdot \frac{n}{(n+n')\cdot \# C'} + p(c')\cdot
\frac{n'}{(n+n')\cdot \# C}. $$
Thus, we have
$$ \sum_{c,c'} p(c,c')= \sum_{c,c'} \frac{p(c) \cdot n}{(n+n')\cdot \# C'} + \sum_{c,c'} 
\frac{p(c')\cdot n'}{(n+n')\cdot \# C} $$
$$ =\frac{n}{(n+n')}  \cdot \sum_c p(c) + \frac{n'}{(n+n')} \cdot\sum_{c'} p(c') 
 = 1 \, . $$
Thus the resulting probabilities
are related by $P_{C\oplus C'} =\lambda P_C + (1-\lambda) P_{C'}$ with $\lambda =n/(n+n')$.
\end{proof}


The counting of ones in the digits of the code words, used to obtain the
probabilities of Lemma~\ref{Psum}, can be viewed as comparing each code
word to the zero word through the Hamming distance. One can refine this
by comparing all the code words with each other through the Hamming
distance. Thus, one can also associate to a code $C$ the pair $(\delta, 1-\delta)$
where $\delta =\min\{d_H(c,c')\,|\, c\neq c' \}/n=d/n$ is the relative minimum
distance. Note that the Shannon information 
$\cI(\delta, 1-\delta)=\delta \log_q \delta + (1-\delta) \log_q (1-\delta)$ and the
associated $q$-ary entropy function 
$H_q(\delta)=\delta \log_q(q-1) -\delta \log_q\delta - (1-\delta) \log_q (1-\delta)$
describe the asymptotic behavior of the volumes of the Hamming balls, 
and determines the position of random codes with respect to the Hamming 
bound~\cite{CoGo90}. It is well known that codes in the SRCE populate the
region of the space of code parameters at and below the Gilbert--Varshamov line
defined in terms of the $q$-ary entropy function~\cite{CoGo90}.


We consider then another possibile construction of a category of codes with a sum and zero object,
which is simply induced by the same structure on pointed sets. 
We will see in the next subsection that this choice has better properties with respect to the assignment of
probabilities to neural codes. Unlike the usual setting of coding theory, we allow here for the possibility
of codes with repeated code words, that is, where some $c,c'\in C$ have zero Hamming distance. While
this is unnatural from the coding perspective as it leads to ambiguous encoding, it is not unreasonable
when thinking of codes that detect firing patterns of neurons, as the possibility exists of two measurements
leading to the same pattern. Thus, we think here of codes as subsets $C\subset \fA^n$ with possible
multiplicities assigned to the code words. We assume the zero word always has multiplicity one. 
Repeated code words arise in the categorical setting we describe
here when coproducts are taken using \eqref{sumcodes2} instead of \eqref{sumcodes1}.
The following is simply the usual categorical structure on finite pointed sets. 


\begin{lem}\label{catcodes}
A symmetric monoidal category ${\rm Codes}_{n,*}$ of pointed codes of length $n$ 
is obtained with set of objects given by  
$[n,k,d]_q$-codes, with fixed alphabet $\fA$ with $\# \fA=q$ and fixed length $n$, 
that contain among their code words the constant $0$-word $c_0=(0,0,\ldots,0)$, and are not
equal to the code $C=\{ c_0, c_1 \}$ consisting only of the constant $0$-word and the constant $1$-word 
$c_1=(1,1,\ldots, 1)$. As maps $f: C \to C'$ we consider
functions mapping the $0$-word to itself. The categorical sum is the wedge sum of pointed sets
\begin{equation}\label{sumcodes2}
C\oplus C':=  C \vee C' = C\sqcup C' / c_0\sim c'_0 
\end{equation}
and the zero object is the code $C=\{ c_0 \}$ consisting only of the $0$-word $c_0$ of length $n$.
\end{lem}


We exclude among the objects of the category of codes the code $C=\{ c_0, c_1 \}$
containing only the constant $0$-word $c_0$ and the constant $1$-word $c_1$ to
ensure that any code that is not just $\{ c_0 \}$ contains at least a word with non-zero
information. We regard the $0$-word as the baseline corresponding to lack of any
spiking activity, and we require that the presence of spiking activity carries some non-trivial
information. 


In the categorical setting we described here, a neural code associated to a network of neurons can be 
viewed as a summing functor $\Phi_E: P(V_{G^*}) \to {\rm Codes}$ where $V_G$ is the set of vertices 
of the network $G$, as discussed in the previous section. The code assigned to a vertex describes
the spiking behavior of that neuron.

\subsubsection{Codes and associated probabilities}\label{ProbCodesSec}

It is convenient here to consider a slightly different version of the category of
finite probabilities, with respect to the versions mentioned earlier. This is
itself a variant, where pointed sets are considered, over a standard construction
of a category of (finite) measure spaces. The morphisms $(f,\Lambda)$ will be defined using 
functions $f: X\to Y$ of finite pointed sets and non-negative weights $\Lambda=\{ \lambda_y \}$ 
on the fibers that serve the purpose of matching the probability measures. More precisely,
we have the following. 


\begin{lem}\label{finprobfibers}
A category $\cP_f$ of finite probabilities with fiberwise measures
as morphisms is obtained by considering as objects the pairs $(X,P_X)$ of a finite pointed 
set $X$ with a probability measure $P_X$ such that $P_X(x_0)>0$ at the base point. 
Morphisms $\phi: (X,P_X)\to (Y,P_Y)$
consist of a pair $\phi=(f,\Lambda)$ of a function $f: X \to Y$ of pointed sets, $f(x_0)=y_0$, with
$f({\rm supp}(P_X))\subset {\rm supp}(P_Y)$, 
together with a collection $\Lambda=\{ \lambda_y \}$ of measures $\lambda_y$ on the fibers $f^{-1}(y)\subset X$,
with $\lambda_{y_0}(x_0)>0$, 
such that $P_X(A)=\sum_{y\in Y} \lambda_y(A\cap f^{-1}(y))\, P_Y(y)$. 
The category has a coproduct and a zero object.
\end{lem}


\begin{proof}
  Note that the fiberwise measures $\lambda_y$ are not assumed to be probability measures.
While in the case of a surjection $f: X\to Y$ we can have $\sum_{x\in f^{-1}(y)} \lambda_y(x)=1$,
in the case of an injection $\iota : X\hookrightarrow Y$ scaling factors $\lambda_y(x)\geq 1$ will
adjust the normalization so that $\sum_x P_X(x)=\sum_{x} \lambda_{\iota(x)}(x) P_Y(\iota(x))=1$
while $\sum_{y\in \iota(X)} P_Y(y) =1-P_Y(Y\smallsetminus \iota(X))\leq 1$. 
Composition of morphisms 
$\phi=(f,\Lambda): (X,P_X)\to (Y,P_Y)$
and $\phi'=(g,\Lambda'): (Y,P_Y) \to (Z,P_Z)$ is
given by $\phi'\circ \phi=(g\circ f, \tilde\Lambda)$ with 
$\tilde\lambda_{g(f(x))} (x)=\lambda_{f(x)}(x) \lambda'_{g(f(x))}(f(x))$. 
We want to show the existence of a unique (up to unique isomorphism)
object $(X,P)\oplus (X',P')$ in $\cP_f$ with morphisms $\psi: (X,P)\to (X,P)\oplus (X',P')$
and $\psi': (X',P')\to (X,P)\oplus (X',P')$ such that for any given 
morphisms $\phi=(f,\Lambda): (X,P)\to (Y,Q)$ and $\phi'=(g,\Lambda'):(X',P')\to (Y,Q)$,
there exists a unique morphism $\Phi: (X,P)\oplus (X',P') \to (Y,Q)$ 
such that the diagram commutes:
$$ \xymatrix{ (X,P) \ar[r]^{\psi\quad\quad} \ar[dr]_{\phi} & (X,P)\oplus (X',P') \ar[d]^\Phi & (X',P') \ar[l]_{\quad\quad\psi'} \ar[dl]^{\phi'} \\ & (Y,Q) & } $$
We take $(X,P)\oplus (X',P')$ to be the object $(X\vee X', \tilde P)$ where $X\vee X'=X\sqcup X'/ x_0\sim x_0'$
and $\tilde P(x)=P(x) \cdot \alpha_{X,X'}$
for all $x\in X\smallsetminus \{ x_0 \}$, $\tilde P(x')=P'(x') \cdot \beta_{X,X'}$ for all 
$x'\in X'\smallsetminus \{ x_0' \}$, and $\tilde P(x_0\sim x_0')=\alpha_{X,X'} P(x_0)+\beta_{X,X'} P'(x_0')$, 
with $\alpha_{X,X'}=N / (N+N')$ with $N=\# X$ and $N'=\# X'$
and $\beta_{X,X'}=1-\alpha_{X,X'}=N'/(N+N')$  so that 
$\sum_{a\in X\sqcup X'}\tilde P(a)=\alpha_{X,X'}\sum_{x\in X\smallsetminus \{ x_0\}} P(x) + \beta_{X,X'} \sum_{x'\in X'\smallsetminus \{ x_0' \}} P'(x')+\tilde P(x_0\sim x_0')=1$.
The morphisms $\psi=(\iota: X \hookrightarrow X\sqcup X', \Lambda=\alpha_{X,X'}^{-1})$ and $\psi'=(\iota': X'\hookrightarrow X\sqcup X', \Lambda'=\beta_{X,X'}^{-1})$ and the induced morphism $\Phi=(F,\tilde\Lambda):
(X\sqcup X', \tilde P) \to (Y,Q)$ given by $F(\iota(x))=f(x)$ and $F(\iota'(x'))=g(x')$ with
$\tilde \lambda_y(x)=\alpha_{X,X'} \cdot \lambda_y(x)$
for $x\in f^{-1}(y)$ and $\tilde \lambda_y(x')=\beta_{X,X'} \cdot \lambda'_y(x')$
for $x'\in g^{-1}(y)$, give a commutative diagram as above. The coproduct constructed in this
way is unique up to unique isomorphism, since if there is another object $(Z,\hat P)$
with morphisms $\hat\psi: (X,P)\to (Z,\hat P)$ and $\hat\psi': (X',P') \to (Z,\hat P)$ that satisfies
the same universal property, there are unique maps 
$\Phi: (X\sqcup X', \tilde P)\to (Z,\hat P)$
and $\hat\Phi: (Z,P)\to (X\sqcup X', \tilde P)$ that make the respective diagrams commute so that
$\Phi \circ \psi=\hat\psi$, $\Phi\circ \psi'=\hat\psi'$, $\hat\Phi\circ \hat\psi=\psi$, and $\hat\Phi\circ \hat\psi'=\psi'$.
The object $(*, 1)$ with a single point with probability one is a zero object, with unique morphism
$(f,\Lambda): (*,1)\to (X,P)$ given by $f(*)=x_0$ and $\lambda_{x_0}(*)=P(x_0)$ and unique
morphism $(f,\Lambda): (X,P) \to (*,1)$ with $f(x)=*$ for all $x\in X$ and $\lambda_*(x)=P(x)$.
\end{proof}


Given a code $C$, we assign a finite probability $P_C$ to the code, as in \eqref{PCprob}, which refines the
binary probability $(p,1-p)$ of \eqref{PCp}. More precisely, we construct $P_C$ as follows.


\begin{defn}\label{PspaceC}
The probability space $P_C$ associated to a binary code $C$ is given by
\begin{equation}\label{PCXC}
P_C(c)=\left\{ \begin{array}{ll} \frac{b(c)}{n (\# C-1)} & c\neq c_0 \\
1-\sum_{c'\neq c_0} \frac{b(c')}{n (\# C-1)} & c=c_0 \end{array}\right.
\end{equation}
with $b(c)$ the number of digits equal to $1$ in the word $c$.
\end{defn} 


\begin{lem}\label{functCXP}
The assignment $C \mapsto P_C$ determines a functor $P: {\rm Codes}_{n,*} \to \cP_f$ 
compatible with sums and zero objects.
\end{lem}


\begin{proof}
  A map of codes $f: C \to C'$ induces a map $\phi=(f,\Lambda): (C,P_C) \to (C', P_{C'})$ with
$\Lambda=\{ \lambda_{c'}(c)\,|\, c\in f^{-1}(c') \}$ given by $\lambda_{f(c)}(c)=\frac{P_C(c)}{P_{C'}(f(c))}$.
This is well defined because by \eqref{PCXC} the only code word with $b(c)=0$ would be the $0$-word
$c_0$ and $b(c)=n$ only for the word $c_1$ with all digits equal to one, so as long as the code $C$ does not
contain only the words $c_0$ and $c_1$, we have both $P_C(c_0)\neq 0$ and $P_C(c)\neq 0$ for all
$c\neq c_0$. The sum of codes is given by the wedge sum of pointed sets \eqref{sumcodes2}. The
associated probability is given by 
$$ P_{C_1\vee C_2}(c)=\frac{b(c)}{n (\#(C_1\vee C_2)-1)} $$
for $c\neq c_0$ and $1-\sum_{c\neq c_0} P_{C_1\vee C_2}(c)$ at the zero word. For $N=\# C_1 -1$
and $N'=\# C_2 -1$, we have $N+N'=\#(C_1\vee C_2)-1$ so that
$$  P_{C_1\vee C_2}(c)=\left\{ \begin{array}{ll} \frac{N}{N+N'} P_{C_1}(c) & c\in C_1\smallsetminus \{ c_0 \} \\
\frac{N'}{N+N'} P_{C_2}(c) & c\in C_2\smallsetminus \{ c_0 \} \\
\frac{N}{N+N'} P_{C_1}(c_0) + \frac{N'}{N+N'} P_{C_2}(c_0) & c=c_0 , \end{array}\right. $$
hence $P_{C_1\vee C_2}$ agrees with the probability $\tilde P$ of the direct sum $(C_1,P_1)\oplus (C_2,P_2)=(C_1\vee C_2, \tilde P)$ as in Lemma~\ref{finprobfibers}. The zero object $C=\{ c_0 \}$ is
mapped to the zero object $(\{ c_0 \},1)$.
\end{proof}

\subsection{Weighted codes and linear relations}\label{LinNeurSec}

In order to illustrate this general framework in a simple example, we show
how a ``linear neuron'' toy model can be fit within the setting described in
the previous subsections.  


Of course, in reality the neuron is non-linear, and the non-linearities can be described in terms of a 
threshold function (such as a sigmoid, or piecewise linear, or step function).
In this subsection we just look at the simplified linear case, while 
we will discuss how to formulate in our setting the case of
non-linear neurons and threshold dynamics in \S \ref{HopfieldSec}.


\begin{lem}\label{WCodes}
A category of weighted codes $\cW{\rm Codes}_{n,*}$ is obtained 
with objects given by pairs $(C,\omega)$ of pointed codes $C$ of length $n$
containing the zero word $c_0$ and a function $\omega: C \to \R$ assigning
a (signed) weight to each code word, with $\omega(c_0)=0$. 
Morphisms $\phi: (C,\omega) \to (C',\omega')$ are pairs $\phi=(f,\Lambda)$
of a pointed map $f: C\to C'$ mapping the zero word to itself and 
$f({\rm supp}(\omega))\subset {\rm supp}(\omega')$, and a collection
$\Lambda=\{ \lambda_{c'}(c) \}_{c\in f^{-1}(c')}$ satisfying 
$\omega(c)=\lambda_{f(c)}(c)\, \cdot \omega'(f(c))$ and $\lambda_{c_0}(c_0)=0$.
The category $\cW{\rm Codes}_{n,*}$ has a sum given by $(C,\omega)\oplus (C', \omega')=
(C\vee C', \omega\vee\omega')$ with $\omega\vee\omega'|_C=\omega$ and
$\omega\vee\omega'|_{C'}=\omega'$ and with zero object $(\{ c_0 \}, 0)$.
\end{lem}

The argument is analogous to the case of the category $\cP_f$, in fact simpler because
in the case of weights instead of probabilities
we do not have the normalization property of probability measures that needs to be
preserved. 


Consider then a pointed directed graph $G^* \in {\rm Func}({\bf 2}, \cF_*)$ as before,
and the categories of summing functors $\Sigma_{\cW{\rm Codes}_{n,*}}(E_{G^*})$
and $\Sigma_{\cW{\rm Codes}_{n,*}}(V_{G^*})$ with the 
source and target functors 
$s,t: \Sigma_{\cW{\rm Codes}_{n,*}}(E_{G^*}) \rightrightarrows  \Sigma_{\cW{\rm Codes}_{n,*}}(V_{G^*})$. 
As discussed earlier, a summing functor $\Phi_G$ in the equalizer of the source and target functors 
$$ \Sigma_{\cW{\rm Codes}_{n,*}}^{\operatorname{eq}}(G):={\rm equalizer}( s,t: \Sigma_{\cW{\rm Codes}_{n,*}}(E_{G^*}) \rightrightarrows  \Sigma_{\cW{\rm Codes}_{n,*}}(V_{G^*}) ) $$
is a summing functor $\Phi_G \in \Sigma_{\cW{\rm Codes}_{n,*}}(E_{G^*})$ with the property that
for all pointed subsets $A\subset V(G^*)$ the conservation law
$\Phi_G( s^{-1}(A)) = \Phi_G(t^{-1}(A))$ holds, which we also write as before as
$\oplus_{s(e)\in A} (C_e, \omega_e)= \oplus_{t(e)\in A} (C_e,\omega_e)$, where the sum is the
categorical sum in $\cW{\rm Codes}_{n,*}$ and $(C_e,\omega_e)=\Phi_G(\{ e, e_* \})$. 


\begin{rem}\label{linneuron}{\rm 
If we assume that the directed graph $G$ has a single outgoing edge at each vertex, 
$\{ e\in E_G \,|\, s(e)=v \}=\{ {\rm out}(v) \}$, then the equalizer condition becomes
\begin{equation}\label{linneu}
 ( C_{{\rm out}(v)}, \omega_{{\rm out}(v)}) =\oplus_{t(e)=v} (C_e,\omega_e), 
\end{equation} 
which is the formulation in our categorical setting of the linear neuron model.
}\end{rem}


In this model, we interpret the directed edges of the network as synaptic connections
between neurons, the code $C_e$ as determined by spiking potentials incoming along
that edge from the neuron at the source vertex $s(e)$, and the weight $\omega_e$ is
a measure of the efficacy of the synapses, depending on physiological properties such as
number of synaptic vescicles in the presynaptic terminal and number of gated channels in the 
post-synaptic membrane, with the sign of $\omega_e$ describing whether the synapse 
is excitatory or inhibitory. In this interpretation, in particular, the sign of $\omega_e(c)$ depends
only on the edge $e$ and not on the code word $c$, so $\omega_e$ has constant excitatory
or inhibitory sign on the entire code $C_e$ and different amplitude on the different code words. 
On codes $C=\vee_{e} C_e$ the sign of $\omega=\vee_e \omega_e$ is no longer constant.

 \subsection{Information measures} \label{InfomeasSec}
 
 The Shannon information of a finite measure $$S(P)=-\sum_{x\in X} P(x) \log P(x)$$
 satisfies the extensivity property 
 $$ S(P')=S(P)+P \, S(Q) $$
 for decompositions over subsystems $P'=(p'_{ij})$ with $p'_{ij}=p_j \cdot q(i|j)$, where
 $$ P\, S(Q) := \sum_j p_j S(Q|j) = - \sum_j p_j \sum_i q(i|j) \log q(i|j). $$
 In fact, extensivity, together with other simple properties completely characterize
 axiomatically the Shannon entropy (Khinchin axioms).
 

\begin{defn}\label{thindef} 
A thin category $\cS$ is a category where, for any two objects $X,Y \in {\rm Obj}(\cS)$,
the set ${\rm Mor}_{\cC}(X,Y)$ consists of at most one morphism. 
\end{defn}


Up to equivalence
a thin category $\cS$ is the same as a partially ordered set (poset). Up to isomorphism
a thin category is the same as a preordered set (proset), which satisfies the same
properties as a partial order except for asymmetry (the property that $X\leq Y$ and 
$Y\leq X$ implies $X=Y$). We will write thin categories in the form $(S,\leq)$, or $(S,\geq)$ 
for the opposite thin category.

 
 \begin{lem}\label{surjS}
 Let $\cP_{f,s}$ be the category of finite probabilities with fiberwise measures,
 where we only consider morphisms $(f,\Lambda)$ with $f:X \to Y$ a surjection
 and where the $\lambda_y(x)$ for $x\in f^{-1}(y)$ are probability measures on
 the fibers. Consider the real numbers $(\R,\geq)$ as a thin category with an object 
 for each $r\in \R$ and a single morphism $r\to r'$ if and only if $r\geq r'$.
 The Shannon entropy is a functor $S: \cP_{f,s} \to \R$.
 \end{lem}
 
\begin{proof}
  In the case of a morphism $(f,\Lambda):(X,P) \to (Y,Q)$ in the category $\cP_{f,s}$
 where the map $f: X \to Y$
 is a surjection and the fiberwise measures are probabilities $\Lambda=\{ \lambda_y(x)\,|\,
 x\in f^{-1}(y) \}$ on each fiber, we have a special case of the extensivity property
 with $P(x)=\lambda_{f(x)}(x) \, Q(f(x))$ and we obtain
 $$ S(P)=-\sum_{y\in Y} \sum_{x\in f^{-1}(y)} \lambda_y(x)\, Q(y) \log (\lambda_y(x)\, Q(y) ) $$
$$  = -\sum_{y\in Y} (\sum_{x\in f^{-1}(y)} \lambda_y(x)) \, Q(y) \log Q(y) - \sum_{y\in Y} Q(y) 
 \sum_{x\in f^{-1}(y)} \lambda_y(x)\, \log (\lambda_y(x)) $$
 $$ = S(Q) + \sum_{y\in Y} Q(y) \, S(\Lambda | y) = S(Q) + Q \, S(\Lambda).  $$
 In particular, this implies that for these morphisms we have $S(P) \geq 
 S(Q)$, and the difference
 $S(P)-S(Q) = \sum_{y\in Y} Q(y) \, S(\Lambda | y)$ measures the information loss 
 along the morphism $(f,\Lambda):(X,P) \to (Y,Q)$.
\end{proof}
 
 
 However, when we consider more general morphisms 
 $(f,\Lambda)$ in the category $\cP_f$, where the map $f$
 is not necessarily a surjection and the fiberwise measures
 $\Lambda=\{ \lambda_y(x)\,|\, x\in f^{-1}(y) \}$ are not necessarily
 probabilities, the relation between the Shannon entropies is no longer
 a case of the usual extensivity property and does not always satisfy
 the same simple estimate. For example, consider the case of an
 embedding $j: X \hookrightarrow Y$ so that the values $\lambda_{j(x)}(x)$
 are dilation factors that adjust the normalization of the measure $Q|_{j(X)}$.
 In this case we only have the relation
 $$ S(P)= -\sum_{y\in j(X)}  \lambda_{j(x)}(x) \, Q(j(x)) \log (\lambda_{j(x)}(x) \, Q(j(x)))= $$ $$
 - \sum_{y\in j(X)}  \lambda_{j(x)}(x) \, Q(j(x)) \log Q(j(x)) - \sum_{y\in j(X)} Q(j(x))\, 
 \lambda_{j(x)}(x) \log(\lambda_{j(x)}(x)). $$
 We can still obtain an estimate relating the Shannon entropies $S(P)$ and $S(Q)$,
 though not in the simple form of Lemma~\ref{surjS}.
  
 \begin{lem}\label{infobound}
 Given a summing functor $\Phi_X: \Sigma_{\cP_f}(X)$ for a finite pointed set $X$,
 there exists constants $\lambda_{\min},\lambda_{\max} \geq 1$ depending only
 on $X$ such that $S(\Phi_X(A))\leq \lambda_{\max} S(\Phi_X(A')) -\lambda_{\min}\log\lambda_{\min}$
 for all inclusions $A\subset A'$ of pointed subsets of $X$.
 \end{lem}

\begin{proof}
  The summing functor $\Phi_X: P(X)\to \cP_f$ assigns to pointed subsets
 $A\subset X$ probabilities $P_A$ and to inclusions $j: A \hookrightarrow A'$ morphisms
 $(j,\Lambda): (A,P_A)\to (A',P_{A'})$ with $\Lambda=\{ \lambda_{j(a)}(a) \}_{a\in A}$
 determined by $P_A(a)=\lambda_{j(a)}(a)\, P_{A'}(j(a))$. The functoriality of $\Phi_X$
 ensures that the probabilities $P_A$ and $P_{A'}$ are assigned with the consistency
 condition that $j({\rm supp}(P_A))\subset {\rm supp}(P_{A'})$. 
 We then assign to pointed subsets $A\subset X$ 
 the value of the Shannon entropy $S(P_A)=-\sum_{a\in A} P_A(a) \log P_A(a)$. 
 A morphism in $P(X)$ is given by a pointed inclusion $j: A\hookrightarrow A'$, with
 $\Phi_X(j)=(j,\Lambda)$ the corresponding morphism in $\cP_f$. 
 Consider the inclusions $j_a: \{ * \} \hookrightarrow \{ *, a \}$ for $a\in X$
 and the inclusions $\iota_{a,k} : \{ *, a \}\hookrightarrow \{ *, a \} \vee_{j=1}^k \{*, a_j\}$
 in wedge sums of finite pointed sets. The corresponding morphisms in $\cP_f$ have dilation
 factors $\lambda(j_a)\geq 1$ and $\lambda(\iota_{a,k})\geq 1$ where these are the dilation
 factors of the embeddings in the coproduct of $\cP_f$ as discussed in Lemma~\ref{finprobfibers}.
 Any $j: A\hookrightarrow A'$ inclusion of finite pointed subsets of $X$ is a composition of
 these maps, hence its scaling factors are products of these factors. Thus, the bounds
 $\lambda_{\min}=\min \lambda_{j(a)}(a)$ and $\lambda_{\max}=\max \lambda_{j(a)}(a)$ over
 $a\in X$ and over all possible morphisms $j: A\hookrightarrow A'$ in $P(X)$ satisfy
 $\lambda_{\min},\lambda_{\max} \geq 1$. The Shannon entropy satisfies
 $$ S(P)=- \sum_{j(a) \in j(A)} \lambda_{j(a)}(a)\, P_{A'}(j(a)) \log (\lambda_{j(a)}(a)\, P_{A'}(j(a)) ) $$
 $$ \leq - \lambda_{\max} \sum_{a'\in A'} P_{A'}(a')\log (P_{A'}(a')) -\lambda_{\min}\log \lambda_{\min}. $$
 In particular, $S(P_A)\leq \lambda_{\max} S(P_{A'})$.
\end{proof}
 
 In \S \ref{GammaNetInfoSec} below we discuss a better way of assigning probabilities and information
 structures to codes that bypasses the problem described here, and gives a 
 good functorial construction that leads to information measures naturally
 associated to networks and their neural codes.
 
 \subsubsection{Category of simplices} As preliminary notation, we
 recall the construction of the category $\Delta$ of simplicial sets, which
 we will be using frequently in the rest of the paper, starting in the next
 subsection, \S \ref{GammaNetInfoSec}.
 
 
 Denote by  $[n]$ for $n=0,1,2, ... $ the totally ordered subset
of integers $[n]=\{0, \ldots, n\}$.
 The simplex category $\triangle$ (not to be confused with the category
 $\Delta$ that we define below) has objects the sets $[n]$ and morphisms
 the nondecreasing maps $f: [n] \to [m]$.
 
 
 Morphisms are generated by two classes of maps (see \cite{GeMa03}, pp.~14--15): 
 $\partial^i_n$ and $\sigma^i_n$, respectively given by the increasing injection $[n-1] \to [n]$
not taking the value $i$, and the nondecreasing surjection $[n+1] \to [n]$ taking the value $i$ twice.
Faces and degeneracies satisfy the relations
$$
\partial^j_{n+1}   \partial^i_{n} = \partial^i_{n+1}  \partial^{j-1}_{n}  \quad \text{for} \quad i < j ;
$$
$$
\sigma^j_{n-1}   \sigma^i_{n+1} =   \sigma^i_n  \sigma^{j+1}_{n+1}  \quad \text{for}\quad i \le j ;
$$
$$
\sigma^j_{n-1}\partial^i_n =\left\{ 
\begin{array} {ll}
\partial^i_{n-1}\sigma^{j-1}_{n-2}  & \text{for}\quad  i<j \, ,\\
id_{[n-1]} & \text{for}\quad  i\in \{j,j+1\} \, , \\
\partial^{i-1}_{n-1}\sigma^{j}_{n-2} & \text{for}\quad  i>j+1 .
\end{array}\right.
$$


A simplicial object of a category $\cC$ is a 
functor $\triangle^{\operatorname{op}} \to \cC$. In particular,
a simplicial set is a functor $\triangle^{\operatorname{op}} \to {\rm Sets}$ and
a pointed simplicial set is a functor $\triangle^{\operatorname{op}} \to {\rm Sets}_*$
to pointed set.


In the following we will always denote by $\Delta$ and $\Delta_*$ the
categories of simplicial sets and of pointed simplicial sets
$$ \Delta:={\rm Func}(\triangle^{\operatorname{op}}, {\rm Sets}), \ \ \ \ \  \Delta_* :={\rm Func}(\triangle^{\operatorname{op}}, {\rm Sets}_*) $$
with morphisms given by natural transformations of the functors. 


The classical description of morphisms in $\triangle$ via generators (``$i$-th face maps'',
``$i$-th degeneracy maps'')  and relations recalled above produces
explicit description of simplicial sets and their topological realizations. 


The objects $[n]$ of $\triangle$ are realized by the standard simplices,
denoted by $\Delta_{n}\subset \R^{n+1}$, namely the $n$-dimensional topological space
$$
\Delta_n : =\{ (x_0,\dots , x_n)   | \sum_{i=0}^n x_i=1, x_i\ge 0\} .
$$

\subsection{Summing functors and information measures}\label{GammaNetInfoSec}

In this section we consider again the formalism of network summing functors introduced
in \S \ref{SummingSec}, but we focus on the associated information structure,
rather than on computational architectures as in \S \ref{GammaNetCompSec}. 
We start with a review of the cohomological information formalism. 

\subsubsection{Cohomological information theory}\label{CohomInfoSec}

We adopt here the point of view of \cite{BauBen1}, \cite{BauBen2}, and especially \cite{Vign},
on a cohomological formulation of information measures. This will allow us to significantly
improve the provisional construction described in \S \ref{ProbCodesSec} of probabilities
assigned to networks via the corresponding neural codes. The main problem with the
construction we described in \S \ref{ProbCodesSec} is that the category of probabilities
we used does not have sufficiently good properties, with respect to information measures.
This was shown in \S \ref{InfomeasSec}: while the Shannon entropy is functorial on the
category of finite probability spaces with surjections with fiberwise probabilities as morphisms (as shown
in Lemma~\ref{surjS}), it is not functorial on the category of finite probabilities of Lemma~\ref{finprobfibers},
which is the target of the functors from codes discussed in \S \ref{ProbCodesSec}. This is
because, to have a sum and a zero object in this category, we need to allow for morphisms that
are not surjections and fiberwise measures that are not probabilities, over which
the Shannon entropy is not a monotone function (see Lemma~\ref{infobound}). Note, however,
that this still determines a symmetric monoidal structure that can be used as a category of
resources. 


To remedy this problem we now give a more refined construction, which uses 
network summing functors with a target category that is
an abelian category describing probability data, as introduced in \cite{Vign}. 


The most important aspects we want to retain of this general formalism of information structures and probabilities
are the fact that there is a suitable category of random variables and functors $\cQ$ from this category to simplicial
sets that assign to a random variable $X$ a corresponding simplicial set of probabilities $\cQ_X$. There is then a functor
$\cM$ to vector spaces, that associates to $P\in \cQ_X$ the real vector space of $P$-measurable functions. These vector
spaces are in turn used to construct cochain complexes using a Hochschild-type resolution and Hochschild coboundary.
This cochain complex is designed so that its cohomology describes classical information functionals. 
In a somewhat more detailed form, we summarize briefly the relevant parts of the setting of \cite{Vign} 
that we need for our purposes. 


\begin{itemize}
\item A {\em finite information structure} $(S,M)$  is a pair of a thin category $S$, as in Definition~\ref{thindef}  
(the observables) and a functor $M: S \to \cF$ to the category of finite sets.
\item The category $S$ has objects $X\in {\rm Obj}(S)$ given by random variables with values in a
finite probability space and a morphism $\pi: X\to Y$ if the random variable $Y$ is coarser than $X$
(values of $Y$ are determined by values of $X$), with the property that, if there are morphisms 
$X\to Y$ and $X\to Z$ then $YZ=Y\wedge Z$ (the random variable given by the joint measurement of
$Y$ and $Z$) is also an object of $S$. 
\item The category $S$ has a terminal object ${\bf 1}$ 
given by the random variable with value set $\{ * \}$ a singleton.
\item The functor $M: S \to \cF$ maps a random variable $X$ to the finite set given by
its range of values $M_X$ and morphisms $\pi: X\to Y$ to surjections $M(\pi) : M_X \to M_Y$. 
The value set $M_{X\wedge Y}$ is a subset of $M_X \times M_Y$.
\item The category $\cI\cS$ of finite information structures has objects the pairs $(S,M)$
as above and morphisms $\varphi: (S,M)\to (S',M,')$ given by pairs $\varphi=(\varphi_0,\varphi^\#)$ 
of a functor $\phi_0: S \to S'$ and a natural transformation $\phi^\#: M\to M'\circ \phi_0$ such that
$\phi_0({\bf 1})={\bf 1}$ and $\phi_0(X\wedge Y)=\phi_0(X)\wedge \phi_0(Y)$ whenever $X\wedge Y$ is
an object in $S$, and such that for all $X$ the morphism $\phi^\#_X: M_X \to M'_{\phi_0(X)}$ is
a surjection. 
\item The category $\cI\cS$ has finite products $(S\times S', M\times M')$ with objects
pairs $(X,X')$ of random variables with value set $M_X \times M'_{X'}$ and coproducts
$(S\vee S', M\vee M')$ with objects ${\rm Obj}(S\vee S')={\rm Obj}(S)\vee {\rm Obj}(S')=
{\rm Obj}(S)\sqcup {\rm Obj}(S')/{\bf 1}_S\sim {\bf 1}_{S'}$ and value set $M_X$ or $M'_{X'}$
if $X\in {\rm Obj}(S)$ or $X'\in {\rm Obj}(S')$.
\item A {\em probability functor} $\cQ: (S,M) \to \Delta$ assigns to each object $X$ a 
simplicial set $\cQ_X$ of probabilities on the set $M_X$ (which is a subset of the
simplex $\Delta_{M_X}$ of all probability distributions on $M_X$) and to morphisms
$\pi: X \to Y$ the morphism $\pi_*: \cQ_X \to \cQ_Y$ with $\pi_*(P)(y)=\sum_{x\in \pi^{-1}(y)} P(x)$.
\item For each $X\in {\rm Obj}(S)$ there is a {\em semigroup} 
$\cS_X=\{ Y \in  {\rm Obj}(S)\,|\, \exists \pi: X \to Y \}$
with the product $Y\wedge Z$, and a {\em semigroup algebra} $\cA_X:=\R[\cS_X]$.
\item There are associated {\em contravariant functors} $\cM(\cQ): (S,M) \to {\rm Vect}$ 
that assign to objects $X\in {\rm Obj}(S)$
and probabilities $P_X \in \cQ_X$ the vector space of real-valued (measurable) functions on $(M_X, P_X)$ 
and to a morphism $\pi: X \to Y$ the map $\cM(\cQ)(\pi): f \mapsto f\circ \pi_*$.
\item There is an {\em action} $\sigma_\alpha$ of the semigroup $\cS_X$ on $\cM(\cQ_X)$ by 
$$ \sigma_\alpha(Y): f \mapsto Y(f)(P_X)=\sum_{y\in M_Y\,:\, Y_*P_X(y)\neq 0} (Y_* P_X(y))^\alpha 
\, f(P_X|_{\pi^{-1}(y)}) $$
for $Y\in \cS_X$ and for an arbitrary $\alpha >0$, with $Y_*P_X(y)=P_X(Y=y)$ the marginal law. 
\item There is an $\cA_X$-{\em module} structure $\cM_\alpha (\cQ_X)$ on $\cM(\cQ_X)$, determined by
the semigroup action $\sigma_\alpha$.
\item The category $\cA$-Mod of modules over the sheaf of algebras $X\mapsto \cA_X$ is an abelian category.
\item There is a sequence $\cB_n(X)$ of free $\cA_X$-modules generated by symbols 
$[X_1\, \vert \ldots \vert \, X_n]$
with $\{ X_1,\ldots, X_n \}\subset \cS_X$, and with boundary maps $\partial_n: \cB_n \to \cB_{n-1}$ of the
Hochschild form
\begin{equation}\label{Bnpartial}
\begin{array}{rl}
\partial_n [X_1\, \vert \ldots \, \vert X_n] = & X_1 \, [X_2\, \vert \ldots \vert \, X_n] \\
+& \displaystyle{ \sum_{k=1}^{n-1} (-1)^k
[X_1\, \vert \ldots \vert \, X_k X_{k+1} \, \vert \ldots \vert \, X_n]} \\[4mm] + & (-1)^n [X_1\, \vert \ldots \vert X_{n-1}].
\end{array} 
\end{equation}
The modules $\cB_n(X)$ give a projective bar resolution of the trivial $\cA_X$-module.
\item There is a functor $C^\bullet(\cM_\alpha(\cQ)): (S,M) \to {\rm Ch}(\R)$ to the 
category of {\em cochain complexes},
that assigns to $X\in {\rm Obj}(S)$ a cochain complex $(C^\bullet(\cM_\alpha(\cQ_X)),\delta)$ with $C^\bullet(\cM_\alpha(\cQ_X))^n=\Hom_{\cA_X}(\cB_n(X),\cM_\alpha(\cQ_X))$ (that is, natural transformations 
of functors $\cB_n \to \cM_\alpha(\cQ)$ compatible with the $\cA$-action) and with 
coboundary $\delta$ given by the Hochschild-type coboundary
\begin{equation}\label{Fcoubound}
\begin{array}{rl}
\delta(f) [X_1\, \vert \ldots \vert \, X_{n+1}]=& X_1(f) [X_2\, \vert \ldots \vert \, X_{n+1}] \\
+ & \displaystyle{ \sum_{k=1}^n (-1)^k f [X_1\, \vert \ldots \vert \, X_k X_{k+1} \, \vert \ldots \vert \, X_{n+1}]
}\\[4mm]
+ & (-1)^{n+1} f [X_1\, \vert \ldots \vert X_n] . 
\end{array} 
\end{equation}
\item One writes $C^\bullet((S,M), \cM_\alpha(\cQ)):=(C^\bullet(\cM_\alpha(\cQ_X)),\delta)$ and
$H^\bullet((S,M), \cM_\alpha(\cQ))$ for the resulting cohomology. The zeroth cohomology is $\R$
when $\alpha=1$ and zero otherwise. In the case of the first cohomology, any non-trivial $1$-cocycle
is locally a multiple of the Tsallis entropy
$$ S_\alpha[X](P)=\frac{1}{\alpha -1} \left( 1- \sum_{x\in M_X} P(x)^\alpha \right), $$
for $\alpha\neq 1$ or of the Shannon entropy for $\alpha=1$. The higher cohomologies 
similarly represent all possible higher mutual information functionals. 
\end{itemize}


This functorial construction can be used to map networks to an {\em abelian} 
category of informational resources. 
According to what we discussed earlier in this section, we want an assignment of informational
resources to networks that factors through an intermediate category of codes (or weighted codes). Thus,
we revisit here the construction of \S \ref{InfomeasSec}, using the more sophisticated setting of
cohomological information recalled above.

\subsubsection{Network summing functors and information}\label{NetSumInfoSec}

We now return to the category of codes ${\rm Codes}_{n,*}$ introduced in
Lemma~\ref{catcodes} and the category of network summing functors
$\Sigma^{\operatorname{eq}}_{{\rm Codes}_{n,*}}(G)$. We show that there is
an associated category of network summing functors obtained by mapping the
summing functors $\Phi\in \Sigma^{\operatorname{eq}}_{{\rm
Codes}_{n,*}}(G)$ to summing functors in $\Sigma_{\cA{\text{\rm -Mod}}}(G)$,
with $\cA{\text{\rm -Mod}}$ the abelian category of sheaves of $\cA_X$-modules
as in \cite{Vign}, and to summing functors in $\Sigma_{{\rm Ch}(\R)}(G)$
with values in cochain complexes. Summing functors in these categories
satisfy the inclusion-exclusion relations of \S \ref{ExInclSec}.


\begin{lem}\label{CodesIS}
There is a contravariant functor $\cI: {\rm Codes}_{n,*}\to \cI\cS$ 
from the category ${\rm Codes}_{n,*}$ to 
the category $\cI\cS$ of finite information structures
that maps the coproduct $C\vee C'$ in ${\rm Codes}_{n,*}$ to the coproduct $(S,M)\vee (S',M')$ in $\cI\cS$.
\end{lem}

\begin{proof}
  Given a code $C\in {\rm Codes}_{n,*}$, with $\# C$ code words of length $n$ including the base
point given by the $0$-word $c_0$, consider the set $\cI(C)=S^C$ of all random variables $X: C \to \R$ with
values in a finite subset of $\R$ and with $X(c_0)=0$. One should think of such a variable as a
probabilistic assignment of weights to the code words.  A morphism in ${\rm Codes}_{n,*}$ is a function
$f: C \to C'$ that maps the $0$-word $c_0$ to itself. For $X'\in S^{C'}$ let $\cI(f)(X')=X\in S^C$ be given
by $X=X'\circ f: C \to \R$. The object ${\bf 1}$ is the random variable that maps the whole code to $0$
and $\cI(f)({\bf 1})={\bf 1}$. Whenever $X'Y'=X'\wedge Y'$ is an object in $S^{C'}$ we have
$X'Y' \circ f=X'\circ f\wedge Y' \circ f$ an object in $S^C$. 
We define the map on the value sets as the projection $\pi_f: M_{X'}= M_{X' \circ f}$
that maps $m\in M_{X'}$ to itself if $m=X'(f(c))$ for some $c\in C$ and to $0$ otherwise. Note that $0$ is
always an element of both $M_{X'}$ and $M_{X' \circ f}$ because of the $0$-word. 
The coproduct of codes $C\vee C'$ is obtained from the disjoint union of the two codes by identifying the
respective $0$-words. Under the functor $\cI$, the code $C\vee C'$ is mapped to the category
$S^{C\vee C'}$ of random variables with finite range $X^\vee: C\vee C' \to \R$ that map the $0$-word to $0$.
Such a random variable $X^\vee$ applied to code words in $C$ determines a random variable in $S^C$
and applied to words in $C'$ determines a random variable in $S^{C'}$, and in turn is determined by
such random variables, which necessarily agree on the $0$-word. The pair of ${\bf 1}_{S^C}$ and ${\bf 1}_{S^{C'}}$
gives ${\bf 1}_{S^{C\vee C'}}$.  Thus, we have $S^{C\vee C'}=S^C \vee S^{C'}$.
\end{proof}


\begin{lem}\label{FQsum}
Under the functor $\cM\cQ: \cI\cS \to \cA$-Mod, the product
$(S,M)\times (S',M')$ maps to the tensor product $\cM_\alpha(\cQ)\otimes \cM_\alpha(\cQ')$ 
of $\cA$-modules and the coproduct $(S,M)\vee (S',M')$ in $\cI\cS$
maps to the sum $\cM_\alpha(\cQ)\oplus \cM_\alpha(\cQ')$ of $\cA$-modules.
\end{lem}

\begin{proof}
  As shown in \S 2.12 of \cite{Vign}, at the level of the probability functors $\cQ, \cQ'$ we
have $\cQ\times \cQ': (S,M)\times (S',M') \to \Delta$ with $(\cQ\times \cQ')_{(X,X')}$ the simplicial
set given by probabilities on $M_X \times M_{X'}$ that are products $P(x,x')=P(x)P'(x')$, so that
$(\cQ\times \cQ')_{(X,X')}\simeq \cQ_X \times \cQ'_{X'}$, while $\cQ\vee \cQ': (S,M)\vee (S',M') \to \Delta$
is defined on $X\in {\rm Obj}(S)$ as $\cQ_X$ and on $X'\in {\rm Obj}(S')$ as $\cQ'_{X'}$. When we
consider the vector space of measurable functions we then obtain
$\cM_\alpha((\cQ\times \cQ')_{(X,X')})=\cM_\alpha(\cQ_X \times \cQ'_{X'})\simeq \cM_\alpha(\cQ_X)\otimes
\cM_\alpha(\cQ'_{X'})$. Similarly, the vector space of functions on the simplicial sets obtained
from $\cQ\vee \cQ'$ splits as a direct sum of $\cM_\alpha(\cQ_X)$ for $X\in {\rm Obj}(S)$ and 
$\cM_\alpha(\cQ'_{X'})$ for $X'\in {\rm Obj}(S')$. 
\end{proof}


The following is a direct consequence of the previous lemmas.

\begin{cor}\label{SumCAMod}
Composition with the functor $\cM\cQ\circ \cI$ maps summing functors 
$\Phi\in \Sigma^{\operatorname{eq}}_{{\rm Codes}_{n,*}}(G)$ to summing functors
$\cM\cQ(\cI(\Phi))\in \Sigma^{\operatorname{eq}}_{\cA{\text{\rm -Mod}}}(G)$. 
\end{cor}


Similarly, we can consider composition with the functor that assigns to 
a finite information structure the corresponding cochain complex $C^\bullet((S,M),\cM_\alpha(\cQ))$
and its cohomology $$H^\bullet((S,M),\cM_\alpha(\cQ))\, . $$ 

\begin{prop}\label{SumCCh}
Let $\cK:=C^\bullet(\cM_\alpha(\cQ)): \cI\cS \to {\rm Ch}(\R)$ be the functor that maps finite information
structures to their information cochain complex $$(S,M)\mapsto C^\bullet((S,M),\cM_\alpha(\cQ)).$$
Composition with $\cK\circ \cI$ maps summing functors 
$\Phi\in \Sigma^{\operatorname{eq}}_{{\rm Codes}_{n,*}}(G)$ to summing functors $\cK(\cI(\Phi))\in \Sigma^{\operatorname{eq}}_{{\rm Ch}(\R)}(G)$,
with $\cK(\cI(\Phi))(G')=C^\bullet((S,M)^{G'},\cM_\alpha(\cQ))$ where we write $(S,M)^{G'}:=\cI(\Phi)(G')$ for
$G'\subset G$.
These satisfy the inclusion-exclusion property of \S \ref{ExInclSec}, namely for $G_1,G_2\subset G$ there is a short
exact sequence of cochain complexes 
$$ 0 \to \cK(\cI(\Phi))(G_1\cap G_2) \to 
\cK(\cI(\Phi))(G_1) \oplus \cK(\cI(\Phi))(G_2) \to 
\cK(\cI(\Phi))(G_1\cup G_2) \to 0, $$ 
with 
$$ \begin{array}{l} 
\cK(\cI(\Phi))(G_1\cap G_2)=C^\bullet((S,M)^{G_1\cap G_2},\cM_\alpha(\cQ)) \\
\cK(\cI(\Phi))(G_i)=C^\bullet((S,M)^{G_i},\cM_\alpha(\cQ)) \\
\cK(\cI(\Phi))(G_1\cup G_2)=C^\bullet((S,M)^{G_1\cup 
G_2},\cM_\alpha(\cQ))  ,
\end{array}
$$
hence a corresponding long exact sequence of information cohomologies.
\end{prop}

\begin{proof}
  Since $\Phi$ is a summing functor in $\Sigma^{\operatorname{eq}}_{{\rm Codes}_{n,*}}(G)$ the
value of $\Phi$ on a subnetwork $G'\subset G$ reduces to the sum, in the category
${\rm Codes}_{n,*}$ of the codes $C_e=\Phi(e)$ associated to the edges $e\in E(G')$,
that is, the coproduct $\bigvee_{e\in E(G')} C_e$. Thus, given $G_1, G_2\subset G$
we have $\Phi(G_1\cap G_2)=\bigvee_{e\in E(G_1\cap G_2)} C_e$ as a subsummand
of both $\Phi(G_1)$ and $\Phi(G_2)$, and each of these in turn gives a subsummand
of $\Phi(G_1\cup G_2)$. Applying Lemma~\ref{FQsum} we then obtain an exact sequence
of $\cA$-modules 
$$ 0\to \cM_\alpha(\cQ^{G_1\cap G_2}) \to \cM_\alpha(\cQ^{G_1}) \oplus \cM_\alpha(\cQ^{G_2})
\to \cM_\alpha(\cQ^{G_1\cup G_2}) \to 0. $$
The $\cA$-modules $\cB_n$ are projective, hence $\Hom_\cA(\cB_n, \cdot)$ is an exact
functor, hence we obtain the short exact sequence of cochain complexes. 
\end{proof}

\subsubsection{Other functorial maps to information structures}\label{InfoChSetsSec}

In the previous subsection we focused on the category $\cC={\rm Codes}_{n,*}$ as 
we have done previously in \S \ref{GammaCodesSec}, and a functor $\cI: {\rm Codes}_{n,*} \to \cI\cS$
from codes to information structures. The same construction and the result of Proposition~\ref{SumCCh}
can be generalized to other categories $\cC$ (as target of the summing functors), together
with a functor $\cI: \cC \to \cI\cS$ with the property that the sum $C_1\oplus C_2$ in $\cC$ 
maps to the coproduct $(S,M)^{C_1}\vee (S,M)^{C_2}$ of information structures. 

\begin{cor}\label{SumCCh2}
Consider summing functors $\Phi\in \Sigma^{\operatorname{eq}}_\cC(G)$. Given a functor $\cI: \cC \to \cI\cS$
preserving coproducts, the composition $\cK\circ \cI$ with $\cK=C^\bullet(\cM_\alpha(\cQ))$
maps summing functors $\Phi\in \Sigma^{\operatorname{eq}}_\cC(G)$ to summing functors $\cK(\cI(\Phi))\in 
\Sigma^{\operatorname{eq}}_{{\rm Ch}(\R)}(G)$ with $\cK(\cI(\Phi))(G')=C^\bullet((S,M)^{G'}, \cM_\alpha(\cQ))$
satisfying the inclusion-exclusion property as in Proposition~\ref{SumCCh}.
\end{cor}


In particular, one can consider the case where $\cC=\cF_*$ is the category of
finite pointed set. As we will see in \S \ref{GammaGeneralSec}, this case 
corresponds to the Gamma-space that is the embedding
of $\cF_*$ in $\Delta_*$, whose spectrum is the sphere spectrum. In this case,
the functor $\cI: \cF_* \to \cI\cS$ maps a finite pointed set $A\in \cF_*$ to
the information structure $(S,M)^A$ with ${\rm Obj}(S^A)$ the random variables
$X: A \to \R$ with $X(a_0)=0$ at the basepoint $a_0\in A$. This satisfies
$(S,M)^{A\vee A'}=(S,M)^A \vee (S,M)^{A'}$. In the case where the sets $A$
describe the sets of vertices $V_{G_*}$ or edges $E_{G_*}$ of a network, we identify
the resulting $C^\bullet((S,M)^{G'}, \cM_\alpha(\cQ))$ and its cohomology
with the measuring of information content of the subnetwork $G'\subset G$.
In the more general case of other categories $\cC$ as in Corollary~\ref{SumCCh2}, the 
information complex $\cK(\cI(\Phi))(G')=C^\bullet((S,M)^{G'}, \cM_\alpha(\cQ))$
measures the information content of the resources $\Phi(G')\in \cC$ assigned to the 
subnetwork $G'$.

\subsection{Codes and simplicial sets}\label{CodesNerveSec}

We discuss here how the simplicial sets associated to binary
(convex) neural codes through the associated open covering and its nerve
fit in the setting of information structures we introduced in \S \ref{GammaNetInfoSec}.
The convexity hypothesis for a code $C\subset \F_2^n$ consists of the requirement
that the code words $c\in C$ can be realized as intersection patterns of a family 
$\{ U_1, \ldots, U_n \}$ of convex open sets in some Euclidean space $\R^d$,
see \cite{CGJMORSY}. 


More precisely, in this setting, we have a code $C$ with $N=\#C$ code words of $n$ letters
each, with alphabet $\{ 0,1 \}$. We consider a collection $\{ U_\nu \}_{\nu=1}^n$
of open sets  (receptive fields) associated to the $n$ neurons $\nu$. For each
code word $c\in C$ we consider the letters $c_\nu=1$. These are neurons
that simultaneously fire in the reading represented by the code word $c$, hence
receptive fields that overlap. This means that we have an intersection 
$\cap_{\nu\,:\, c_\nu=1} U_\nu$ associated to each code word $c\in C$. 
One then considers a simplicial set associated to the code given by the 
nerve $\cN(\cU(C))$ of the collection $\cU(C)=\{ U_\nu \}$. This has a $k$-simplex for
every non-empty $(k+1)$-fold intersection. We write these as $\Delta_c$
for the simplex associated to the intersection $\cap_{\nu\,:\, c_\nu=1} U_\nu$.
The code $C$ is convex if the $U_\nu$ are convex. 


\begin{lem}\label{NUCprobQ}
Given a binary convex code $C$, there is a finite information structure $(S,M)$ and a
probability functor $\cQ$ for which there is a random variable $X$ in ${\rm Obj}(S)$ 
such that $\cQ_X=\cN(\cU(C))$ is the nerve of the collection of open coverings $\cU(C)$
associated to the code $C$.
\end{lem}

\begin{proof}
  Given a code $C$ as above, we write $C^\vee$ for the transpose code
that has $n$ code words with $N$ letters each. We write the code
words of $C^\vee$ as $\nu=(\nu_c)_{c=1}^N$.  Consider then the set
of real-valued random variables $X: C\times C^\vee \to \R$. In particular,
consider the case of 
$$ X(c,\nu)=X_c(\nu)=\left\{ \begin{array}{ll} 0 & c_\nu=0 \\ \alpha_\nu & c_\nu=1 \end{array}\right. $$
where $\alpha_\nu \neq 0$ and $\alpha_\nu \neq \alpha_{\nu'}$ for $\nu\neq \nu'$. 
Consider a probability functor $\cQ$, in the sense recalled in \S \ref{GammaNetInfoSec} 
mapping random variables $X$ to simplicial sets $\cQ_X \subset \Delta_{M_X}$, that maps
$X: C \times C^\vee \to \R$ to the simplicial set $\cQ_X =\cup_{c\in C} \Delta_{M_{X_c}}$.  
Note that the simplex $\Delta_{M_{X_c}}$ is just the simplex on 
a number of vertices equal to $\# \{ \nu\,:\, c_\nu =1 \}$. 
\end{proof}


We have obtained in this way a
realization of the nerve simplicial set $\cN(\cU(C))$ through the information structures
and probability functors construction of \S \ref{GammaNetInfoSec}.

\subsubsection{Nerves of coverings and functoriality}\label{NerveUFunctorSec}

We can also consider the question of whether the assignment of the simplicial set
$\cN(\cU(C))$ to a code $C$ is functorial with respect to an appropriate choice of
morphisms of codes. 


In Lemma~\ref{catcodes} we defined the category ${\rm Codes}_{n,*}$ of binary codes with
objects that are binary codes that include the zero word and morphisms that are maps of
pointed sets between them. Here we consider a subcategory on the same objects with a
subclass of morphisms. For simplicity we will neglect
base points, and work with an un-based version ${\rm Codes}_n$ of the category of 
binary codes.


Since we want to think of our codes as neural codes that detect
the spiking activity of a population of neurons over a span of time subdivided into basic
intervals, we can regard codes as maps
\begin{equation}\label{CodeMapEq}
 C: X \times T_n \to \{ 0, 1 \}, 
\end{equation} 
where $X\in \cF$ is a finite set and $$T_n=\{ [t_0,t_0+\Delta t], [t_0+\Delta t,t_0+2\Delta t], \ldots, [t_0+(n-1)\Delta t, t_0+ n \Delta t] 
\}$$ is the set of basic
intervals, identified with $T_n =\{ 1,2 , \ldots, n \}$. Thus, the set $T_n$ is fixed and dependent
only on the choice of $n\in \N$. Code words in $C$ are given by
\begin{equation}\label{codewordMap}
 c_x = C(\{ x\}\times T_n) \, , \ \ \ \text{ for } x\in X \, . 
\end{equation} 


\begin{prop}\label{CatCodeMaps}
Let ${\rm Codes}'_n$ be the category of codes with objects the maps as in \eqref{CodeMapEq}
and with morphisms $f\in {\rm Mor}_{{\rm Codes}'_n}(C,C')$, for $C: X \times T_n \to \{ 0, 1 \}$ and $C': X'\times T_n \to \{ 0,1 \}$, 
given by maps $f: X\to X'$ that fit in a commuting diagram
$$ \xymatrix{ X\times T_n \ar[rr]^{f\times \operatorname{id}} \ar[rd]_{C} & & X'\times T_n \ar[ld]^{C'} \\ & \{ 0,1 \} \, . & } $$
There is a functor $F: {\rm Codes}'_n \rightarrow {\rm Codes}_n$ that
is faithful when restricting to codes that have no repeated code words. The map 
$$ \cN\cU: {\rm Codes}_n\to \Delta, \ \ \ \  C\mapsto \cN(\cU(C)), $$ 
that assigns to a code the simplicial set given by the nerve of the covering $\cU(C)$ determined by the code
defines a functor $\cN\cU\circ F: {\rm Codes}'_n \to \Delta$.
\end{prop}

\begin{proof}
  We identify an object of ${\rm Codes}'_n$ with an object of ${\rm Codes}_n$,
by assigning to the map $C: X \times T_n \to \{ 0, 1 \}$ the set of code words
$C=\cup_{x\in X} c_x=\cup_{x\in X} C(\{ x\}\times T_n)$.
Given a map $f: X\to X'$ we obtain a morphism $\phi_f: C \to C'$ in ${\rm Codes}_n$
by setting
$$ \phi_f(c)=\phi_f(C(\{ x\}\times T_n)) :=C'(\{ f(x) \}\times T_n) \, . $$
In other words, the morphism $\phi_f$ places the word $c_x$ of $C$ in the position $f(x)$ in $C'$. Indeed,
since $C'\circ (f,id)=C$, these code words agree, $c'_{f(x)}=c_x$, as binary words of length $n$. 
Suppose we only consider codes that have no repeated words (which implies we also consider 
only injective maps $f: X\to X'$). The identity $\phi_f(c)=\phi_g(c)$ for all $c\in C$ means that the code words  
$c'_{f(x)}=C'(\{ f(x) \}\times T_n)=c_x$ and $c'_{g(x)}=C'(\{ g(x) \} \times T_n)=c_x$ are the same for all $x\in X$. 
If $f(x)\neq g(x)$ for some $x\in X$, the code $C'$ has repeated words. 
Thus, in the case of codes with no repeated words, we obtain a faithful functor
$F: {\rm Codes}'_n \hookrightarrow {\rm Codes}_n$ that realizes ${\rm Codes}'_n$ as a subcategory
of the category of codes ${\rm Codes}_n$.
To check the functoriality of the assignment $C\mapsto \cN(\cU(C))$, we can describe the
simplicial set $\cN(\cU(C))$ in the following way. The set $\cN(\cU(C))_0$ of vertices of $\cN(\cU(C))$ is given by
the subset of $X$
$$ \cN(\cU(C))_0 =\{ x\in X \, |\, C(x,i)=1, \, \text{ for some } i \in T_n \}\, .$$
The set $\cN(\cU(C))_k$ of $k$-simplices is given by the set
$$ \cN(\cU(C))_k =\left\{ \sigma= \{ x_0, \ldots, x_k \} \subset X \,|\, \exists i \in T_n \, \text{ such that } \, C(x,i)=1, \,\, \forall x\in \sigma \right\} \, . $$
Then we can associate to a morphism $f\in {\rm Mor}_{{\rm Codes}'_n}(C,C')$ an induced simplicial map $f_*: \cN(\cU(C)) \to \cN(\cU(C'))$ by setting
$$ f_*: \cN(\cU(C))_k \to \cN(\cU(C'))_k\, , \ \ \ \   \sigma=\{ x_0, \ldots, x_k \} \mapsto f_*(\sigma)=\{ f(x_0), \ldots, f(x_k) \}\, . $$
Indeed, if there is an $i\in T_n$ such that $C(x,i)=1$ for all $x\in \sigma$, then $C'(f(x),i) =C(x,i)$ by our choice of morphisms,
so that we also have $f_*(\sigma)=\{ f(x_0), \ldots, f(x_k) \}\in \cN(\cU(C'))_k$.
\end{proof}

\subsection{Transition systems, codes, and information structures}\label{TransCodesInfoSec}

We describe here a functorial mapping from the category $\cC$ of transition systems to the category ${\rm Codes}_{n,*}$  of codes,
describing codes generated by the automata in $\cC$, and its composition with the functor ${\rm Codes}_{n,*}$  
to the category $\cI\cS$ of finite information structures, as in Lemma~\ref{CodesIS}.



We consider again the category $\cC$ of concurrent/distributed computational architectures
given by transition systems, as in \cite{WiNi95}, recalled in \S \ref{CompResSec} 
and \S \ref{GammaNetCompSec}. Also let $\cI\cS$ denote the category of finite
information structures of \cite{Vign}, recalled in \S \ref{GammaNetInfoSec}. 


As discussed in \cite{WiNi95}, in the category $\cC$ of transition systems 
$\tau=(S,\iota,\cL,\cT)$ one usually assumes that the set $\cT$ of transitions
always contains also the ``idle transitions'' of the form $(s,\star, s)$ with a
special label symbol $\star \in \cL$, which describe the case where the system
at the state $s\in S$ does not update to a new state. 


Recall that, given an automaton $\tau=(S,\iota,\cL,\cT)$, the formal language $\bL(\tau)$
recognized by the automaton consists of all the sequences
of composable transitions in the automaton $\tau$, of arbitrary finite length, 
$$ (s_0,\ell_1,s_1)(s_1,\ell_2,s_2)\cdots (s_{n-1},\ell_n,s_n), \ \ \ \text{ with } \ \ 
s_0=\iota. $$


\begin{lem}\label{funcCIS}
There is a contravariant functor $\cJ: \cC \to \cI\cS$ that assigns to a transition system
$\tau=(S,\iota,\cL,\cT)$ the finite information structure $\cJ(\tau)=(S,M)^{\bL(\tau)}$, where
$\bL(\tau)$ is the language of the automaton $\tau$, and the category $(S,M)^{\bL(\tau)}$
has objects the random variables $X: \bL(\tau) \to \R$ with finite range that map
to $0$ the language word consisting of the idle transition $(\iota,\star,\iota)$.
\end{lem}

\begin{proof}
  A morphism $\phi:\tau \to \tau'$ consists of a pair $\phi=(\sigma,\lambda)$ 
of a function $\sigma: S \to S'$ with
$\sigma(\iota)=\iota'$ and a (partially defined) function $\lambda: \cL \to \cL'$
such that, if $(s,\ell, s') \in \cT$ and $\lambda(\ell)$ is defined, then one has
$(\sigma(s),\lambda(\ell),\sigma(s'))\in \cT'$. Let us consider here, for
simplicity, the case where $\lambda$ is globally defined. 
Such a morphism determines a function $\bL(\tau)\to \bL(\tau')$, by
identifying words in the language $\bL(\tau)$ with composable finite sequences
of transitions in $\tau$ and mapping such a sequence via $(\sigma,\lambda)$ to
a corresponding sequence of composable transitions in $\tau'$, that is, to
a word in the language $\bL(\tau')$. When including idle transitions,
one requires that morphisms $\phi=(\sigma,\lambda)$ in $\cC$ not only
have $\sigma(\iota)=\iota'$ but also $\lambda(\star)=\star'$, hence they
map the idle transition $(\iota,\star,\iota)$ to the idle transition $(\iota',\star',\iota')$
and the word consisting of a concatenation of $n$ idle transitions at the
initial state is then mapped to itself.  One then obtains a morphism
$\cJ(\phi): (S,M)^{\bL(\tau')} \to (S,M)^{\bL(\tau)}$ by precomposition with $\phi$.
\end{proof}


\begin{lem}\label{funcCCodes}
For all $n\in \N$, there is a functor $C_{\bL,n}: \cC \to {\rm Codes}_{n,*}$ 
from the category of transition systems $\cC$ to the category of pointed binary codes,
obtained by assigning to a system $\tau=(S,\iota,\cL,\cT)$ the set $\cW_n(\tau)\subset \bL(\tau)$
of words of length $n$ in the automaton language $\bL(\tau)$, and then mapping the set
$\cW_n(\tau)$ to a binary code $C_{\tau,n}$ of length $n$, with code words
$c(w)$, for $w\in \cW_n(\tau)$ given by
\begin{equation}\label{codetau}
 c(w)_i =\left\{ \begin{array}{ll} 0 & w_i =(s,\star,s) \, \text{ for some } s\in S \\
1 & w_i\neq (s,\star,s) \, \forall s\in S \, , \end{array} \right. 
\end{equation}
detecting whether $w_i$ is the idle transition or not.
\end{lem}

\begin{proof}
  Note that the code $C_{\tau,n}$ contains the zero word $c_0$ as the image of the
word in $\cW_n(\tau)$ consisting of a concatenation of $n$ idle words $(\iota,\star,\iota)$.
The code detects whether, in each of the $n$ discrete time intervals $\Delta t$, the system
$\tau$ has moved from its current state to another state or has idled in the current state
without activity. A morphism $\phi=(\sigma,\lambda): \tau \to \tau'$ induces a map
$\phi: \bL(\tau) \to \bL(\tau')$ that maps $\cW_n(\tau)$ to $\cW_n(\tau')$ and the word
given by the concatenation of $n$ idle transitions $(\iota,\star,\iota)$ to the
concatenation of $n$ idle transitions $(\iota',\star',\iota')$. It therefore induces
a corresponding map $C_{\bL,n}(\phi): C_{\tau,n} \to C_{\tau',n}$ that maps the
code word $c(w)$ to the code word $c(\phi(w))$ mapping the zero word to itself.
\end{proof}


The next statement is then a direct consequence of Lemma~\ref{funcCIS}, Lemma~\ref{funcCCodes},
and Lemma~\ref{CodesIS}. 

\begin{prop}\label{funcCCodesIS}
Let $\cI: {\rm Codes}_{n,*} \to \cI\cS$ be the contravariant functor from codes
to finite information structures constructed in Lemma~\ref{CodesIS}. Let
$(S,M)^{C_\bL(\tau)}=\cI(C_{\bL,n}(\tau))$ be the image of an object $\tau=(S,\iota,\cL,\cT)$
of $\cC$ under the composition $\cI\circ C_{\bL,n}$ with the functor $C_{\bL,n}$ of
Lemma~\ref{funcCCodes}. Let $(S,M)^{\bL(\tau)}=\cJ(\tau)$
with the functor $\cJ$ as in Lemma~\ref{funcCIS}. The category $(S,M)^{C_\bL(\tau)}$
is the subcategory of $(S,M)^{\bL(\tau)}=\cJ(\tau)$ whose objects are the  
random variables $X: \bL(\tau)\to \R$ with finite range such that, when restricted to $\cW_n(\tau)\subset \bL(\tau)$
factor through the code $C_{\tau,n}=C_{\bL,n}(\tau)$,
$$ \xymatrix{ \cW_n(\tau) \ar[rr]^{X|_{\cW_n(\tau)}} \ar[rd] & & \R \\  & C_{\tau,n} \ar[ru] } $$
\end{prop}


Equivalently, the random variables $X: \bL(\tau)\to \R$ with finite range that are 
in the subcategory $(S,M)^{C_\bL(\tau)}$
are those whose value on words in $\cW_n(\tau)$ depends only on which transitions in these words 
are or are not idle, but does not depend on the specific non-idle transitions. This means that we
can regard the set $\cW_n(\tau)$ of words of length $n$ in the automaton language $\bL(\tau)$
as a natural refinement of the binary code $C_{\tau,n}$. In terms of networks of neurons, the
binary code represents the neural code obtained by only retaining the information on whether
a certain neuron in the network is firing or not during each of the $n$ time intervals $\Delta t$,
while the set $\cW_n(\tau)$ also encodes more specific information on the output of the active
neurons, with each interval of time $\Delta t$ corresponding to a transition in the corresponding
automata that simulate the computational activity of the neurons. 


Note that there are different possible ways of constructing computational models 
of individual neurons, in the form of automata and computational
architectures. For example, in \cite{Mar-new} the categorical Hopfield equations
introduced in this paper are analyzed in the case where computational models
of the neuron are given by certain deep neural networks as in \cite{BeSeLo}.


In particular, we can then apply the construction of cohomological information as in
\S \ref{GammaNetInfoSec} and \S \ref{IntegInfoSec} either by applying probability 
functors $\cQ$ to $(S,M)^{C_\bL(\tau)}$ or to the larger category $(S,M)^{\bL(\tau)}$.

\subsection{Clique complexes and information structures}\label{CliqueInfoSec}

We have shown in \S \ref{CodesNerveSec} that the nerve simplicial set $\cN(\cU(C))$ of
a (convex) code $C$ can be recovered from the construction
of \S \ref{GammaNetInfoSec} of the simplicial set $\cQ_X$ associated to
a random variable $X$ in the information structure $(S,M)^C=\cI(C)$ obtained from 
a binary code through the functor $\cI$ of Lemma~\ref{CodesIS}, for a particular choice
of the probability functor $\cQ$ and of the random variable.


We show here that in a similar way, for a particular
choice of the probability functor $\cQ$ and the random variable $X$,
the simplicial set given by the clique complex
$K(G)$ of the network $G$ can be recovered from the construction of $\cQ_X$
for $X$ in $(S,M)^{\bL(\tau_{G})}$, where $\tau_G=\Upsilon(\Phi)(G)$, for some
$\Phi\in \Sigma_{\cC'}(G)$ and $\Upsilon(\Phi)\in\Sigma_{\cC'}^{\operatorname{prop}}(G)$ obtained
by grafting as in Proposition~\ref{compfunctorprop}
of \S \ref{CompArchSec}. This shows
that the homotopy types obtained from the simplicial sets $\cQ_X$ encompass
both the usual homotopy types $\cN(\cU(C))$ detecting the nontrivial topological information
carried by the receptive fields of neural codes and also the homotopy
types $K(G)$ that detect the amount of non-trivial topology present in the activated
network.
 
 
 \begin{prop}\label{KGfromQX}
 Consider the composition $\cJ \circ \Upsilon$ of the functors $\cJ$ of Lemma~\ref{funcCIS}
 with the functor $\Upsilon$ of Proposition~\ref{compfunctorprop}. There is a choice of a
 probability functor $\cQ$ and of a random variable $X$ in the finite information
 structure $(S,M)^{\tau_G}=\cJ \circ \Upsilon(\Phi)(G)$ such that the resulting simplicial set
 $\cQ_X$ is the (directed) clique complex $K(G)$ of the network $G$ (see \S \ref{GammaCliqueSec}).
 \end{prop}

\begin{proof}
  For a network $G$ and a summing functor $\Phi\in \Sigma_{\cC'}(V_G)$, the functor $\cJ \circ \Upsilon$
 determines a finite information structure $(S,M)^{\tau_G}$ whose objects are
 the random variables $X: \bL(\tau_G)\to \R$ with finite range, where $\tau_G$
 is the transition system in $\cC'$ (or $\cC^t$) obtained 
 through the grafting procedure described in \S \ref{CompArchSec} applied to the systems $\tau_v=\Phi(v)$. 
 For simplicity we consider the case where $G$ itself is acyclic.
 The case for more general directed $G$ is treated as in \S \ref{CompArchSec} by
 considering strong connected components $G_i$ and the condensation acyclic graph $\bar G$
 and is similar. For $\omega$ a topological ordering of the vertices of $G$ as described 
 in \S \ref{CompArchSec}, we can write any word in the automaton language $\bL(\tau_{G})$
 as a sequence  
\begin{equation}\label{wes}
w_{i_0} e_{i_0} \cdots w_{i_{k-1}} e_{i_{k-1}} w_{i_k}, \ \ \  \text{ for some } k\in \N, \ \ \ 
\text{ with } \ w_\ell \in \tau_{v_\ell}=\Phi(v_\ell)
\end{equation} 
and with the $e_{i_r}$ given by edges in $G$, 
with vertices along the path satisfying $v_i \leq v_j$ in the order $\omega$ for $i\leq j$. 
In other words, a sequence of transitions in the automaton $\tau_{G}$ consists of a
sequence that alternates transitions in $G$ (along the directed edges of a path in $G$) 
and transitions inside the automata $\tau_v$ associated to the vertices along the path.  
For $\sigma=\{ v_{i_0}, \ldots, v_{i_k} \}$ we write $\sigma \in {\rm supp}(X)$ if $X$
 takes non-zero values on all the words \eqref{wes}. We write $X|_\sigma$ for the
 restriction of the random variable $X$ to the set of words of the form \eqref{wes} for
 the ordered sequence of vertices in $\sigma$. We consider a probability
 functor $\cQ$ given by $\cQ_X=\cup_{\sigma \in {\rm supp}(X)} \Delta_{M_{X|_\sigma}}$.
 We restrict then to those random variables $X: \bL(\tau_{G}) \to \R$ that are supported 
 on the subset of words in $\bL(\tau_{G})$ of the form \eqref{wes}, where the
 set $\sigma=\{ v_{i_0}, \ldots, v_{i_k} \}$ of vertices is a $k$-clique of $K(G)$. We write $\Delta_\sigma$
 for the $k$-simplex in $K(G)$ that corresponds to this clique.
 Note that by construction $\Delta_\sigma$ is in fact a directed clique in the ordering $\omega$. 
 We further consider, among these random variables, an $X: \bL(\tau_{G}) \to \R$ such
 that $X$ takes on exactly $k+1$ different non-zero values on each 
 set of words \eqref{wes} for each $k$-clique $\sigma$. For such a random variable, we then
 obtain that $\cQ_X=\cup_\sigma \Delta_{M_{X|_\sigma}}=K(G)$ 
 is the (directed) clique complex of $G$.
\end{proof}

\section{Categorical Hopfield dynamics}\label{HopfieldSec}

The setting we described in the previous sections for modeling neural information
networks, based on categories of network summing functors and symmetric monoidal categories 
of systems and resources, should be regarded as a static setting, like the kinematic description of a physical
system, the overall configuration space, while we did not yet introduce an adequate modeling of dynamics.
This is the topic we discuss in this section. Our model is based on the traditional way
of describing dynamics of networks in terms Hopfield networks, where nodes have a dynamics
governed by excitatory or inhibitory synaptic connections with certain thresholds. 


It is important to note that the threshold-linear dynamics of Hopfield networks, which
is what we formulate here in our categorical setting, is a {\em non-linear} model of
the neuron, unlike the linear model we discussed briefly in \S \ref{LinNeurSec}. 


Formulating a Hopfield network type of dynamics directly in the setting
of categories of summing functors makes it possible to simultaneously include in the
dynamics all the different levels of structures we have been analyzing in the previous
sections, with their functorial relations: the network together with its associated codes and
weights, the associated computational systems, the associated resources and constraints,
both metabolic and informational. All of the structure evolves then according to an overall
dynamics that functions in functorially related ways at the various different levels.

\subsection{Continuous and discrete Hopfield dynamics}\label{ContDiscrHopfSec}

Typically, the Hopfield network models are either formulated in a discrete form
with binary neurons and the dynamics in the form
\begin{equation}\label{HopfDisc}
\nu_j(n+1) =\left\{ \begin{array}{ll} 1 & \text{if } \sum_k T_{jk} \nu_k(n) + \eta_j >0 \\[2mm] 0 & \text{otherwise,}
\end{array}\right. 
\end{equation}
or in the continuum form with neurons firing rates as variables and a threshold-linear dynamics
of the form
\begin{equation}\label{HopfCont}
\frac{dx_j}{dt} = - x_j + \left( \sum_k W_{jk} x_k  + \theta_j \right)_+
\end{equation}
where $W_{jk}$ are the real-valued connection strengths, $\theta_j$ are constant external inputs, 
and 
\begin{equation}\label{maxthreshold}
(\cdot)_+=\max\{ 0, \cdot \}
\end{equation}
is the threshold function that introduces the non-linearities in the equation. 
For a detailed analysis of the
dynamics of the continuum Hopfield networks see \cite{CuGeMo}, \cite{CuLaMo}, \cite{MoCurto}. 


Here we consider a version of the Hopfield networks dynamics that can be formulated in
a categorical setting and that can be applied to the setting of categories of network
summing functors that we described in the previous sections.

\subsection{Categorical threshold non-linearity}\label{CatNonlinSec}

A main step in constructing the categorical version of the Hopfield dynamics is to have an appropriate way
of describing the non-linearities through a threshold function. We do this using the measuring monoids
$(R,+,\succeq,0)$ associated to the symmetric monoidal categories of resources $(\cR,\otimes,{\mathbb I})$,
as recalled in \S \ref{MeasSemigrSec}.


We assume here that $\cC$ is a symmetric monoidal category, which we write additively with $\oplus$ and $0$,
in order to maintain in the following the similarity of notation with the classical threshold function \eqref{maxthreshold}.
Let $\rho: \cC \to \cR$ be a monoidal functor from the category $\cC$ to a symmetric monoidal category
of resources, as discussed in the previous section and let $(R, +, \succeq,0)$ be the preordered
monoid associated to the category $\cR$. 


One could just assume here, for simplicity, that $\cR=\cC$. We allow for 
another $\cR$ to express the possibility that the threshold in $\cC$ is measured
with respect to another type of resources $\cR$ that is related to $\cC$ functorially.
For example, we may be interested in viewing the dynamics at the level of a 
category of codes, with a threshold measured in terms of information associated
to codes functorially as in \S \ref{GammaCodesSec}. 


\begin{prop}\label{thresholdProp}
Let $\cC$ and $\cR$ be unital symmetric monoidal categories with a monoidal functor $\rho: \cC \to \cR$ as above. 
Let $\hat\cC$ denote the category with the same objects as $\cC$ and with morphisms the invertible morphisms of $\cC$. 
There is a threshold endofunctor $(\cdot )_+ : \hat\cC \to \hat\cC$ that
acts on objects as
\begin{equation}\label{Cplus}
 ( C )_+ =\left\{ \begin{array}{ll} C & \text{if }\, [\rho(C)]\succeq 0 \,\, \text{ in } \, (R, +, \succeq,0) \\[2mm]
0 & \text{otherwise.} \end{array} \right. 
\end{equation}
Composition with this threshold endofunctor induces an endofunctor of the categories of summing
functors $\Sigma_\cC(X)$, for finite pointed sets $X$.
\end{prop}

\begin{proof}
  The class $[\rho(C)]$ in $R$ only depends on the isomorphism class $[C]$, as the functor
$\rho: \cC \to \cR$ induces a corresponding semigroup homomorphism. Thus, if $\phi: C\to C'$ is an
isomorphism, the image $(\phi)_+$ is either $\phi$ itself if $[\rho(C)]=[\rho(C')]\succeq 0$ or the identity
morphism ${\id}_0$ otherwise. This determines $(\cdot )_+$ as an endofunctor of $\hat\cC$. 
Note that $(\cdot )_+$ is in general not an endofunctor of $\cC$, and also that $(\cdot )_+$ need not
be a monoidal functor. Suppose given a summing functor $\Phi \in \Sigma_\cC(X)$. Since we are
working here in the setting of unital symmetric monoidal categories, rather than categories with
sums and zero object, we define $\Sigma_\cC(X)$ as in Definition~\ref{SumPhiSymmMon}. Thus,
$\Phi$ is defined by the collection of objects $\{ \Phi(x) \}_{ x\in X\smallsetminus \{ * \}}$ 
of $\cC$. Thus, we can assign to $\Phi$ a new summing functor $( \Phi )_+$ which is determined by 
the values $( \Phi(x) )_+$ in $\cC$ for $x\in X$. 
Morphisms $\phi: \Phi \to \Psi$ in the category of summing functors are a collection 
$\phi_x: \Phi(x)\to \Psi(x)$ of isomorphisms in $\cC$. Composing with the
threshold endofunctor $(\cdot )_+$ of $\hat\cC$ then gives the corresponding isomorphisms 
$(\phi_x)_+: (\Phi(x))_+\to (\Psi(x))_+$,
hence the corresponding invertible natural transformation $(\phi)_+:  (\Phi)_+ \to (\Psi)_+$. 
\end{proof}


In the case where $\cC$ is a commutative monoidal category, the argument above can be
adapted to the other possible definition of summing functors, as in Definition~\ref{SumFunctCatDef}
and Corollary~\ref{PhiPtsCor}.

\subsection{Discrete Hopfield dynamics in categories of summing functors}\label{DiscrHopfSec}

As above, let $\cC$ be a symmetric monoidal category, written additively with $\oplus$ and $0$,
and let $\rho: \cC \to \cR$ be a monoidal functor from the category $\cC$ to a symmetric monoidal category
of resources, as discussed in the previous section and let $(R, +, \succeq)$ be the preordered
semigroup associated to the category $\cR$. For simplicity of notation, we will write $r_C \in R$ for
the class $[\rho(C)]$ used in our definition of the threshold functor \eqref{Cplus}.


Let $\Sigma_\cC(X)$ be the category of summing functors. For a directed graph $G$, we focus
here on the subcategory of the category $\Sigma_\cC(G)$ of network summing functors given by
the equalizer $\Sigma_\cC^{\operatorname{eq}}(G)$ of the source and target functors 
$s,t: \Sigma_\cC(E_{G^*})  \rightrightarrows   \Sigma_\cC(V_{G^*})$.  In principle, the construction
we present below can be adapted, mutatis mutandis, to other subcategories of network summing
functors, but we focus here on discussing only one case. For simplicity of
notation we just write $\Sigma_\cC(E)$ and $\Sigma_\cC(V)$ for these two
categories of summing functors. We can then define a dynamical system with threshold-dynamics on 
$\Sigma_\cC^{\operatorname{eq}}(G)$ in the following way. 


Let $\cE(\cC)={\rm Func}(\cC,\cC)$ be the category of monoidal endofunctors of $\cC$,
with morphisms given by natural transformations. The sum of
endofunctors is defined pointwise by $(F \oplus F')(C)=F(C)\oplus F'(C)$ for
all $C\in {\rm Obj}(\cC)$.  


Assume given a graph $G$ and $E=E_{G^*}$ as above.
Let $\cP(E)\times \cP(E)$ be the product category with objects given by
pairs of objects $(A,B)$ with pointed subsets $A\subset E$ and $B\subset E$
and morphisms given by pairs of inclusions $A\hookrightarrow A'$ and
$B\hookrightarrow B'$. 


Let $T: \cP(E)\times \cP(E) \to \cE(\cC)$ be a functor satisfying the summing properties
$T_{A\cup A',B}=T_{A,B} \oplus T_{A',B}$,
for $A\cap A'=\{ e_* \}$ in $E_{G^*}$ and for all $B\in \cP(E)$, and
$T_{A,B\cup B'}=T_{A,B} \oplus T_{A,B'}$ for $B\cap B'=\{ e_* \}$ and for all $A$. 
In particular, we write $T_{ee'}$ for the case where $A=\{ e, e_* \}$ and $B=\{ e', e_* \}$.
By the same argument as in Lemma~\ref{PhiPtsLem}, the endofunctors $T_{ee'}$ completely 
determine $T: \cP(E)\times \cP(E) \to \cE(\cC)$ because of the summing properties. 


Let $\Sigma^{(2)}_{\cE(\cC)}(E)$ denote the category of functors $T: \cP(E)\times \cP(E) \to \cE(\cC)$
with the summing properties as above, with morphisms given by the invertible natural transformations.
Similarly, we define $\Sigma^{(2)}_{\cE(\cC)}(V)$ for $V=V_{G^*}$ with source and target
functors $s,t: \Sigma^{(2)}_{\cE(\cC)}(E) \rightrightarrows \Sigma^{(2)}_{\cE(\cC)}(V)$ given by
$T^s_{A,B}=T_{s^{-1}(A),s^{-1}(B)}$ and $T^t_{A,B}=T_{t^{-1}(A),t^{-1}(B)}$, for $A,B \in \cP(E)$.
Let $\Sigma^{(2)}_{\cE(\cC)}(G)$ denote the equalizer of the functors $$s,t: \Sigma^{(2)}_{\cE(\cC)}(E) \rightrightarrows \Sigma^{(2)}_{\cE(\cC)}(V).$$


\begin{defn}\label{CatHopfDef}
Let $\Phi_0 \in \Sigma_\cC^{\operatorname{eq}}(G)$ be an initial choice
of a summing functor $\Phi_0: \cP(E) \to \cC$ with conservation law at vertices. We write
$X_e(0):=\Phi_0(e)$ where $\Phi_0(e)$ stands for the object in $\cC$ that is the 
image under $\Phi_0$ of the pointed subset $\{ e, * \}$ of $E_{G^*}$. The
choice of a functor $T: \cP(E)\times \cP(E) \to \cE(\cC)$ as above, together with
the initial $\Phi_0 \in \Sigma_\cC^{\operatorname{eq}}(G)$ determine a dynamical system
\begin{equation}\label{CatHopf}
X_e(n+1)= X_e(n)\oplus \left( \oplus_{e'\in E} T_{ee'}(X_{e'}(n)) \oplus \Theta_e \right)_+
\end{equation}
where $\Theta_e=\Psi(e)$ are the values at $\{ e, * \}$ of a fixed summing functor 
$\Psi\in \Sigma_\cC^{\operatorname{eq}}(G)$ and with $(\cdot )_+$ the threshold functor of Proposition~\ref{thresholdProp}. 
\end{defn}


\begin{lem}\label{lemPhin}
For $T: \cP(E)\times \cP(E) \to \cE(\cC)$ in the equalizer $\Sigma^{(2)}_{\cE(\cC)}(G)$ and 
$\Phi_0: \cP(E) \to \cC$ and $\Psi: \cP(E) \to \cC$ in the equalizer $\Sigma_\cC^{\operatorname{eq}}(G)$, 
the dynamics \eqref{CatHopf} defines a sequence $\Phi_n$ of summing functors in $\Sigma_\cC^{\operatorname{eq}}(G)$.
\end{lem}

\begin{proof}
  If $r_{X_e(n)}$ never satisfies the threshold condition then the dynamics is
trivial and just gives the constant $\Phi_0$ functor. Assuming a non-trivial dynamics,
the right-hand side of \eqref{CatHopf} defines the values $\Phi_{n+1}(e)$ at the subsets 
$\{ e, * \}$ of $E_{G^*}$ of the new functor $\Phi_{n+1}$. Indeed, we have shown in
Proposition~\ref{thresholdProp} that the threshold functor is an endofunctor of the
category of summing functors, so the right-hand side determines a unique summing functor,
with values completely specified by the $\Phi_{n+1}(e)$, through the summing property
$\Phi_{n+1}(A) =\oplus_{e\in A} \Phi_{n+1}(e)$. We need to check that the resulting
$\Phi_{n+1}$ still satisfies the conservation law at vertices, so that it defines a summing
functor in the equalizer $\Sigma_\cC^{\operatorname{eq}}(G)$. We have 
$$ \bigoplus_{s(e)=v}  \Phi_{n+1}(e) = \bigoplus_{s(e)=v} \Phi_n(e) \oplus \bigoplus_{e'\in E} 
\bigoplus_{s(e)=v} T_{ee'} (\Phi_n(e')) \oplus \bigoplus_{s(e)=v} \Psi(e), $$
The first and last term on the right-hand side are respectively equal to 
$\oplus_{t(e)=v} \Phi_n(e)$ and $\oplus_{t(e)=v} \Psi(e)$. 
Since  $T: \cP(E)\times \cP(E) \to \cE(\cC)$ is in the equalizer $\Sigma^{(2)}_{\cE(\cC)}(G)$, 
the endofunctors $T_{ee'}$ of $\cC$ satisfy 
$\oplus_{s(e)=v} T_{ee'}(C)= \oplus_{t(e)=v} T_{ee'}(C)$ for all $C\in {\rm Obj}(\cC)$, hence
also the second term in the above sum is equal to
$\oplus_{e'\in E}  \oplus_{t(e)=v} T_{ee'} (\Phi_n(e'))$, hence we obtain that 
$$ \bigoplus_{s(e)=v}  \Phi_{n+1}(e) =  \bigoplus_{t(e)=v}  \Phi_{n+1}(e) $$
which implies that $\Phi_{n+1}$ is in the equalizer $\Sigma_\cC^{\operatorname{eq}}(G)$.
\end{proof}


We should think of the equation \eqref{CatHopf}  as the categorical 
version of a finite-difference
form of the Hopfield network equations
\begin{equation}\label{HopfDiff1}
 \frac{x_j(t+\Delta t)-x_j(t)}{\Delta t} = (\sum_k T_{jk} x_k(t) +\theta_j)_+\, , 
\end{equation}
where for simplicity we can assume discretized time intervals $\Delta t =1$.
Usually, in the Hopfield network dynamics, one introduces an additional ``leak term'' $-x_j(t)$ 
on the right-hand side of the equation, to ensure that a neuron firing rate would decay
exponentially to zero if the threshold term is zero, so that the corresponding difference
equation would look like
\begin{equation}\label{HopfDiff2}
 \frac{x_j(t+\Delta t)-x_j(t)}{\Delta t} =-x_j(t) + (\sum_k T_{jk} x_k(t) +\theta_j)_+\, . 
\end{equation} 
An analog of equation \eqref{HopfDiff1} (with $\Delta t =1$) in the categorical setting would be of
the form 
\begin{equation}\label{CatHopf2}
X_e(n+1)= \left( \oplus_{e'\in E} T_{ee'}(X_{e'}(n)) \oplus \Theta_e \right)_+ \,
\end{equation}
for which again the result of Lemma~\ref{lemPhin} holds. A categorical analog
of \eqref{HopfDiff2} for $\Delta t << 1$ can be formulated as 
\begin{equation}\label{CatHopf3}
X_e(n+1) \oplus X_e(n)= \left( \oplus_{e'\in E} T_{ee'}(X_{e'}(n)) \oplus \Theta_e \right)_+ \, .
\end{equation}
In this case however, one cannot directly apply the argument of Lemma~\ref{lemPhin}
anymore. One can still
seek solutions of \eqref{CatHopf3} where all the $\Phi_n$ are in $\Sigma_\cC^{\operatorname{eq}}(G)$,
if they exist. 


Note that, in the case where the symmetric monoidal category $\cC$ of resources 
has a zero object, one does have projection maps 
$X_e(n+1) \oplus X_e(n) \to X_e(n+1)$ and $X_e(n+1) \oplus X_e(n) \to X_e(n)$,  
obtained by applying the unique morphism $X_e(n) \to 0$ and $X_e(n+1) \to 0$. 
However, this does not
suffice to extend the argument of Lemma~\ref{lemPhin} to this case.  Moreover,
if $0$ is a zero object, then the threshold-nonlinearity becomes trivial and the equation reduces  
to a linear dynamics. 


For the purpose of this discussion, we will only consider the equation of the
form \eqref{CatHopf2}, where we can incorporate a diagonal term 
so as to include the case \eqref{CatHopf}, so that Lemma~\ref{lemPhin} applies.


\begin{lem}\label{lemNerveDyn}
The categorical Hopfield network dynamics \eqref{CatHopf}, \eqref{CatHopf2} 
induces a discrete dynamical system $\tau$ on the simplicial set given by the nerve
$\cN( \Sigma_\cC^{\operatorname{eq}}(G) )$ and its realization, the classifying space 
$| \cN( \Sigma_\cC^{\operatorname{eq}}(G) ) |=B\Sigma_\cC^{\operatorname{eq}}(G)$. 
\end{lem}

\begin{proof}
  The functoriality of the nerve construction, seen as a functor $\cN: {\rm Cat} \to \Delta$
from the category of small categories to the category of simplicial sets implies that the
endofunctor $\cT$ that assigns to an object $\Phi$ in  $\Sigma_\cC^{\operatorname{eq}}(G)$ the object
$\cT(\Phi)$ determined by the equation \eqref{CatHopf} or \eqref{CatHopf2} induces a
simplicial self-map $\cT_\cN$ of the nerve $\cN(\Sigma_\cC^{\operatorname{eq}}(G))$ and a corresponding self-map $\cT_B$ of
the realization $B \Sigma_\cC^{\operatorname{eq}}(G)$ as a topological space. Thus, the categorical dynamical
system \eqref{CatHopf} or \eqref{CatHopf2} determines a classical discrete dynamical system on
the topological space $B \Sigma_\cC^{\operatorname{eq}}(G)$ given by the orbits under the iterates $\cT_B^n$.
\end{proof}

We will return to comment  more extensively on this topological model of the categories
of summing functors and the dynamics in \S \ref{GammaGeneralSec} below.

\subsection{Category of weighted codes and ordinary Hopfield dynamics}\label{DynWCodesSec}

The goal of the very general categorical form of Hopfield dynamics introduced in
the previous section is to model dynamics of different types of resources associated to
a network. For this reason, we have formulated the equations \eqref{CatHopf}, \eqref{CatHopf2}
in such a way that the dynamical variable is an assignment of resources of type $\cC$ to
a network, that is, a summing functor. This setting is very general in the sense that the
equations allow for an arbitrary choice of an initial assignment $\Phi_0$, a constant term $\Psi$ 
(which is a choice of another summing functor) and an endofunctor $T$ that generates the dynamics. 


Since we want this broad setting to be a generalization of the usual Hopfield equations
on networks, we need to check a basic consistency with the original equations, namely we
need to show that those can be re-obtained as a {\em special case} of the categorical
Hopfield dynamics described above, for a very special choice of the category $\cC$ and
the data of the equation. 


Thus, we now check that, in the case where the category $\cC$ is a version of the category of weighted
codes considered in \S \ref{GammaCodesSec}, with a particular choice of the functor $T$ in the
categorical Hopfield equation, the categorical Hopfield dynamics 
recovers the usual Hopfield network dynamics (in a discretized finite-difference form)
on associated total weights. To this purpose we restrict to the case with only non-negative weights,
which in the resulting Hopfield network dynamics would be interpreted as activity levels.


\begin{defn}\label{Codesplus}
Let $\cW{\rm Codes}_{n,*}^+$ be the category of weighted codes, where we only consider
non-negative weights, that is, objects $(C,\omega)$ have $\omega(c)\geq 0$ for all $c\in C$
and morphisms $\phi=(f,\Lambda): (C,\omega) \to (C',\omega')$ with the
weights satisfying $\sum_{c\,:\,f(c)=c'}\lambda_{c'}(c)\leq 1$, for all $c'\in C'$. 
These conditions are well behaved under composition of morphisms.
\end{defn}


\begin{lem}\label{alphafunct}
The assignment $\alpha(C,\omega)=\sum_{c\in C} \omega(c)$ defines a
functor $$\alpha: \cW{\rm Codes}_{n,*}^+\to \R,$$ where we view $(\R,\leq)$ as a thin category,
compatible with sums.
\end{lem}

\begin{proof}
  For $\phi=(f,\Lambda) : (C,\omega)\to (C',\omega')$ a morphism in $\cW{\rm Codes}_{n,*}^+$ 
we have $\alpha(C,\omega)=\sum_{c\in C} \omega(c)=\sum_{c\in C} \lambda_{f(c)}(c)\, \omega'(f(c))
 \leq \sum_{c'\in C'} \omega'(c')=\alpha(C',\omega')$, hence
$\alpha(\phi)$ is the unique morphism in $(\R,\leq)$ between $\alpha(C,\omega)$ and
$\alpha(C',\omega')$. The functor $\alpha$ maps the sum $(C,\omega)\oplus (C',\omega')=(C\vee C',
\omega\vee \omega')$ to $\alpha(C,\omega)+\alpha(C',\omega')\in \R$, and maps the
object $(\{c_0\},0)$ to $0\in \R$.
\end{proof}


In order to distinguish, in our setting, between inhibitory and excitatory effects in the
Hopfield dynamics \eqref{CatHopf}, \eqref{HopfDiff1} and \eqref{HopfDiff2}, we can consider the possibility
of a term $T: \cP(E)\times \cP(E) \to \cE(\cC)$ in the equation that has values either
in the category $\cE(\cC)$ of endofunctors of $\cC$ (excitatory case) or in the category
$\cE^o(\cC)$ of {\em contravariant} endofunctors, determined by a collection of functors 
$T_{e,e'}: \cC^{\operatorname{op}} \to \cC$ (inhibitory case). More precisely, we can consider the
following setting.


\begin{defn}\label{Tinhiblin}
Let $\cC=\cW{\rm Codes}_{n,*}^+$ and 
let $T: \cP(E)\times \cP(E) \to \cE(\cC)$ and $T^o: \cP(E)\times \cP(E) \to \cE^o(\cC)$ be,
respectively, functors in the equalizers $\Sigma^{(2)}_{\cE(\cC)}(G)$ and $\Sigma^{(2)}_{\cE^o(\cC)}(G)$,
where $\cE(\cC)$ and $\cE^o(\cC)$ are, respectively, the categories of covariant and contravariant endofunctors of $\cC$.
The functors $T$ and $T^o$ are, respectively, linear-excitatory and linear-inhibitory if 
for all $e,e'\in E$ there is a covariant or contravariant endofunctor, $\tau_{ee'}$ and $\tau^o_{ee'}$, respectively,
of the thin category $(\R,\leq)$ such that the diagrams of functors commute
$$ \xymatrix{ \cC \ar[r]^{T_{ee'}} \ar[d]_{\alpha} & \cC \ar[d]^{\alpha} \\
(\R,\leq) \ar[r]^{\tau_{ee'}} & (\R,\leq) } 
\ \ \ \ \ 
\xymatrix{ \cC^{\operatorname{op}} \ar[r]^{T^o_{ee'}} \ar[d]_{\alpha} & \cC \ar[d]^{\alpha} \\
(\R,\geq) \ar[r]^{\tau^o_{ee'}} & (\R,\leq) } 
$$
where $\tau_{ee'}$ and $\tau^o_{ee'}$ act linearly on $\R$, 
$\tau_{ee'}(r)=t_{ee'}\cdot r$ for some $t_{ee'}\in \R^*=\R\smallsetminus \{0\}$ and for all $r\in \R$, and similarly for $\tau^o_{ee'}$ and 
the corresponding $t^o_{ee'}$, where by covariance/contravariance $t_{ee'}>0$ and $t^o_{ee'}<0$.
\end{defn}


We focus on the linear-inhibitory case. The excitatory case is analogous.
Linear-inhibitory functors $T$ also satisfy the following properties.

\begin{lem}\label{Msemihom}
Let $\cC=\cW{\rm Codes}_{n,*}^+$ with $\rho: \cC \to \cR$ a functor to a symmetric monoidal 
category of resources and $(R,+,\succeq, 0)$ the associated monoid. 
Assume that there exists a measuring monoid homomorphism $M: (R,+, \succeq, 0) \to (\R,+,\geq, 0)$
satisfying $M\circ \rho =\alpha: \cC \to \R$, and such that $M(r)\geq 0$ in $\R$ iff $r\succeq 0$ in $R$. 
A linear-inhibitory functor $T^o: \cP(E)\times \cP(E) \to \cE^o(\cC)$ satisfies
\begin{enumerate}
\item By contravariance of $\tau^o_{ee'}$ and linearity, all the multiplicative factors satisfy $t^o_{ee'}<0$.
\item For any object $(C,\omega)\in {\rm Obj}(\cC)$ such that $r_{\rho(C,\omega))}\succeq 0$ in $(R,+,\succeq, 0)$,
and for all $e,e'\in E$, we have $0\succeq r_{\rho(T^o_{ee'}(C,\omega))}$ in $(R,+,\succeq,0)$.
\item The ratio $M r_{\rho(T^o_{ee'}(C,\omega))}/ M r_{\rho(C,\omega))}$ is independent of the object $(C,\omega)$
and equal to $t^o_{ee'}<0$.
\end{enumerate}
\end{lem}

\begin{proof}
  By contravariance of $\tau^o_{ee'}$ we have $\tau^o_{ee'}(r)\geq \tau^o_{ee'}(s)$ when $r\leq s$,
hence if $\tau^o_{ee'}(r)=t^o_{ee'}\cdot r$ is linear, the multiplicative factor satisfies $t^o_{ee'}<0$.
The measuring homomorphism preserves the order relation so $r_{\rho(C,\omega))}\succeq 0$
implies $0\succeq r_{\rho(T^o_{ee'}(C,\omega))}$ since $M r_{\rho(C,\omega))} \geq 0$ implies
$M r_{\rho(T^o_{ee'}(C,\omega))} =\alpha T^o_{ee'}(C,\omega) =\tau^o_{ee'} \alpha (C,\omega) =t^o_{ee'}\cdot M r_{\rho(C,\omega))}\leq 0$. The ratio $$M r_{\rho(T^o_{ee'}(C,\omega))}/ M r_{\rho(C,\omega))} = \alpha T^o_{ee'}(C,\omega)/ \alpha(C,\omega)=t^o_{ee'}$$ is independent of the object $(C,\omega)$.
\end{proof}


\begin{lem}\label{weightsHop}
Let $\rho: \cW{\rm Codes}_{n,*}^+ \to \cR$ be a functor to a symmetric monoidal category
of resources, with $(R,+,\succeq,0)$ the associated semigroup, with a measuring semigroup homomorphism
$M: (R,+, \succeq,0) \to (\R,+,\geq,0)$ as in Lemma~\ref{Msemihom}. Let 
$T^o: \cP(E)\times \cP(E) \to \cE^o(\cC)$ be a linear-inhibitory functor in the equalizer 
$\Sigma^{(2)}_{\cE^o(\cC)}(G)$. Let $\Theta_e$ in \eqref{CatHopf} be such that
$\theta_e=\alpha(\Theta_e) >0$. The Hopfield dynamics
\eqref{CatHopf} on $\Sigma_{\cW{\rm Codes}_{n,*}^+}^{\operatorname{eq}}(G)$ induces the finite differences
Hopfield network equation on the total weights 
\begin{equation}\label{eqwHop}
\alpha_{n+1}(e)=\alpha_n(e)+ \left( \sum_{e'} t^o_{ee'}\, \alpha_n(e') + \theta_e \right)_+ \, ,
\end{equation}
with inhibitory connections $t^o_{ee'}<0$ and with $(x)_+=\max\{ 0, x \}$.
\end{lem}

\begin{proof}
  Given a summing functor $\Phi: \cP(E) \to \cW{\rm Codes}_{n,*}^+$, with
$(C_e,\omega_e)=\Phi(e)$, we define the total weight as a functor
$\alpha_\Phi: \cP(E) \to \R$,  with
$\alpha_\Phi(A)=\sum_{e\in A} \sum_{c\in C_e} \omega_e(c)$, so that 
$\alpha(A\cup A')=\alpha(A)+\alpha(A')$ for
$A\cap A'=\{ e_* \}$ and with $\alpha_\Phi(j: A\hookrightarrow A')$ 
a morphism in $(\R,\leq)$ since $\alpha_\Phi(A)\leq \alpha_\Phi(A')$ under the
assumption that all the weights are non-negative. 
The total weight $\alpha_\Phi: \cP(E) \to \R$ is the composite 
of the functor $\Phi: \cP(E) \to \cW{\rm Codes}_{n,*}^+$ with the functor 
$\alpha: \cW{\rm Codes}_{n,*}^+\to \R$ of Lemma~\ref{alphafunct}.
Similarly, we associate to functors $T^o: \cP(E)\times \cP(E)\to \cE^o(\cW{\rm Codes}_{n,*}^+)$ 
and $\Phi_0 : \cP(E)\to \cW{\rm Codes}_{n,*}^+$ the composites
$\tau =\alpha\circ T^o: \cP(E)\times \cP(E)\to \R$ and $\alpha_0 =\alpha\circ \Phi_0 : \cP(E)\to \R$. 
By applying the functor $\alpha: \cW{\rm Codes}_{n,*}^+\to \R$ to the equation \eqref{CatHopf} 
we then obtain an equation of the form
$$ \alpha_{n+1}(e) = \alpha_n(e)+ \left( \sum_{e'} \alpha(T^o_{ee'}(\Phi_n(e'))) + \theta_e \right)_+, $$
where $\theta_e=\alpha(\Theta_e) >0$. (The positivity of $\alpha(\Theta_e)$ is assumed in
order to have a non-trivial dynamics.)
The hypothesis of linearity of $T$ ensures that $\alpha(T^o_{ee'}(\Phi_n(e')))=\tau^o_{ee'} \alpha(\Phi_n(e'))=
t^o_{ee'} \alpha_n(e')$.
The condition that  $\sum_{e'} T^o_{ee'}(\Phi_n(e')) + \Theta(e) \succeq 0$ in $(R,+\succeq)$ is 
satisfied iff there is a morphism in the monoidal category $\cR$ of resources 
from $\rho(\sum_{e'} T^o_{ee'}(\Phi_n(e')) + \Theta(e))$ to the unit of $\cR$. By the properties
of the measuring semigroup homomorphism $M$ this condition is satisfied iff 
$\alpha(\sum_{e'} T^o_{ee'}(\Phi_n(e')) + \Theta(e))\geq 0$ in $\R$, hence it matches the
condition that $\sum_{e'} t^o_{ee'}\, \alpha_n(e') + \theta_e\geq 0$, so that we obtain the
equation \eqref{eqwHop}.
\end{proof}

\section{Gamma-spaces and Gamma networks} \label{GammaGeneralSec}

In the previous sections we have assigned resources in a category $\cC$ to networks through a category
of network summing functors $\Sigma_\cC(G)$, or some suitable subcategory. As we discussed
in \S \ref{GammaNetCompSec}, these categories of network summing functors are obtained as simple
modifications of the original definition of \cite{Segal} of categories of summing functors $\Sigma_\cC(X)$, 
for $X$ a finite pointed set. We have interpreted such categories $\Sigma_\cC(G)$ of network summing
functors as a configuration space of all possible consistent assignments of resources of type $\cC$ to 
subnetworks of the network $G$. In \S \ref{HopfieldSec} we have also described how to introduce a form
of dynamics on this configuration space, through our categorical formulation of the Hopfield equations. 


As we observed in Lemma~\ref{lemNerveDyn}, this categorical configuration space with the
associated categorical dynamical system has a topological model provided by the nerve of the
category of summing functors, together with the induced dynamics, given by a discrete dynamical
system on a topological space. The latter can then be studied by the usual tools of dynamics on
topological spaces.


Thus, while the category $\Sigma_{\cC}(X)$ of summing functors represents the parameterizing space 
of all consistent assignments of resources to a system and its subsystems, 
the nerve $\cN(\Sigma_{\cC}(X))$ of the category $\Sigma_{\cC}(X)$
organizes the data of these assignments of resources to subsets in a topological structure that
keeps track of all equivalence relations between them, determined by the invertible natural transformations
that are the morphisms of $\Sigma_{\cC}(X)$ and their compositions.   Thus, we view the
topological space $\cN(\Sigma_{\cC}(X))$ as an actual geometric incarnation of our
configuration space $\Sigma_{\cC}(X)$.


Note that the geometric realization $|\cN(\cA)|$ of the nerve of a category $\cA$ is the classifying
space $B\cA$ of the category. This can be described (see \cite{Weiss}) as parameterizing 
sheaves of $\cA$-sets with representable stacks, where an $\cA$-set is a functor from $\cA^{\operatorname{op}}$
to ${\rm Sets}$ and it is representable if it is of the form $F_A: B\mapsto \Hom_\cA(B,A)$.


In the case of categories of summing functors $\Sigma_{\cC}(X)$ for finite pointed sets $X$,
the equivalence of categories between $\Sigma_{\cC}(X)$ and $\hat\cC^n$ with $\# X = n+1$,
shows that the nerves $\cN(\Sigma_{\cC}(X))$, when considered for all possible $X$,
describe topological information about the category $\cC$ in the form of a delooping of the
infinite loop space given by (a completion of) the classifying space $B\cC$, see \cite{Carlsson}
for a detailed discussion of this delooping construction. The point we want to stress here is that
the collection of the nerves $\cN(\Sigma_{\cC}(X))$, for finite pointed sets $X$, only encode
topological information about the category $\cC$. This changes, however, when we consider
network summing functors in $\Sigma_{\cC}(G)$, as these also contain information on the structure of the network and subnetworks.
We will describe in this section the original Segal construction of Gamma-spaces, which
accounts for the collection of the nerves $\cN(\Sigma_{\cC}(X))$ and their relations under
maps of pointed sets, and we will introduce a corresponding notion of Gamma networks that
is based instead on the nerves $\cN(\Sigma_{\cC}(G))$ for finite directed graphs $G$. 

\subsection{Gamma-spaces}\label{GammaSegalSec}

A {\em Gamma-space} (see \cite{Segal}) is a functor $\Gamma: \cF_* \to \Delta_*$ from the category $\cF_*$ 
of finite pointed sets to the category $\Delta_*$ of pointed simplicial sets. 


In the original construction of Segal~\cite{Segal}, the source category 
of Gamma-spaces was
taken to be the category (called $\Gamma^0$ in \cite{Segal} and identified here with $\cF_*^{\rm op}$) where the
objects are finite pointed sets as in $\cF_*$ but with morphisms given by the preimages
under a map of pointed sets. This means that for pointed finite sets $X$ and $Y$ 
a morphism $\phi: Y \to X$ is a collection $\{ S_y \}_{y\in Y}$ of subsets of $X$, 
given by $S_y=f^{-1}(y)$, for a map of pointed sets $f: X\to Y$ (a morphism in $\cF_*$). 
However, we follow here the later use (see for instance the discussion in \S XIV.3 of \cite{Richter}) and
we define Gamma-spaces as functors from the opposite of this category, for which we
use the same notation $\cF_*$ that we used in the previous sections, which is just the category 
of finite pointed sets with base-point-preserving maps. 
Working with this version of Gamma-spaces as {\em covariant} functors
of pointed maps will be more convenient for us. 


It is shown in \cite{Segal} that to any category $\cC$ with a categorical sum and a zero object, one can associate
a Gamma-space $\Gamma_\cC : \cF_* \to \Delta_*$, which assigns to a pointed set $X$ the nerve
$\cN (\Sigma_\cC(X))$ of the category $\Sigma_\cC(X)$ of summing functors. 


Note that with this choice of $\cF_*$ rather than Segal's $\Gamma^0$ as the source 
category of a Gamma-space,
the morphism $\Gamma_\cC(f)$ associated to a map of pointed finite sets $f: X\to Y$ is obtained using
the pushforward map $f_*: \Sigma_\cC(X)\to \Sigma_\cC(Y)$  on summing functors defined by setting
\begin{equation}\label{fPhiB}
 f_* \Phi (B) =\Phi (f^{-1}(B\smallsetminus \{ * \}) \cup \{ * \} ), \ \ \ \text{ for } B\in P(Y), 
\end{equation} 
with $\Phi\in \Sigma_\cC(X)$, so that $f_* \Phi : P(Y)\to \cC$ is a summing functor, see \S XIV.4 of \cite{Richter}.


As we will discuss more
in details in \S \ref{GammaSpectraSec}, it is also shown in \cite{Segal} that a Gamma-space
$\Gamma: \cF_* \to \Delta_*$ extends to an endofunctor $\Gamma: \Delta_* \to \Delta_*$ and the latter
determines an associated spectrum with spaces $X_n=\Gamma (S^n)$ and structure maps 
$S^1\wedge \Gamma(S^n)\to \Gamma(S^{n+1})$. More generally, one can consider categories $\cC$ that are 
unital symmetric monoidal categories. 
It was shown in \cite{Tho95} that the Segal construction of $\Gamma$-spaces,
seen as a functor $\Gamma: \cM\to \bS$ 
from the category $\cM$ of small symmetric monoidal categories
to the category $\bS$ of connective spectra determines an equivalence of categories 
between the localization of the first category, obtained by
inverting those morphisms that are sent to weak homotopy equivalences, and
the stable homotopy category of connective spectra. Additionally, by this result of \cite{Tho95}, all 
connective spectra can be obtained from
Gamma-spaces. Moreover,  the smash product of spectra has a very natural and simple 
description in terms of Gamma-spaces,  as shown in \cite{Lyk99}. 


The $\Gamma$-space construction is functorial. A strict symmetric monoidal functor $\rho: \cC \to \cC'$ of small symmetric
monoidal categories induces a functor $\rho : \Sigma_\cC(X) \to \Sigma_{\cC'}(X)$
between the respective categories of summing functors given by composition
$(\Phi_X: P(X)\to \cC) \mapsto \rho\circ \Phi_X: P(X) \to \cC'$. The fact that $\rho$ is strict
shows the summing property is preserved under $\Phi_X \mapsto \rho\circ \Phi_X$.
This functor in turn determines a natural transformation $\rho: \Gamma_{\cC}\to \Gamma_{\cC'}$ of
the corresponding $\Gamma$-spaces.


The construction of $\Gamma$-spaces $\Gamma_\cC$ was extended from the case of categories
$\cC$ with sums and zero object as in \cite{Segal} to the case of unital symmetric monoidal categories in
\cite{Tho95}, \cite{Tho82}. In this more general setting, $\Gamma_\cC$ is first defined as a pseudo-functor
$\Gamma_\cC: \Gamma^0 \to {\rm Cat}$ that assigns to a finite set $X$ its category of summing functors $\Sigma_\cC(X)$ as
in Definition~\ref{SumPhiSymmMon}. This is a pseudo-functor since 
compatibility with composition of morphisms and identity morphisms is only satisfied up to canonical
isomorphisms, involving the associators, unitors, and braiding, see the Appendix of \cite{Tho82}. One then
obtains an actual functor by applying the Kleisli construction of \cite{Street}.
We will not discuss this case in detail, but we refer the reader to the Appendix of \cite{Tho82} for a
more precise treatment. 


For our purposes we only need to know that the $\Gamma$-space formalism applies to
unital symmetric monoidal categories and that,
for example, a functor  $\rho: \cC \to \cC'$ as above 
from a symmetric monoidal category of
computational architectures to another associated category of resources, 
induces a corresponding natural transformation $\rho: \Gamma_{\cC} \to \Gamma_{\cC'}$ of
the associated $\Gamma$-spaces. Note that here one needs to pay attention to the
distinction between lax monoidal functors and strict monoidal functors as we mentioned
in \S \ref{CatSumFunctSec}, in relation to the setting considered in \cite{Tho82} for the
$2$-category of unital symmetric monoidal categories.

\subsection{Gamma-spaces as endofunctors of simplicial sets}\label{GammaSpectraSec}

The extension of a Gamma-space $\Gamma_\cC: \cF_* \to \Delta_*$ to an
endofunctor $\Gamma_\cC: \Delta_* \to \Delta_*$ is obtained in the following way.


Given a functor $F: \cC^{\operatorname{op}}\times \cC \to \cD$, a {\em cowedge} for $F$ is
a dinatural transformation (a natural transformation for both entries of $F$)
from  $F$ to the constant functor on an object $D\in {\rm Obj}(\cD)$, that is,
a family of morphisms $h_A: F(A,A)\to D$ such that, for all morphisms $f:A\to B$
in $\cC$ one has a commutative diagram
$$ \xymatrix{  F(B,A) \ar[r]^{F(f,A)} \ar[d]_{F(B,f)} & F(A,A) \ar[d]^{h_A} \\
F(B,B) \ar[r]_{h_B} & D } $$


The {\em coend} ${\rm coend}(F)$
is an initial object in the category of {\em cowedges} for $F$, that is, for every
morphism $f:A\to B$ there is a unique arrow ${\rm coend}(F)\to D$ that
gives a commutative diagram
$$ \xymatrix{  F(B,A) \ar[r]^{F(f,A)} \ar[d]_{F(B,f)} & F(A,A) \ar[d]_{\omega_A} \ar[ddr]^{h_A} & \\
F(B,B) \ar[drr]_{h_B} \ar[r]^{\omega_B} & {\rm coend}(\cF) \ar[dr] & \\
& & D } $$
It is customary to use for the coend the notation 
$$ \int^{C\in \cC} F(C,C) := {\rm coend}(F). $$


Let $[n]=\{ 0, \ldots, n\}$ denote the finite pointed set in $\cF_*$ with $n+1$ elements. Given a pointed simplicial
set $K$ with $K_n$ the pointed set of $n$-simplexes of $K$, the extension 
of a Gamma-space
$\Gamma_\cC: \cF_* \to \Delta_*$ to an endofunctor of $\Delta_*$ is given by the coend
\begin{equation}\label{coendGamma}
\Gamma_\cC: K \mapsto \int^{[n]\in \cF_*} K_n \wedge \Gamma_\cC ([n]).
\end{equation}
The smash product $K_n \wedge \Gamma_\cC ([n])$ has the effect of attaching a copy of the
simplicial set $\Gamma_\cC ([n])$ to each element of the set $K_n$, and the coend takes
care of the fact that these attachments are made compatibly with the face and degeneracy maps
of the simplicial set $K$.  By comparison with the geometric realization of the pointed
simplicial set $K$, where one takes the coend
\begin{equation}\label{geomreal}
 | K | = \int^{[n]\in \cF_*} K_n \wedge \Delta_n, 
\end{equation} 
we see that in \eqref{coendGamma} the functor $\Gamma_\cC$ acts on the simplicial set $K$ by replacing
all the $n$-simplexes $\Delta_n$ of $K$ with copies of $\Gamma_\cC ([n])$. 


The spectrum associated to the Gamma-space $\Gamma_\cC: \cF_* \to \Delta_*$
is then the collection of $X_n=\Gamma_\cC(S^n)$ with $S^n$ the $n$-sphere, with the
structure maps $S^1\wedge \Gamma_\cC(S^n) \to \Gamma_\cC(S^{n+1})$.

\subsection{Gamma-spaces and homotopy types}\label{HtpyTypesSec}

By Proposition~4.9 of \cite{BouFrie}, the endofunctor $\Gamma_\cC: \Delta_*\to \Delta_*$
determined by a Gamma-space $\Gamma_\cC$ preserves weak homotopy equivalences,
hence it descends to a map of homotopy types. 
Given a Gamma-space, we associate to any pointed simplicial set a collection of
homotopy types defined as follows.


\begin{defn}\label{htpytypesdef} {\rm 
Let $\Gamma_\cC$ be the Gamma-space associated to a category $\cC$
with $\Gamma_\cC(X)=\cN(\Sigma_\cC(X))$ for a finite pointed set $X$.
Consider its extension $\Gamma_\cC:\Delta_*\to \Delta_*$ to an endofunctor
of pointed simplicial sets as above.
Given a pointed simplicial set $K$, the family of homotopy types associated to $K$ by $\Gamma_\cC$
is the collection of pointed simplicial sets $\{ \Gamma_\cC(\Sigma^n(K)) \}_{n\in \N}$ up to weak
homotopy equivalence, with
$\Sigma^n(K)$ the $n$-fold suspension. We refer to this collection of homotopy types as
the ``representation of $K$ under $\Gamma_\cC$'' and in particular to the homotopy type
$\Gamma_\cC(K)$ as the ``primary representation''.}
\end{defn}


The point of view we have in mind here is to view a Gamma-space $\Gamma_\cC$, seen
as an endofunctor of $\Delta_*$ as a machine that encodes input simplicial sets
(or input homotopy types) into output simplicial sets (output homotopy types) where
the encoding is done via a combination of the input data with data from the category
of resources $\cC$. 


This can be seen more precisely by comparing, as in \S \ref{GammaSpectraSec} above,
the two coend constructions of \eqref{coendGamma} and \eqref{geomreal}. If we 
have an input simplicial set, which we think of in  terms of its realization
$$ | K | = \int^{[n]\in \cF_*} K_n \wedge \Delta_n\, , $$
the Gamma-space $\Gamma_\cC$ transforms it into the simplicial set
$$ \Gamma_\cC(K)= \int^{[n]\in \cF_*} K_n \wedge \Gamma_\cC ([n]), $$
where we have substituted, as basic building blocks, the simplices $\Delta_n$ with
the simplicial sets $\Gamma_\cC ([n])$, which now depend on the category $\cC$. 


The following subsection provides examples of how this encoding of homotopy
types into other homotopy types via a Gamma-space $\Gamma_\cC$ affects their
topological complexity. We will return to interpret this in terms of our model
of neural information networks in \S \ref{HTCodesInfoSec}.

\subsection{Spectra and homotopy types}\label{SpectraHTSec}

We analyze here a few examples of input data $K$ and how these simplicial
data are encoded into the $\Gamma_\cC(K)$ and $\Gamma_\cC(\Sigma^n(K))$
by a Gamma-space $\Gamma_\cC$. The purpose of this choice of examples 
is to illustrate how the encoding by $\Gamma_\cC$ preserves certain properties
of connectedness. Indeed the statements presented in this section can be
regarded as illustrating the principle that the presence of non-trivial homotopy
groups in the output representation detects the presence of non-trivial
homotopy groups in the input, at least within certain ranges. 


The specific examples are chosen so that the input data are certain simplicial sets
associated to networks. The reason for this choice will become more evident in 
\S \ref{GammaNetSec} and \ref{HTCodesInfoSec}.

\subsubsection{Gamma-space representation of clique complexes}\label{GammaCliqueSec}


Suppose given an undirected graph $G$, which we assume has no looping edges and no parallel edges. 
The {\em clique complex} (clique simplicial set) $K(G)$ is the simplicial complex
obtained from $G$ by filling with an $n$-simplex each $n$-clique in $G$, that is,
each subgraph $\Delta_n$ of $G$ that is a complete graph on $n+1$ vertices. 


In the case of a directed graph $G$, one can similarly consider a {\em directed clique complex}
$K(G)$ (as in \cite{Hess}, \cite{Giu}, \cite{MasVil}) where an $n$-simplex is added to an $n$-clique
of the graph $G$ only when the $n$-clique is directed. Here one also assumes no looping edges
and no parallel edges. Parallel edges are anyway collapsed to a single edge in the clique
construction.) Thus, the skeleta are given by $Sk_\ell(K(G))=\cup_{n\leq \ell} K(G)_n$, with the set
of $n$-simplexes given by
$$ K(G)_n=\{ (v_0,\ldots,v_n)\,|\, v_i\in V_G\, \text{ such that } \forall i<j, \, \, \exists e_{ij}\in E_G  \}, $$
where $e_{ij}$ is a directed edge with $s(e_{ij})=v_i$ and $t(e_{ij})=v_j$. 
In particular, a directed $n$-clique $\sigma=(v_0,\ldots,v_n)$ as above is an $n$-clique
(complete graph on $n+1$ vertices) such that there is a single source and a single sink vertex 
and an ordering of the vertices such that if $v_i < v_j$ there is a directed path of edges from 
$v_i$ to $v_j$, see \cite{MasVil}. (The no looping edges condition ensures that the single sink
property holds.) This $K(G)$ is also referred to as the {\em directed flag complex}.

Here and elsewhere in this paper we will consider constructions that give rise
to simplicial complexes, and we will then consider associated simplicial sets. While a
simplicial complex has unordered vertices hence it does not directly define a simplicial set,
which requires an ordering, one can use the nerve of the poset of simplices to obtain, functorially, 
an associated simplicial set, whose geometric realization is homeomorphic to the realization of
the barycentric subdivision of the simplicial complex. We will use here the notation $K(G)$ for both 
the simplicial complex and the simplicial set obtained in this way. 


Under the endofunctor of simplicial sets defined by the Gamma-space, the clique simplicial
sets $K(G)$ associated to directed networks $G$ (or to subnetworks of a fixed network) are mapped to the
simplicial set $\Gamma_\cC(K(G))$ obtained as in \eqref{coendGamma} by gluing to each directed $n$-clique  
$\Delta_n$ of $G$ a copy of the simplicial set $\Gamma_\cC(\Delta_n)$. 


\begin{prop}\label{htpygrps}
Let $\Gamma_\cC$ be the Gamma-space associated to a category $\cC$,
extended to an endofunctor of pointed simplicial sets. Let $K(G)$ be the clique complex
of a directed graph. Suppose that the simplicial set
$K(G)$ is $m$-connected for some $m\geq 0$.
Then its primary representation $\Gamma_\cC(K(G))$ is also $m$-connected.
Moreover, if $X_n=\Gamma_\cC(S^n)$ is the spectrum determined by the Gamma-space,
and $X_n\wedge K(G)$ is $\ell$-connected for some $\ell\leq 2m+n+3$, then
$\Gamma_\cC(\Sigma^n(K(G)))$ is also $\ell$-connected. 
\end{prop}

\begin{proof}
  Let $K$ be a simplicial set and $\Gamma$ be an endofunctor
of simplicial sets given by a Gamma-space. By Corollary~4.10 of
\cite{BouFrie}, if $K$ is $m$-connected for some $m\geq 0$, 
then so is $\Gamma(K)$. Moreover, 
if $K, K'$ are connected simplicial sets, it follows from Proposition~5.21 
of \cite{Lyk99} that if $K'$ is $m$-connected and $K$ is $n$-connected, then the map
$\Gamma(K')\wedge K \to \Gamma(K'\wedge K)$ is $2m+n+3$-connected,
hence it induces an isomorphism on homotopy groups $\pi_i$ with $i< 2m+n+3$
and a surjection on $\pi_{2m+n+3}$.  When applied to $K'=S^n$ with
$\Gamma_\cC(S^n)=X_n$, this gives the second part of the statement.
\end{proof}


It follows from Proposition~\ref{htpygrps} that nontrivial homotopy groups of 
$\Gamma_\cC(\Sigma^n(K(G)))$ imply 
corresponding nontrivial homotopy groups for $X_n$ and $K(G)$.
Note that the converse implication does not hold: nontrivial homotopy groups
of $X_n$ and $K(G)$ do not necessarily imply nontrivial corresponding homotopy
groups of $\Gamma_\cC(\Sigma^n(K(G)))$ under the map
$X_n\wedge K(G) \to \Gamma_\cC(\Sigma^n(K(G)))$ as in Proposition~\ref{htpygrps}.


This shows that enough non-trivial topology is required in the clique
complex $K(G)$ to generate enough non-trivial topology in
the simplicial sets $\Gamma_\cC(K(G))$ and that enough non-trivial
topology in both the clique complex $K(G)$ and the $K$-theory
spectrum of the category $\cC$ are needed to generate enough 
non-trivial topology in the simplicial sets $\Gamma_\cC(\Sigma^n(K(G)))$.
Thus, a sufficiently rich class of homotopy types produced by the Gamma-space
can be obtained as representation of an ``activated subnetwork'' $G'\subset G$  only 
if both $K(G')$ and the spectrum $X_n$ of the Gamma-space have sufficiently
rich homotopy types. (We will return to this interpretation more precisely in 
\S \ref{GammaNetSec} and \ref{HTCodesInfoSec}.)
The existence of such non-trivial homotopy-type representations 
constrains both the topology of the clique complex of the activated network and the 
$K$-theory spectrum of the category $\cC$.

\subsubsection{The case of random graphs}\label{randomgraphsec}

In the case of a non-oriented graph $G$ with no multiple edges and no looping edges, 
we can still define the clique complex
$K(G)$ as the simplicial complex with all complete subgraphs of $G$ as its
simplices, as we noted at the beginning of \S \ref{GammaCliqueSec}. 
Note that topologically the case of directed and non-directed graphs
can behave differently, since the forgetful functor from directed to ordinary graphs
does not preserve homotopy groups. Here it is more convenient to work with
ordinary graphs as we will be using results on random graphs that are proven in
that setting. Again the goal here is to provide a class of examples relevant to
the discussion in \S \ref{GammaNetSec} and \ref{HTCodesInfoSec} below.


A detailed analysis of the topology of clique complexes of random graphs (in the
non-directed sense specified above) is given in \cite{Kahle}. We only refer here 
to the results of \cite{Kahle} that are immediately relevant in our context. 


\begin{prop}\label{ERcase}
Let $\Gamma_\cC$ be the Gamma-space associated to a category $\cC$,
extended to an endofunctor of pointed simplicial sets.
Let $G$ be an Erd\H{o}s--R\'enyi graph $G=G(N,p)$, where $N=\# V(G(N,p))$
and $0<p<1$ is the probability with which edges are independently inserted. 
\begin{enumerate}
\item Let $p=p(N)$ be a function of the form
\begin{equation}\label{ERp}
p=\left( \frac{(2k+1)\log N + \omega(N)}{N} \right)^{1/(2k+1)}
\end{equation}
where $\omega(N)\to \infty$. Then the simplicial set
$\Gamma_\cC(K(G(N,p)))$ is almost always $k$-connected.
\item If $G(N,p)$
is such that $p^{k+1} N \to 0$ but $p^k N \to \infty$, then $\Gamma_\cC(\Sigma^n(K(G(N,p))))$ 
is almost always homotopy equivalent to the space $X_{k+n}$ of the spectrum
of $\Gamma_\cC$. 
\end{enumerate}
\end{prop}

\begin{proof}
  It is shown in Theorem~3.4 of \cite{Kahle} that for $p$ as in \eqref{ERp} with $\omega(N)\to \infty$
then the clique simplicial complex $K(G(N,p))$ is almost always $k$-connected. This means
that the probability that $K(G(N,p))$ is $k$-connected, with $p=p(N)$ as in \eqref{ERp},
goes to $1$ when $N\to \infty$. By Theorem~3.5 of \cite{Kahle}, if $G(N,p)$
is such that $p^{k+1} N \to 0$ but $p^k N \to \infty$, then $K(G(N,p))$
almost always retracts onto a sphere $S^k$, hence $\Gamma_\cC(K(G(N,p)))$
is homotopy equivalent to the space $X_k=\Gamma_\cC(S^k)$ of the spectrum
of $\Gamma_\cC$ and similarly for $\Gamma_\cC(\Sigma^n(K(G(N,p))))\simeq \Gamma_\cC(S^n\wedge S^k)=X_{n+k}$.
\end{proof}


The two cases for random graphs described in Proposition~\ref{ERcase} represent
situations where for sufficiently large probability $p$ the (non-oriented) clique simplicial complex 
$K(G(N,p))$ and its image $\Gamma_\cC(K(G(N,p)))$ have no non-trivial topology
up to level $k$, or the situation where the topology of $\Gamma_\cC(\Sigma^n(K(G(N,p))))$
exactly captures the topology of the $K$-theory spectrum of the category $\cC$ at level $k$.

\subsubsection{Feedforward networks}\label{FeedSec1}

Another explicit case we want to consider, which will be relevant for the
discussion in \S \ref{IntegInfoSec}, is the case of a feedforward network $G$, 
in particular in the form of multilayer perceptrons. 


The topology of directed clique complexes for feedforward networks was
analyzed in \cite{Chowd}. The kind of networks considered in \cite{Chowd} 
are fully-connected feedforward neural networks, that is, multilayer perceptrons. 
The work of \cite{Chowd} also analyzes a different kind of topological invariant,
given by the path homology, but for our purposes it is the clique complex
that is most relevant.


The result of \cite{Chowd} on the case of the directed clique complex
is based on the simple observation that a multilayered perceptron does
not have any ``skip connections'', that is, any edges that connect a node in
a layer at level $i$ to a node in a layer at level $i+j$ with $j\geq 2$. 
In particular, this means that there cannot be any cliques of order $j\geq 2$.
In particular, this means that the topology of the clique simplicial set $K(G)$ is just
the topology of $G$ itself, with possible nontrivial homotopy groups only in
degree zero and one. 


Thus, the case of feedforward networks is essentially trivial from the
point of view of the possible homotopy types  $\Gamma_\cC(\Sigma^n(K(G)))$,
as these depend only on the number of loops of $G$ and on the
$K$-theory spectrum of $\cC$ without any higher-rank contributions from $K(G)$.


The fact that feedforward networks behave poorly in this
respect, in the sense that they do not generate interesting
homotopy types when mapped through a Gamma-space
is interesting. Indeed, it is well known that 
feedforward networks also behave
poorly with respect to measures of informational complexity like integrated
information. The relation to integrated information will be discussed in 
\S \ref{IntegInfoSec}.

\subsection{Gamma networks}\label{GammaNetSec}

In the previous sections we have simply used Gamma-spaces $\Gamma_\cC$, as functors from
finite pointed sets (and, by extension, from pointed simplicial sets) to pointed simplicial sets to discuss how
topological properties of certain types of input simplicial sets arising from networks are
mapped under these functors. However, as discussed at the beginning of this section, the functors
$\Gamma_\cC$ only depend on the target category $\cC$ and the topology of its classifying space $B\cC$.


For our purposes, we need to generalize the notion of Gamma-space so that it also encodes
data from networks. We do this through our previously discussed notion of network
summing functors. 


As in the previous sections, we identify finite directed graphs $G$ with objects in the category of functors
$\cG={\rm Func}({\bf 2}, \cF)$, with $\cF$ the category of finite sets, and pointed finite
directed graphs $G_*$ as objects in $\cG_*={\rm Func}({\bf 2}, \cF_*)$.


\begin{defn}\label{GammaNetDef}
A {\em Gamma network} is a functor 
$$ \cE: {\rm Func}({\bf 2}, \cF_*) \to \Delta_* \, .$$
\end{defn}


As in the case of Gamma-spaces, we can see that categories $\cC$ with sum
and zero object (or more generally unital symmetric monoidal categories) are
a source of Gamma networks. In particular, we focus here on two constructions of
Gamma networks that use the data of a category $\cC$ of resources. The first
construction uses a Gamma-space, together with a functor from graphs to simplicial sets,
while the second construction replaces categories of summing functors with
categories of network summing functors. In the first case (see Lemma~\ref{cliqueGamma})
we first assign to a network its clique complex and then use that as input for a Gamma
space, while in the second (see Lemma~\ref{CGammaNetLem}) one takes the network
directly as input of a Gamma network. An advantage of the latter is that it does not
require first to perform a clique decomposition, which is computationally complicated. 
On the other hand it is preferable to assign resources to cliques, for example
in the setting discussed in \S \ref{CliqueInfoSec}.


\begin{lem}\label{cliqueGamma}
There is a covariant functor $K: \cG \to \Delta$ (or $K: \cG_* \to \Delta_*$ in the pointed case)
that assigns to a graph $G$ its clique simplicial set $K(G)$. Given a category of resources $\cC$ and
the associated Gamma-space $\Gamma_\cC$, seen as an endofunctor
$\Gamma_\cC: \Delta_* \to \Delta_*$, we obtain by precomposition a Gamma network of the form
\begin{equation}\label{KGammanet}
 \cE_\cC^K := \Gamma_\cC\circ K : \cG_* \to \Delta_* \, . 
\end{equation} 
\end{lem}


This Gamma network takes graphs as input. If we work with directed graphs $\cG_*={\rm Func}({\bf 2}, \cF_*)$
then we consider the directed clique complex $K(G)$, while if we consider non-directed graphs then we take
as $K(G)$ the non-directed clique complex. The construction works similarly in both cases. The simplicial
set $\cE_\cC^K(G)$ that we obtain associated to the graph is the coend
$$ \cE_\cC^K(G)= \int^{[n]\in \cF_*} K(G)_n \wedge \Gamma_\cC([n]), $$
namely, as observed in the previous section, it is the simplicial set obtained by gluing in a copy of
$\Gamma_\cC([n])$ at every $n$-simplex of $K(G)$, that is, at every $n$-complete graph in $G$. 
In other words, given a graph $G$, we consider a decomposition of $G$ into cliques. The clique covering
problem for a graph is computationally NP-hard but an optimal partition into cliques can be found in polynomial 
time for graphs with bounded clique-width~\cite{Espela}. If $X\subset V_G$ is a subset of vertices
corresponding to one of the cliques in the decomposition, we consider all possible assignments of
resources of type $\cC$ to the nodes in this clique. This is described by the category of summing
functors $\Sigma_\cC(X)$. The output simplicial set $\cE_\cC^K(G)$ is obtained by considering
the geometric model $\cN(\Sigma_\cC(X))$ of each of these configuration spaces of resource
assignments, and gluing them together according to the way the cliques fit together in the graph $G$
(and the corresponding simplices in the clique complex $K(G)$).


We can then interpret the examples discussed in \S \ref{SpectraHTSec} as describing
how a Gamma network of the form \eqref{KGammanet} encodes an input of the form
$K(G)$ (typically the activated subnetwork of a given network, in response to an
external stimulus) into a new homotopy type $\cE^K_\cC(G)$ that reflects to some
extent the connectivity properties of $K(G)$ but that also reflects the topology of the
category $\cC$ describing the type of resources that the network carries. 


We describe another class of interesting Gamma networks, that also depend
on a category of resources $\cC$. These are obtained in the same way as the
classical Gamma-spaces, but replacing summing functors with network summing functors.


\begin{lem}\label{CGammaNetLem}
Let  $\cC$ be a category of resources and, for
$G\in \cG_*$, let $\Sigma_\cC(G)$ denote the associated category of network summing functors
as in Definition~\ref{SummingNetDef}, with invertible natural transformations as morphisms. 
The assignment 
\begin{equation}\label{GammanetSum}
 G \mapsto \cE_\cC(G)=\cN (\Sigma_\cC(G)) 
\end{equation} 
determines a Gamma network.
\end{lem}

\begin{proof}
  The construction works exactly as the original case of the 
Gamma-spaces
$\Gamma_\cC: \cF_*\to \Delta_*$ recalled in \S \ref{GammaSegalSec}, namely,
given a natural transformation $\alpha: G \to G'$ between functors $G,G'\in {\rm Func}({\bf 2},\cF)$,
we take $\alpha_* \Phi: P(G')\to \cC$, for $\Phi\in \Sigma_\cC(G)$, to be defined as
$\alpha_* \Phi (H) = \Phi(\alpha^{-1}(H))$, for  $H\in P(G')$, 
where $\alpha^{-1}(H): {\bf 2}\to \cF$ is the functor with $V_{\alpha^{-1}(H)}=\alpha_V^{-1}(H)$
and $E_{\alpha^{-1}(H)}=\alpha_E^{-1}(H)$ and source and target morphisms induced by
those of $G$. Note that if we write everything in terms of the associated pointed graph $G_*$,
then $\alpha_* \Phi: P(G'_*)\to \cC$ is defined as in \eqref{fPhiB}. 
\end{proof}


The class of Gamma networks obtained as in Lemma~\ref{CGammaNetLem} model a
somewhat different idea about how networks generate associated homotopy types, with
respect to the construction of Lemma~\ref{cliqueGamma}. In the cases of
Lemma~\ref{cliqueGamma} there is an underlying functorial construction from graphs
to simplicial sets, at the level of {\em input} of the Gamma-space (through the clique complex,
or in principle through other relevant constructions of a similar nature). On the other hand,
in the construction of Lemma~\ref{CGammaNetLem} the input is only the network itself
and the Gamma network $\cE_\cC$ assigns to it the nerve of the category of network
summing functors (or a suitably chosen subcategory). Thus, in the first case a network
is first decomposed into cliques and the configuration space of assignment of resources
is built from the resources associated to the individual cliques through a gluing procedure, while in the
second case there is no a priori decomposition of the network and the resulting
configuration space counts all assignments of resources according to the choice of
the type of network summing functors used. These two examples illustrate possible
different viewpoints that can be used separately or combined (the smash product of
Gamma networks is still a Gamma network as for Gamma-spaces), depending on
the type of model of networks with resources that one wants to consider.  


The way one should interpret this, in terms of the model of networks with
resources that we are describing, is the following. There is an overall network $G$
with an associated configuration space describing all the assignments of
resources of a given type $\cC$ to the network. On this configuration space 
there is a way of describing the dynamics that governs such assignments of
resources. When responding to an external stimulus, only a certain
subnetwork $G'\subset G$ becomes activated. This means that the actual
configuration space involved in describing the response to a given stimulus
is a subset of the overall configuration space, which is determined by the
value on this subnetwork $G'$ of the appropriate Gamma network functor, $\cE_\cC(G')$.
Thus, this is a way to account in a consistent way for a setting where the actual 
network (or part of network) involved varies according to the stimulus. 

\subsubsection{Gamma networks, codes, and nerves of coverings}\label{GammaNetNervesSec}

We present here another example of Gamma networks, of the type described in Lemma~\ref{cliqueGamma},
but with a different functor from networks to simplicial sets, based on associated neural codes and
nerves of coverings. For simplicity we do not explicitly introduce base points.


\begin{defn}\label{DeltaGcat}
Let $\cG:={\rm Func}({\bf 2}, \cF)$ be the category of finite directed graphs and let $\cC$
be a category of resources.
Let $\Delta_{\cG,\cC}$ denote the category with objects given by pairs $(G,\Phi)$ with
$G\in {\rm Obj}(\cG)$ and $\Phi\in \Sigma_\cC(V_G)$. Morphisms $\alpha \in {\rm Mor}_{\Delta_{\cG,\cC}}((G,\Phi),(G',\Phi'))$
are morphisms $\alpha: G\to G'$ in $\cG$ (natural transformations in ${\rm Func}({\bf 2}, \cF)$) 
such that $\Phi'(\alpha_V(v))=\Phi(v)$, with $\alpha_V: V_G \to V_{G'}$ the natural transformation 
$\alpha$ at the object $V\in {\bf 2}$.
\end{defn}


We consider here in particular the case where $\cC={\rm Codes}_n$. We write $\cC'$ for the category
of codes ${\rm Codes}_n'$ discussed in Proposition~\ref{CatCodeMaps}.


By Lemma~\ref{PhiPtsLem} a summing functor $\Phi \in \Sigma_\cC(V_G)$ is completely
determined by the assignment of an object $\Phi(v)$ for each $v\in V_G$. The objects
$\Phi(v)$ are (binary) codes $C_v$ of length $n$. If we think of the set of vertices $V_G$ of the
network as neurons and of codes as neural codes generated by spiking activity of neurons
over a fixed set $T_n$ of $n$ basic time intervals, then we can restrict our attention to the
case where $\Phi(v)$ consists of a single code word $c_v$ with binary entries describing 
whether the neuron $v$ is spiking or not during each time interval in $T_n$. (If we
want to include base points, then we would have two code words $\Phi(v)=\{ c_v, c_0 \}$, 
with $c_0$ the zero word. We will ignore base points to simplify the discussion.)


\begin{defn}\label{elemPhi}
We refer to summing functors $\Phi\in \Sigma_{{\rm Codes}_n}(V_G)$ with the property that
$\Phi(v)=c_v$ consists of a single binary code word of length $n$ as {\em elementary}. 
We write $\Delta'_{\cG,{\rm Codes}_n}$ for the subcategory of $\Delta_{\cG,\cC}$ with
objects $(G,\Phi)$ where the summing functor $\Phi\in \Sigma_{{\rm Codes}_n}(V_G)$ is
elementary.
\end{defn} 


\begin{lem}\label{GPhiCode}
With $\cC={\rm Codes}_n$ and $\cC'={\rm Codes}_n'$ as above, 
there is a functor $C: \Delta'_{\cG,\cC} \to \cC'$ that assigns to
a pair $(G,\Phi)$ with $\Phi$ elementary, the map
$C: V_G\times T_n \to \{ 0, 1 \}$ with $C(v,i)=\Phi(v)_i$, the $i$-th letter
of the binary code word $c_v=\Phi(v)$.
\end{lem}

\begin{proof}
  Consider a morphism $\alpha: (G,\Phi)\to (G',\Phi')$ in $\Delta'_{\cG,\cC}$.
Since we assume both $\Phi$ and $\Phi'$ are elementary, and we have $\Phi=\Phi'\circ\alpha_V$,
we obtain that the morphism $\alpha$ and the induced map $\alpha_V: V_G\to V_{G'}$
give a morphism of the category $\cC'={\rm Codes}_n'$, since $C=C'\circ \alpha_V$. 
\end{proof}


We obtain in this way another example of Gamma network, similar to the
case discussed in Lemma~\ref{cliqueGamma}. This is a more general form
of Gamma networks, where we allow the input
category to be given by $\Delta'_{\cG,\cC}$ instead of just $\cG$, so that
the choice of an elementary $\Phi\in \Sigma_\cC(V_G)$ is assumed here as
part of the input data. The following statement is a direct consequence of
Lemma~\ref{GPhiCode} and Proposition~\ref{CatCodeMaps}.


\begin{prop}\label{NUPhiGamma}
The composite $\cN\cU\circ C$ of the functor $C$ of Lemma~\ref{GPhiCode} and
the functor $\cN\cU: {\rm Codes}_n' \to \Delta$ of Proposition~\ref{CatCodeMaps},
gives a functor $\Xi=\cN\cU\circ C : \Delta'_{\cG,\cC} \to \Delta$. Composition with any Gamma-space
$\Gamma_\cR$, associated to a category of resources $\cR$, determines a Gamma network
$$ \cE^\Xi_\cR= \Gamma_\cR \circ \Xi: \Delta'_{\cG,\cC} \to \Delta\, . $$
\end{prop}


Note that incorporating a choice of a summing functor $\Phi\in \Sigma_\cC(V_G)$ as
part of the input data is consistent with settings such as our Hopfield equations, where
solutions depend on the choice of a summing functor specifying the initial condition
for the evolutionary equation.

\section{Gamma networks and integrated information} \label{IntegInfoSec}

Integrated information was introduced in neuroscience as a measurement
of causal influence structures and informational complexity in neuronal 
networks \cite{BaTon}, \cite{Tono}. In neuroscience, integrated information
was proposed as a possible quantitative measurement of consciousness.
(For a general discussion of this point of view on consciousness, see
\cite{Koch}, \cite{MasTon}.) There are several slightly different versions of 
integrated information: for a comparative analysis, see \cite{MeSeBa}.
We adopt here the geometric version of integrated information developed
in \cite{OizTsuAma}, based on information geometry~\cite{AmaNag}, which we
recall in \S \ref{InfoGeomSec}.


Our main results in this section are the construction of a cohomological form of
integrated information, and using this to show that there is a way to keep track
of the change of integrated information along the orbits of our categorical
Hopfield dynamics, and under composition of a probability functor on
random graphs with a Gamma-space (with the latter seen as an
endofunctor of simplicial sets). We show that composition with a Gamma-space
increases integrated information by an amount describable in terms of
Shannon entropy.

\subsection{Information geometry and integrated information}\label{InfoGeomSec}

The geometric version of integrated information of \cite{OizTsuAma}
is constructed in the following way. Suppose given a stochastic dynamical
system, where the state of the system at (discrete) time $n$ is described
by a set of random variables $\{ X_i=X^{(n)}_i \}_{i=1}^N$ which
correspond to a partition of the system into $N$ subsystems, and the state at time
$n+1$ by a set $\{ Y_i=X^{(n+1)}_i \}_{i=1}^N$. The full system including all
the mutual influences between these two sets of variables, understood in
a statistical sense, is described by a probability distribution $P(X,Y)$.
Integrated information is meant to capture the difference between 
this distribution and an approximation $Q(X,Y)$ where only certain
kinds of mutual influences are retained. These are usually taken to
be the interdependencies between the variables at the same time
and between each $X_i$ and the corresponding $Y_i$, while one
removes the dependencies of the $Y_i$ from the $X_j$ with $j\neq i$.
More precisely this condition of removal of dependencies is 
described by the requirement that the measure $Q(X,Y)$ satisfies,
for all $i=1,\ldots,N$ of the given partition, the condition
\begin{equation}\label{disconnQ}
Q(Y_i |X) = Q(Y_i | X_i)  .
\end{equation}


The discrepancy between $P(X,Y)$ and $Q(X,Y)$ is measured by their
Kullback--Leibler divergence 
\begin{equation}\label{KLPQ}
{\rm KL}(P(X,Y)||Q(X,Y))=\sum_{x,y} P(x,y) \log \frac{P(x,y)}{Q(x,y)},
\end{equation}
where $(x,y)$ varies over the set of values of $(X,Y)$, which we assume
finite here.
 

The best approximation to the full system probability $P(X,Y)$ by a 
measure $Q(X,Y)$ in the class of measures satisfying \eqref{disconnQ}
can be described using information geometry. Given a partition $\lambda$
$$ \{ (X,Y) \}=\sqcup_{i=1}^N \{ (X_i,Y_i) \}  $$
of the random variables $X,Y$,
one considers the space $\Omega_\lambda$ of all probability measures $Q(X,Y)$ that satisfy the
constraint \eqref{disconnQ} for the partition $\lambda$.  For a given $P(X,Y)$, a minimizer 
$Q_\lambda(X,Y)\in \Omega_\lambda$ 
of the Kullback--Leibler divergence \eqref{KLPQ} is obtained via the {\em projection
theorem} of information geometry~\cite{AmaNag}. 


The setting of information geometry that is used for obtaining geometrically the
minimizer probability
\begin{equation}\label{QstarXY}
 Q^*_\lambda(X,Y)= {\rm argmin}_{Q \in \Omega_\lambda}\,\, {\rm KL}(P(X,Y)||Q(X,Y))
\end{equation}
is summarized as follows (see \S 3.2 and \S 3.4 and in particular Theorem~3.8 and Corollary~3.9 of
\cite{AmaNag}).


A {\em divergence function} is a function $D(P||Q)$ on pairs of probability distributions (which
we assume finite here), with the property that the quadratic term $g^{(D)}$ in the expansion
$$ D(P+\xi || P+\eta)\sim \frac{1}{2} \sum_{i,j} g_{ij}^{(D)}(P) \xi^i \eta^j + \text{ higher order terms} $$
is positive definite, that is, a Riemannian metric, and the cubic term 
$$ h^{(D)}_{ijk} = \partial_i g^{(D)}_{jk} + \Gamma^{(D)}_{jk,i} $$
determines a connection $\nabla^{(D)}$ with Christoffel symbols $\Gamma^{(D)}_{ij,k}=\Gamma^{(D)}_{ji,k}$.
Similarly, the dual divergence $D^*(P||Q):=D(Q||P)$ determines the same metric $g^{(D^*)}=g^{(D)}$
and a connection $\nabla^{(D^*)}$ that is dual to $\nabla^{(D)}$ under $g^{(D)}$. The duality condition
for connections $\nabla, \nabla^*$ with respect to a metric $g$ means that, for any triple of vector fields 
$V,W,Z$, one has $Z\, g(X,Y)=g(\nabla_Z X, Y)+ g(X, \nabla^*_Z Y)$. In particular, in \S 3.2 of \cite{AmaNag}
conditions are given under which, for a smooth function $f(x)$, the expression
\begin{equation}\label{Df}
D_f(P||Q)=\sum_i P_i \, f(\frac{Q_i}{P_i})
\end{equation}
defines a divergence, with the associated metric $g^{(D_f)}$ proportional to 
the Fisher--Rao information metric $g_{FR}$ (see Theorem~2.6 of \cite{AmaNag}).  
In particular, for $f(x)=x\log x$ one has $D_f(P||Q)={\rm KL}(Q||P)$
and for $f(x)=-\log(x)$ one has $D_f(P||Q)={\rm KL}(P||Q)$.


Suppose given the triple $(g^{(D_f)}, \nabla^{(D_f)}, \nabla^{(D^*_f)})$ associated to
a divergence $D_f$ as above. One can consider, in the space of probabilities $P$, 
either $\nabla^{(D_f)}$-geodesics or $\nabla^{(D^*_f)}$-geodesics, that is, paths $\gamma(t)$ that are
solutions to the geodesic equation
$$ \ddot{\gamma}(t)^k +\sum_{ij} \Gamma^k_{ij}(\gamma(t))\,\, \dot{\gamma}^i(t) \dot{\gamma}^j(t)=0, $$
with $\Gamma^k_{ij}$ the Christoffel symbols of the corresponding connection. 


An important property of the divergence functions $D(P||Q)$ is the Pythagorean relation (Theorem~3.8 of
\cite{AmaNag}). Namely, if $P,Q,R$ are three probability distributions, consider the $\nabla^{(D)}$-geodesic
from $P$ to $Q$ and the $\nabla^{(D^*)}$-geodesic from $Q$ to $R$. If these two geodesics meet orthogonally
at $Q$, then the divergences satisfy the {\em Pythagorean relation}
\begin{equation}\label{Pytha}
D(P||R)=D(P||Q) + D(Q||R).
\end{equation}
A consequence of this relation is the orthogonal projection theorem of information geometry (Corollary~3.9
of \cite{AmaNag}). Namely, given $P$ and a submanifold $\Omega$ of the space of probabilities, a
point $Q^*\in \Omega$ satisfies
$$ Q^*= {\rm argmin}_{Q\in \Omega} \, D(P||Q) $$
if and only if the $\nabla^{(D)}$-geodesic
from $P$ to $Q^*$ meets $\Omega$ orthogonally at $Q^*$.


Consider the minimizer probability \eqref{QstarXY} obtained as above.
Then the geometric integrated information, for a given partition $\lambda$, is defined as
\begin{equation}\label{IIPlambda}
{\rm II}_\lambda(P(X,Y)):= {\rm KL}(P(X,Y)||Q^*_\lambda(X,Y)) =\min_{Q\in \Omega_\lambda} {\rm KL}(P(X,Y)||Q(X,Y)),
\end{equation}
with a further minimization over the choice of the partition,
\begin{equation}\label{IIP}
{\rm II}(P(X,Y)):= \min_\lambda {\rm KL}(P(X,Y)||Q^*_\lambda(X,Y)) =
\min_{Q\in \cup_\lambda \Omega_\lambda} {\rm KL}(P(X,Y)||Q(X,Y)).
\end{equation}
The partition $\lambda$ realizing the minimum is referred to as the ``minimal information
partition''. Note that this notion is slightly different in other versions of the integrated
information where one minimizes in information measure over partitions with a 
normalization factor that corrects for the asymmetry between the sizes of the pieces
of the partition; see \cite{MeSeBa} for a comparative discussion of these different
versions. 


It is customary to use the letter $\Phi$ to denote integrated information (also referred
to as the $\Phi$-function). However, since in this paper we have been using the letter
$\Phi$ for our summing functors, we will use the notation of \eqref{IIP} for
integrated information.

\subsection{Feedforward networks and integrated information}\label{FeedSec2}

To see an explicit and relevant example of the behavior of integrated information,
consider again the case of a feedforward network with the architecture of a
multilayer perceptron as in \S \ref{FeedSec1}. The fact that feedforward networks
behave poorly with respect to integrated information was discussed in \cite{BaTon},
using a slightly different form of integrated information. We show here that indeed,
with the notion of geometric integrated information of \cite{OizTsuAma} we also
see a similar phenomenon.

\begin{lem}\label{IIMLP}
Let $G$ be a multilayer perceptron. Consider the set $S$ of binary random variables $X: V_G \to \{ 0,1 \}$ 
on the nodes $V_G$, detecting whether a node is activated or not. The network is subject to
a dynamics that updates the state $X(v)$ of a node $v$ through a function 
$$ X_{t+1}(v)=\sigma(X_t(v')\,|\, \exists e\in E_G\, : \, v'=s(e), v=t(e)) $$
of the $X_t(v')$ at all vertices that feed into $v$. Let $P(X_t, X_{t+1})$ be their joint
probability distribution. There is a partition $\lambda$ of $S$, with $X_i=X|_{S_i}$ such 
that the distribution $P(X_t, X_{t+1})$ satisfies $P(X_{t+1,i}| X_t)=P(X_{t+1,i}| X_{t,i})$,
hence the integrated information vanishes, ${\rm II}(P(X_t,X_{t+1}))=0$.
\end{lem}

\begin{proof}
  Consider the input nodes $v_1,\ldots, v_r$ of the multilayer perceptron $G$.
These nodes have outgoing edges to the next layer nodes but no incoming edges 
from inside the system. If the state $X(v_i)$ of the input nodes is assigned at $t=0$,
it remains unchanged during the rest of the time evolution. Thus, we can choose
a partition $\lambda$ of the set $S$ into $2^r$ subsets determined by the possible
values of $X(v_i)$ at the input nodes $v_1,\ldots, v_r$.  All these subsets $S_i$ are
preserves by the time evolution. Thus, the probability $P(X_{t+1,i}| X_t)$ of those
variables $X_{t+1,i}$ in $S_i$ given the state $X_t$ at time $t$ only depends on
$X_{t,i}$ as only these variables have causal influence under the time evolution
on the $X_{t+1,i}$. So we have $P(X_{t+1,i}| X_t)=P(X_{t+1,i}| X_{t,i})$, hence
the probability distribution $P(X_t, X_{t+1})$ already lies in the manifold $\Omega_\lambda$,
hence ${\rm II}_\lambda(P(X_t,X_{t+1}))=0$.
\end{proof}


Note that the source of the vanishing of integrated information for multilayered perceptrons
is different from the source of the vanishing of the topological invariants in \S \ref{FeedSec1}.
Here the fact that ${\rm II}(P(X_t,X_{t+1}))=0$ is caused by the input nodes that do not get
any incoming input from the rest of the system, while in \S \ref{FeedSec1} the vanishing
of the higher $\pi_i(K(G))$ of the clique complex is caused by the lack of skip connections
between layers.

\subsection{Kullback--Leibler divergence and information cohomology}\label{KLintifoSec}

Like the Shannon entropy, the Kullback--Leibler divergence can be interpreted as a $1$-cocycle in 
information cohomology, see \S 3.7 of \cite{Vign2}. 


Just like the Tsallis entropy provides a one-parameter family of entropy functionals 
that recover the Shannon entropy for $\alpha\to 1$, a similar one-parameter
deformation of the Kullback--Leibler divergence can be defined as
\begin{equation}\label{KLalpha}
{\rm KL}_\alpha(P || Q)=\frac{1}{1-\alpha} \sum_i P_i \left( (\frac{P_i}{Q_i})^{1-\alpha} -1 \right).
\end{equation}
This clearly satisfies ${\rm KL}_\alpha(P||Q)\to {\rm KL}(P||Q)=\sum_i P_i \log(\frac{P_i}{Q_i})$ for $\alpha\to 1$.


Consider information
structures $(S,M)$ and $(S',M')$ and a joint random
variable $(X,Y)$ with values in a finite set $M_{XY}\subset M_X \times M'_Y$,
where $X\in {\rm Obj}(S)$ and $Y\in {\rm Obj}(S')$. Also consider a pair of
probability functors $\cQ: (S,M)\times (S',M') \to \Delta$ and $\cQ': (S,M)\times (S',M')\to \Delta$,
where the simplicial sets $\cQ_{(X,Y)}$ and $\cQ'_{(X,Y)}$ are subsimplicial sets
of the full simplex $\Delta_{M_{XY}}$.


Consider then the contravariant functor $\cM^{(2)}(\cQ,\cQ'): (S,M)\times (S',M') \to {\rm Vect}$
that maps $(X,Y)\mapsto \cM^{(2)}(X,Y)$ to the vector space of
real-valued (measurable) functions on the simplicial set of probabilities
$\cQ_{(X,Y)} \times \cQ'_{(X,Y)}$. For $X\in {\rm Obj}(S), Y\in {\rm Obj}(S')$, the semigroup $\cS_{(X,Y)}$
acts on $\cM^{(2)}(X,Y)$ by 
\begin{equation}\label{alphaF2}
((X',Y')\cdot f)(P,Q)=\sum_{(x',y')\in M_{X'Y'}} P(x',y')^\alpha 
Q(x',y')^{1-\alpha} \, f((P,Q)|_{(X',Y')=(x',y')}),
\end{equation}
for $(X',Y')\in \cS_X$ and $(P,Q)\in \cQ_{(X,Y)} \times \cQ'_{(X,Y)}$, and with 
$\{(X',Y')=(x',y')\}=\pi^{-1}(x',y')$ under the surjection $\pi: M_{(X',Y')}\to M_{(X,Y)}$ determined by
the morphism $\pi: (X',Y')\to (X,Y)$ (which exists by the definition of the semigroup $\cS_{(X,Y)}$).
This gives $\cM^{(2)}(\cQ,\cQ')$ a structure of $\cA$-module, which we denote by 
$\cM^{(2)}_\alpha(\cQ,\cQ')$. It is then shown in \S 3.7 of \cite{Vign2} that 
the Kullback--Leibler divergence \eqref{KLalpha} is a $1$-cocycle in the resulting
cochain complex $(C^\bullet(\cM^{(2)}_\alpha(\cQ,\cQ')), \delta)$.

\subsection{Cohomological integrated information}\label{IIHSec}

Consider the setting as in the previous subsection, with $\cQ:(S,M)\times (S',M') \to \Delta$ 
a given probability functor and $\cQ'_\lambda:(S,M)\times (S',M') \to \Delta$ a probability functor
with the property that, for all $(X,Y)$ with $X\in {\rm Obj}(S)$ and $Y\in {\rm Obj}(S')$,
the simplicial set $\cQ'_{\lambda,(X,Y)}$ is contained in the subspace $\Omega_{\lambda,(X,Y)}\subset \Delta_{M_{XY}}$
\begin{equation}\label{MXY}
 \Omega_{\lambda,(X,Y)}=\{ Q(X,Y)\in  \Delta_{M_{XY}} \,|\, Q(Y_i |X) = Q(Y_i | X_i)\,   \} 
\end{equation} 
as in \eqref{disconnQ}, for a partition $\lambda$ of $S=\sqcup_{i=1}^N S_i$ and 
$S'=\sqcup_{i=1}^N S_i'$ so that $X_i \in {\rm Obj}(S_i)$ and $Y_i \in {\rm Obj}(S'_i)$.


Given $P(X,Y)\in \cQ_{(X,Y)}$, let $Q^*_\alpha(X,Y)\in \cQ'_{\lambda,(X,Y)}$ be obtained by taking
\begin{equation}\label{argminQXYlambda}
Q^*_{\alpha,\lambda}(X,Y):= {\rm argmin}_{Q \in \cQ'_{\lambda,(X,Y)}}\,\, {\rm KL}_\alpha(P(X,Y)||Q(X,Y)) \, .
\end{equation}
as in \eqref{QstarXY} and
\begin{equation}\label{argminQXY}
Q^*_{\alpha}(X,Y):= {\rm argmin}_\lambda \, {\rm KL}_\alpha(P(X,Y)||Q^*_{\alpha,\lambda}(X,Y)) \, .
\end{equation}


In the case where $\alpha=1$, the minimizer $Q^*_{1,\lambda}(X,Y)$ can be determined as recalled
above, through the orthogonal projection method of information geometry for the divergence 
$D(P||Q)={\rm KL}(P||Q)$. The case of $\alpha\neq 1$ can also be treated similarly, using a
divergence $D_f(P||Q)$ with $f(x)=\frac{1}{\alpha-1} (x^{\alpha-1}-1)$, with the general formalism
for the information geometry orthogonal projection theorem recalled in \S \ref{InfoGeomSec} above
(see \S 3.4 of \cite{AmaNag}).


The following result is then a direct consequence of the result
of \S 3.7 of \cite{Vign2} recalled in the previous subsection.


\begin{prop}\label{KLcohom}
The minimizer \eqref{argminQXY} determines a probability functor 
$$\cQ^*_\alpha: (S,M)\times (S',M') \to \Delta$$  
$$ \cQ^*_{\alpha,(X,Y)}:=\{ (P,Q^*_\alpha)\in \cQ_{(X,Y)}\times \cQ'_{(X,Y)}\,|\, Q^*_\alpha
={\rm argmin}_{\lambda, Q \in \cQ'_{\lambda,(X,Y)}}\,\, {\rm KL}_\alpha(P||Q) \}, $$
and a contravariant functor $\cM^{(2)}(\cQ,\cQ^*_\alpha):(S,M)\times (S',M') \to {\rm Vect}$
that maps $(X,Y)$ to the vector space of real-valued (measurable) functions
on $\cQ^*_{\alpha,(X,Y)}$. The action \eqref{alphaF2} restricted to $(P,Q^*_\alpha)\in \cQ^*_{\alpha,(X,Y)}$
gives $\cM^{(2)}(\cQ,\cQ^*_\alpha)$ the structure of an 
$\cA$-module $\cM^{(2)}_\alpha(\cQ,\cQ^*_\alpha)$, hence we obtain a
cochain complex $(C^\bullet(\cM^{(2)}_\alpha(\cQ,\cQ^*_\alpha)),\delta)$. 
\end{prop}


\begin{defn}\label{IIcohom}
The cohomological integrated information $${\rm IIH}^*(\cQ):={\rm IIH}^*((S,M)\times (S',M'),\cM^{(2)}_\alpha(\cQ,\cQ^*_\alpha))$$
is the cohomology of the cochain complex $(C^\bullet(\cM^{(2)}_\alpha(\cQ,\cQ^*_\alpha)),\delta)$ obtained
as in Proposition~\ref{KLcohom}.
\end{defn}


In particular, the usual geometric integrated information of \eqref{IIP} is identified 
with an element of the cohomological integrated information, which corresponds
to the $1$-cocycle given by the Kullback--Leibler divergence, in the case $\alpha=1$.
One interprets then the rest of the cohomological integrated information as measures
of the difference between $P(X,Y)$ and its best approximation $Q^*_\alpha(X,Y)\in \cQ^*_{(X,Y)}$
when measured using the higher cocycles. These can be seen as relative versions of the
higher mutual information functionals of cohomological information, in the same way as the
Kullback--Leibler divergence can be seen as a relative version, for a pair of measures, 
of the Shannon entropy.

\subsection{Categorical Hopfield dynamics and integrated information}\label{CatDynIIHSec}

We show here that our formulation of Hopfield dynamics allows for a
way of keeping track of the behavior of integrated information along solutions
of the dynamics, namely of the change in integrated information that occurs
in the subsequent steps of the dynamics.


We consider then again the setting we described in \S \ref{HopfieldSec}. 
For a given network $G$, consider a categorical Hopfield dynamics as in
\eqref{CatHopf} (or \eqref{CatHopf2} or \eqref{CatHopf3}) 
on the category $\Sigma_\cC^{\operatorname{eq}}(G)$, with a given initial condition
$\Phi_0\in \Sigma_\cC^{\operatorname{eq}}(G)$ and with a functor $T\in \Sigma^{(2)}_{\cE(\cC)}(E)$
that determines the dynamics, as in \S \ref{HopfieldSec}. As shown in \S \ref{HopfieldSec},
the assignment $\Phi_n \mapsto \Phi_{n+1}$ given by the dynamics is an endofunctor
of $\Sigma_\cC^{\operatorname{eq}}(G)$.


\begin{prop}\label{IIHopfield}
The Hopfield dynamics \eqref{CatHopf} determines a functor 
$$\cT_n: \Sigma^{\operatorname{eq}}_\cC(G)\to \Sigma^{\operatorname{eq}}_\cC(G)^2$$
mapping the initial condition $\Phi_0$ to the pair of summing functors $(\Phi_n, \Phi_{n+1})$. Let $\cI: \cC  \to\cI\cS$
be a functor compatible with coproducts. Composition with the functor $C^\bullet(\cM^{(2)}_\alpha(\cQ,\cQ^*_\alpha))$
and passing to cohomology determines a functor $${\rm IIH}^\bullet_n: \Sigma^{\operatorname{eq}}_\cC(G) \to 
\Sigma_{{\rm GrVect}}(G)\subset {\rm Func}(P(G),{\rm GrVect})$$ 
\begin{equation}\label{IIHn}
{\rm IIH}^\bullet_n(\Phi_0)={\rm IIH}^\bullet((S,M)^{G'}\times (S',M')^{G'},\cM^{(2)}_\alpha(\cQ,\cQ^*_\alpha))
\end{equation}
that assigns to an initial condition $\Phi_0$ the cohomological integrated information of the
network $G$ in the $n$-th step of the Hopfield evolution.
\end{prop}

\begin{proof}
  The functoriality of the assignment $\Phi_0 \mapsto (\Phi_n, \Phi_{n+1})$ follows from
Lemma~\ref{lemPhin}. We then consider the composition $\cI^2\circ \cT_n$, with
$\cI^2: \cC^2 \to \cI\cS^2$. This is a functor $\cI^2\circ \cT_n: \Sigma_\cC^{\operatorname{eq}}(G)\to {\rm Func}(P(G),\cI\cS^2)$
that maps $\Phi_0$ to the functor $G' \mapsto (S,M)^{G'}_n\times (S,M)_{n+1}^{G'} \in {\rm Obj}(\cI\cS^2)$ where
$(S,M)^{G'}_n=\cI(\Phi_n(G))$ and $(S,M)_{n+1}^{G'}=\cI(\Phi_{n+1}(G'))$.  As in Corollary~\ref{SumCCh2} we 
can then compose with the functor $\cK=C^\bullet(\cM^{(2)}_\alpha(\cQ,\cQ^*_\alpha))$ and obtain a functor
$\cK\circ \cI^2\circ \cT_n: \Sigma_\cC^{\operatorname{eq}}(G)\to {\rm Func}(P(G), {\rm Ch}(\R))$
$$ G' \mapsto (C^\bullet((S,M)^{G'}_n\times (S,M)_{n+1}^{G'},\cM^{(2)}_\alpha(\cQ,\cQ^*_\alpha)),\delta). $$
Further passing to cohomology gives ${\rm IIH}\circ\cK\circ \cI^2\circ \cT_n: \Sigma_\cC^{\operatorname{eq}}(G)\to
{\rm Func}(P(G), {\rm GrVect})$
$$ G' \mapsto {\rm IIH}^\bullet((S,M)^{G'}_n\times (S,M)^{G'}_{n+1},\cM^{(2)}_\alpha(\cQ,\cQ^*_\alpha)). $$
We refer to the functor obtained in this way as ${\rm IIH}^\bullet_n(\Phi_0)$.
\end{proof}


\subsection{Integrated information and Gamma networks} \label{IntegInfoGammaNet}

We now consider how to adapt the formalism of information cohomology to deal with
data of networks. This in particular will provide us with a notion of ``random graphs''
that is more general than the usual models such as the Erd\H{o}s--R\'enyi graphs
discussed in Proposition~\ref{ERcase}, based on finite information structures and
probability functors as in \cite{Vign} (see \S \ref{CohomInfoSec} above).
We proceed as in the case of finite information structures of \cite{Vign}.


\begin{defn}\label{graphinfostr} 
A graph information structure consists of a pair
$(S,M)$ of a thin category $S$, defined as in \cite{Vign}, consisting of random variables $X$ with
morphisms describing a ``coarsening'' relation (see our summary of \cite{Vign} in \S \ref{CohomInfoSec})
and a functor 
$$ M : S \to \cG={\rm Func}({\bf 2}, \cF) \, .$$
Probability functors on graph information structures are functors $\cQ: (S\times {\bf 2}, M) \to \Delta$ that 
assign to a pair of random variables $X_E,X_V$ simplicial sets $\cQ_{X_E}$, $\cQ_{X_V}$ of probabilities over the vertex 
sets $M_{X_E}$, $M_{X_V}$, with source and target morphisms.
\end{defn}


\begin{rem}\label{GXrandom} {\rm 
In the definition above, we view the functor $M$ equivalently as an object 
$$ M \in {\rm Func}({\bf 2}\times S, \cF). $$
To a pair $X_E, X_V$ of random variables in $S$, the functor
$M$ assigns sets given by their ranges $M_{X_E}, M_{X_V}$ endowed with source and target maps
$s,t: M_{X_E} \to M_{X_V}$. 
These data determine a directed random graph $G_X$ with these sets as vertices and edges.
Thus, we can identify each pair $(S\times {\bf 2},M)$ with a category $\cG_{(S,M)}$ of random graphs $G_X
\in {\rm Obj}(\cG_{(S,M)})$.
A probability functor $\cQ$ can be seen as a functor $\cQ: \cG_{(S,M)}\to \Delta$, from a category 
of random graphs to simplicial sets. }
\end{rem}


The same construction above can be adapted to the case where the category of finite sets $\cF$ 
is replaced by pointed finite sets $\cF_*$ and the functors $\cQ$ take values in $\Delta_*$.
Proceeding as in Lemma~\ref{cliqueGamma}, we can then consider Gamma networks obtained in the following way.


\begin{lem}\label{QGammaNet}
Let $\cC$ be a category of resources, with an associated Gamma-space $\Gamma_\cC: \Delta_* \to \Delta_*$.
Given a probability functor $\cQ: \cG_{(S,M)}\to \Delta_*$, we obtain an associated Gamma network
\begin{equation}\label{EQCnet}
\cE_\cC^\cQ = \Gamma_\cC \circ \cQ : \cG_{(S,M)}\to \Delta_* \, \ \ \ \  
\cE_\cC^\cQ(G_X)= \Gamma_\cC(\cQ_{G_X}) \, . 
\end{equation}
\end{lem}


Note that we can view $\cE_\cC^\cQ$ itself as a new probability functor, assigning to $G_X\in \cG_{(S,M)}$
the simplicial set $\Gamma_\cC(\cQ_{G_X})$. Thus, we can view the Gamma-space $\Gamma_\cC$ as
an endofunctor of the category of probability functors $\cQ: \cG_{(S,M)}\to \Delta_*$.


Consider then the case of pairs of random variables $(X,Y)$, as in our discussion of
Kullback--Leibler divergence and integrated information in \S \ref{KLintifoSec} and \S \ref{IIHSec}.

\begin{prop}\label{IIHQ2prop}
The joint distribution of a pair of random variables $(X,Y)$ in $(S\times {\bf 2}, M) \times (S'\times {\bf 2}, M')$
determines a subgraph $G_{(X,Y)}$ of the Kronecker product $G_X\times G_Y$. A probability functor
$\cQ: \cG_{(S,M)\times (S,M')} \to \Delta$ has an associated cohomological integrated information
${\rm IIH}^*(\cQ)$ as in Definition~\ref{IIcohom} that measures 
the amount of information in the associated
simplicial set $\cQ_{(X,Y)}$ of probabilities that is not reducible to a decomposition into independent 
subsystems.
\end{prop}

\begin{proof}
  We consider information structures $(S\times {\bf 2}, M)$ and $(S'\times {\bf 2}, M')$,
which correspond, respectively, to categories of random graphs $\cG_{(S,M)}$ and $\cG_{(S,M')}$.
A pair of {\em independent} random variables $(X,Y)\in (S\times {\bf 2}, M) \times (S'\times {\bf 2}, M')$
will correspond to the Kronecker product of the random graphs $G_X\times G_Y$. For a more general
pair $(X,Y)$, the joint distribution will determine a subgraph $G_{(X,Y)}\subset G_X\times G_Y$.
We consider probability functors $\cQ: (S\times {\bf 2}, M) \times (S'\times {\bf 2}, M') \to \Delta$
that assign to a pair of random variables $(X,Y)$ the simplicial set $\cQ_{G_{(X,Y)}}$, which is
a subsimplicial set in the full simplex $\Delta_{X,Y}$ on the set $V_{G_X}\times V_{G_Y}$. As in \S \ref{IIHSec}, we can
then consider those functors $\cQ'_\lambda: (S\times {\bf 2}, M) \times (S'\times {\bf 2}, M') \to \Delta$
with the property that the simplicial set $\cQ'_{\lambda, G_{(X,Y)}}$ is contained in the subspace
$$ \Omega_{\lambda, (X,Y)}=\{ Q(X,Y)\in \Delta_{X,Y}\,|\, Q(Y_i|X)=Q(Y_i|X_i) \}\, , $$
for a partition $\lambda$ of $S=\sqcup_i S_i$ and $S'=\sqcup_i S_i'$ with $X_i\in {\rm Obj}(S_i)$ and $Y_i\in {\rm Obj}(S'_i)$.
We can then proceed as in \S \ref{IIHSec} and minimize the Kullback--Leibler divergence as in \eqref{argminQXYlambda}
and \eqref{argminQXY}. The resulting minimizer determines a probability functor
\begin{equation}\label{GKLmin}
 \cQ_\alpha^* :   \cG_{(S,M)\times (S,M')} \to \Delta \, ,
 \end{equation}
with respect to which one can compute the cohomological integrated information as
the cohomology of the cochain complex $(C^\bullet(\cM_\alpha^{(2)}(\cQ,\cQ^*_\alpha)),\delta)$. 
One obtains in this way a cohomological integrated information
${\rm IIH}^*(\cQ)$ as in Definition~\ref{IIcohom}. 
Assume that the random variables $X,Y$ describe the
activated subnetwork, in response to an external stimulus, at time $t$ and at time $t+1$. Then the
integrated information ${\rm IIH}^*(\cQ)$ described above captures the amount of information in the associated
simplicial set $\cQ_{(X,Y)}$ of probabilities that is not reducible to a decomposition into independent 
subsystems, in which the variables $Y_i$ of a subsystem at time $t+1$ are only correlated to the 
variables $X_i$ of the same subsystem at time $t$. 
\end{proof}


Consider now probability functors $\cQ: \cG_{(S,M)\times (S,M')} \to \Delta$ as in Proposition~\ref{IIHQ2prop}
and the composition $\cE_\cC^\cQ=\Gamma_\cC\circ \cQ$ with a Gamma-space of a category $\cC$ of
resources as in \ref{QGammaNet}. The chain rule
for the Kullback--Leibler divergence (for $\alpha=1$) then allows us to compare the integrated information of $\cE_\cC^\cQ$
and $\cQ$, hence to measure the effect of $\Gamma_\cC$ on integrated information. 



\begin{prop}\label{IIHEQ}
For the functor $\cE_\cC^\cQ: \cG_{(S,M)\times (S,M')} \to \Delta$, the  
Kullback--Leibler divergence $$KL(P(X,Y)||Q^*(X,Y))$$ for $\alpha=1$, with $P\in (\cE_\cC^\cQ(G_{(X,Y)}))_n$ and
$Q^*\in \cQ^*_{\alpha,G_{(X,Y)}}$, for $\cQ^*: \cG_{(S,M)\times (S,M')} \to \Delta$ the KL-minimizer,
is of the form
$$ KL(P(X,Y)||Q^*(X,Y)) = KL(P'(X,Y)||Q^*(X,Y)) + S(P''), $$
where $S$ is the Shannon entropy, and $P(X,Y)=P'(X,Y)\, P''$ with $P'(X,Y)\in \cQ_{G_{(X,Y)}}$ and
$P''$ is a probability in the simplicial sets $\{ \Gamma_\cC([n]) \}_{n\in \N}$.
\end{prop}

\begin{proof}
The image $\cQ_{G_{(X,Y)}}$ is some simplicial set $K$ with $K_n$ the set of $n$-simplexes in the $n$-th skeleton.
Thus we can write a probability $P(X,Y)\in \cQ_{G_{(X,Y)}}$ as $\{ P_\sigma(X,Y) \}_{\sigma \in K_n}$ with each 
$P_\sigma(X,Y)$ a probability in an $n$-simplex $\sigma$.
With the same notation, using the fact that $\cE_\cC^\cQ(G_{(X,Y)})$ is a simplicial set obtained
as the coend of the $(\cQ_{G_{(X,Y)}})_n \wedge \Gamma_\cC([n])$, we can write a probability 
$P(X,Y)\in \cE_\cC^\cQ(G_{(X,Y)})$ as a collection
$$ \{ P_{\sigma,\tau}(X,Y) \,|\, \sigma\in (\cQ_{G_{(X,Y)}})_n \, , \, \tau\in \Gamma_\cC([n])_m \}\, , $$
with $\Gamma_\cC([n])_m$ the set of $m$-simplexes in the skeleton.
Moreover, since the simplicial sets $\Gamma_\cC([n])$ are independent of the random variables $(X,Y)$, we
can further write these as products of independent probabilities
$$ \{ P'_\sigma(X,Y)\, P''_\tau  \,|\, \sigma\in (\cQ_{G_{(X,Y)}})_n  \, , \, \tau\in \Gamma_\cC([n])_m \}\, . $$
The chain rule for the Kullback--Leibler divergence then gives
$$ KL( P_{\sigma,\tau}(X,Y) || Q_{\sigma,\tau}(X,Y) )=\sum P'_\sigma(X,Y)\, P''_\tau \log Q_{\sigma,\tau}(X,Y) $$ $$ -
\sum P'_\sigma(X,Y)\,  \log P'_\sigma(X,Y) - \sum P''_\tau \log P''_\tau $$
$$ =\sum_\tau P''_\tau \, KL(P'(X,Y) || Q_\tau(X,Y)) + S(P'')\, , $$ 
where $P'(X,Y)=\{ P'_\sigma(X,Y) \}$ and $Q_\tau(X,Y)=\{ Q_{\sigma,\tau}(X,Y) \}$. Convexity
of the Kullback--Leibler divergence gives 
$$\sum_\tau P''_\tau \, KL(P'(X,Y) || Q_\tau(X,Y))\geq KL\left(P'(X,Y) || \sum_\tau P''_\tau Q_\tau(X,Y)\right),$$ 
and the minimizer $Q^*(X,Y)$ of $KL(P'(X,Y)||Q'(X,Y))$ over $Q'(X,Y)\in \Omega_{\lambda, X,Y}$
also minimizes $KL(P'(X,Y)||Q(X,Y))$ with respect to $Q(X,Y)\in \Omega_{\lambda, X,Y}$. 
\end{proof}

We can interpret this result as saying that the integrated informations of $\cE_\cC^\cQ$ and of $\cQ$ differ 
by the Shannon entropy of $\Gamma_\cC$, where the latter is understood as the Shannon entropy
functional from the simplicial sets $\{ \Gamma_\cC([n]) \}_{n\in \N}$ to $\R$ (see the similar discussion of
information-loss functionals on Gamma-spaces in \cite{Mar19}).
 
\subsection{Homotopy types, spectra, information and cohomology}\label{HTCodesInfoSec}

In this final subsection we outline some connections between some of the threads developed in the previous 
parts of the paper. In particular, we return to the theme of homotopy types. We start from the viewpoint that
neural codes generate homotopy types, in the form of the nerve simplicial set of an open
covering associated to a (convex) code, as in \cite{Cu17}, \cite{Man15}, and that activated
subnetworks of a given network also generate homotopy types in the form of the associated clique
complexes. We have discussed in \S \ref{CodesNerveSec} and \S \ref{CliqueInfoSec} how both
of these constructions of simplicial sets can be incorporated into the general framework of
information structures discussed in \S \ref{GammaNetInfoSec}. We have also discussed in \S
\ref{SpectraHTSec} and \S \ref{GammaNetSec}
how Gamma networks, especially those obtained as composition $\Gamma_\cC \circ \cQ$
of a classical Gamma-space $\Gamma_\cC$ with a functor $\cQ: \cG \to \Delta$ from a
category of (random) graphs to simplicial sets, transform these homotopy types into new
homotopy types that incorporate topological structure arising from the category of resources $\cC$. 
This has the effect of combining
the simplicial sets $\cQ_X$ obtained from information structures with those obtained
via the spectra associated to Gamma-spaces, into a single object. For example, when the
input simplicial set is the clique complex of the activated part of the network, or the
nerve complex of a neural code, the output through the Gamma network can be thought of as
a total measure of topological
complexity associated to the system and its subsystems {\em together} with the associated category of resources. 
Thus, non-trivial homotopy types coming from these clique complexes $K(G)$ (or from nerves of covering
complexes) is reflected in the non-trivial topology of their
``representation'' under the Gamma-space associated to the category $\cC$, in the non-trivial
homotopy type of the simplicial sets $\Gamma_\cC(\Sigma^n(K(G)))$, in which the homotopy
structure of $K(G)$ is combined with the homotopy structure of the spectrum determined by the
Gamma-space $\Gamma_\cC$, which reflects the contribution of the additional structure that
the network carries, determined by the category $\cC$ of resources, see Proposition~\ref{htpygrps}.


There are two forms
of (co)homology one can associate to this object, as a measurement of its topological
structure: the information cohomology that we discussed in \S \ref{GammaNetInfoSec} 
and \S \ref{IntegInfoSec} and the generalized cohomology determined by the homotopy-theoretic
spectra discussed in \S \ref{SpectraHTSec}. Again we can consider possible combinations of
these two kinds of (co)homological structures that capture both the informational and the
structural sides of the topology of the system. 


The main property of homotopy-theoretic spectra is the fact that they determine
generalized cohomology theories. Given a spectrum $\bS$, the associated
generalized cohomology theory is defined by
$$ H^k(A,\bS):= \pi_k (\Sigma(A) \wedge \bS), $$
for simplicial sets $A$, with $\Sigma(A)$ the suspension spectrum. 


In \S \ref{SpectraHTSec} we considered the spectra $\Sigma(K(G))\wedge \Gamma_\cC$,
where we write here $\Gamma_\cC$ for the spectrum $X_n=\Gamma_\cC(S^n)$
determined by the Gamma-space, together with the map $\Sigma(K(G))\wedge \Gamma_\cC \to
\Gamma_\cC(\Sigma(K(G)))$ as in Proposition~\ref{htpygrps}. These determine the generalized
cohomology $H^\bullet(K(G),\Gamma_\cC)$.


We have seen in Proposition~\ref{KGfromQX} that the simplicial set $K(G)$ given by the
clique complex of the network $G$ can be realized as a special case of our more general
construction of simplicial sets $\cQ_X$ associated to a probability functor $\cQ$ and 
random variables $X$ in the finite information structure functorially associated to a pair
$(G,\Phi)$ of a network and a summing functor $\Phi\in \Sigma_\cC(V_G)$.


Thus, it is also natural to consider the spectrum $\Sigma(K(G))\wedge \Gamma_\cC$ as
a special case of a sheaf of spectra 
$X \mapsto \Sigma(\cQ_X) \wedge \Gamma_\cC$ and the associated generalized
cohomologies $H^\bullet(\cQ_X, \Gamma_\cC)$.


In \S \ref{GammaNetInfoSec} and \S \ref{IntegInfoSec} we have considered 
information cohomologies $H^\bullet( C^\bullet( \cF_\alpha(\cQ_X), \delta) )$.
We can also extend these by considering the more general information
cohomology groups
$$H^\bullet( C^\bullet( \cF_\alpha(\Sigma^k(\cQ_X) \wedge \Gamma_\cC(S^m)), \delta) ).$$
While information cohomology itself is not a generalized cohomology theory
(it cannot be expected in general to satisfy the Steenrod axioms), one can ask the
question of whether a generalized cohomology theory modeled on the
case of the $H^\bullet(\cQ_X, \Gamma_\cC)$ described above can be constructed
that incorporates information measures such as Shannon entropy, Kullback--Leibler
divergence, integrated information, in the way that the information cohomology does 
(see the discussion in \S \ref{IntegInfoGammaNet}). For some more
interpretation of this encoding of homotopy types via Gamma networks see~\cite{Mar21}.


\appendix  

\section{Probabilistic and persistent Gamma-spaces}\label{ProbPersApp}

Throughout the paper we have worked with ``classical'' Gamma-spaces, namely functors $\Gamma_\cC: \cF_* \to \Delta_*$ from finite
pointed sets to finite simplicial sets, as well as with some generalizations, that we referred to as Gamma-networks. Much of what
has been formulated in those terms can be adapted easily to two further variants of the notion of Gamma-space: 
a {\em probabilistic} version of Gamma-spaces already considered
in \cite{Mar19}, which we recall here in \S \ref{ProbGammaSec}, 
and a {\em persistent} version that we introduce in \S \ref{persGamma}.


We decided to present these two variants separately as an appendix, rather than blending them into the main text,
to maintain clarity of exposition. It should be kept in mind though, that both incorporating probabilistic structures
and introducing filtrations that account for the change of topological structure over time, are important features for
viable applications to neuronal networks. Since adapting the results of the paper to probabilistic and persistent
Gamma-spaces does not present technical obstacles, we will not give a detailed account here, beyond briefly
presenting these two notions in \S \ref{ProbGammaSec} and \ref{persGamma}, and their combined form in
\S \ref{probpersGamma}. In \S \ref{Persist1Sec} and \ref{Persist2Sec} we discuss briefly some motivations
for introducing these variants. We also include in this Appendix a brief account of possible variants of the nerve
construction.


Having a persistent version of Gamma-spaces and spectra is useful when one
needs to keep into account dependence on some scale parameter (or more
generally some parameter in an ordered set, such as time) and keep track of when
the topological structures considered undergo changes with respect to that
parameter (for instance when homotopy and homology groups acquire or lose
new generators). Having a probabilistic version allows for considering probabilistic
superpositions of objects and morphisms in the categories involved, for example
when assignment of resources involves a random rather than a simply a
deterministic choice. 

\subsection{Simplicial topology enriched with probabilities}\label{Simplprob}

As in the earlier sections, we write $\Delta$  for the category of simplicial sets ($\Delta_*$ for pointed simplicial sets),
namely the category of functors $S:\triangle^{\operatorname{op}} \to {\rm Sets}$ from the simplex category $\triangle$
to sets (respectively, pointed sets). Enrichments of simplicial structures with probabilities have
been variously considered, for instance in \cite{Che65}, \cite{Mar19}, \cite{CoMa20}.


In the general setting of \cite{Che64}, \cite{Che65}, \cite{Che78}, \cite{MoChe91}, also used in \cite{CoMa20}, one constructs
a category of probability distributions, where objects are triples $(\Omega,\Sigma,P)$ of a set, a
$\sigma$-algebra, and a probability distribution, and with morphisms given by ``transition measures''. 


More precisely, consider pairs $(\Omega,\Sigma)$ with $\Omega$ a set and $\Sigma\subset \cP(\Omega)$
a collection of subsets satisfying
\begin{itemize}
\item[(a)] $\Omega \in \Sigma$.
\item[(b)] If $X,Y \in \Sigma$, then $X\setminus Y \in \Sigma$.
\item[(c)] The union of all elements of any countable subcollection of $\Sigma$ belongs to $\Sigma$.
\end{itemize}


Let $(S,+,0)$ be a commutative semigroup with zero. An $S$-valued (finitely additive) measure 
on $(\Omega ,\Sigma)$ is a map $\mu: \Sigma\to S$ such that $\mu (\emptyset ) =0$
and $\mu (X\cup Y) + \mu (X\cap Y)= \mu (X) +\mu (Y)$.  


A ($\sigma$-additive) {\it probability distribution} $P$ on $(\Omega, \Sigma)$ is a $(\R_+, +,0)$-valued
measure such that $P (\Omega) =1$ and for any countable subfamily
$\{ X_i \}\subset  \Sigma$ with empty pairwise intersections we have
$P(\cup_{i=1}^{\infty} X_i )= \sum_{i=1}^{\infty} P(X_i)$.


A {\em category of probability distributions} is obtained as follows.
Denote by $Cap (\Omega ,\Sigma)$ the set of probability distributions
on the $\sigma$-algebra $(\Omega, \Sigma)$.
Given two such sets $Cap (\Omega_1 ,\Sigma_1)$ and  $Cap (\Omega_2 ,\Sigma_2)$, call
``a transition measure'' $\Pi$ between them a function $\Pi \{* | \omega \}$
upon $\Sigma_2\times \Omega_1$ such that for any fixed $X\in \Sigma_2$, 
$\Pi \{X | \omega_1\}$ is a $\Sigma_1$-measurable function on $\Omega_1$,
and for any fixed $\omega_1\in \Omega_1$, $\Pi\{X | \omega_1\}$ is a probability
distribution on $\Sigma_2$.
A transition measure $\Pi$ determines a map $Cap (\Omega_1,\Sigma_1)\to Cap (\Omega_2,\Sigma_2)$
given by
$$
\Pi P_1 (X_2) := \int_{\Omega_1} \Pi\{ X_2 |\omega_1\} P_1\{d \omega_1\}.
$$
One can then take morphisms between objects $(\Omega_1,\Sigma_1,P_1)$ and $(\Omega_2,\Sigma_2,P_2)$
to be transition measures $\Pi : Cap (\Omega_1 ,\Sigma_1) \to Cap (\Omega_2 ,\Sigma_2)$ such that
$P_2 = \Pi \, P_1$. Note that this definition of the category of probability distributions differs slightly from
\cite{Che65}, \cite{CoMa20}: it has been adapted for compatibility with the setting of \cite{Mar19}.

\subsubsection{Probability distributions on finite sets}\label{FinProbSec}

If $\Omega$ is a finite set, then the collection
of all subsets $X\subseteq \Omega$ is a $\sigma$-algebra, and  probability distributions
on it are in the bijective correspondence with maps $P:\Omega \to [0,1]$ such that $\sum_{x\in \Omega} P(x) =1$.
The transition measures $\Pi \{* | \omega \}$ are simply stochastic matrices with obvious properties.


Thus, the category described above reduces to the category $\cF\cP$ used in \cite{Mar19}
with objects the pairs $(X,P)$ of a finite set and a probability distribution and morphisms $S: (X,P)\to (Y,Q)$
given by stochastic matrices: $S_{yx}\geq 0$ and $\sum_{y\in Y} S_{yx} =1$ for all $x\in X$,
satisfying $Q=S P$. This category $\cF\cP$ is the undercategory ${\mathbb I}/{\rm FinStoch}$
of the category ${\rm FinStoch}$ of stochastic maps of \cite{BaFr14} (see Remark~2.2 of \cite{Mar19}), 
just as the more general category of probability distributions described above is the 
undercategory of the one of \cite{Che65}.


In other words, such distributions can be considered as probabilistic enrichment of the simplex $\Delta_X$
whose vertices are coordinate points in $\R^X$. If we consider the category of 
{\it pointed finite sets} $(X,x)$, morphisms are $(X,x)\to (Y,y)$
in which $X\to Y$  are maps sending $x$ to $y$. A probabilistic enrichment of such
category based on the transition measures 
$\Pi \{* | \omega \}$ is described as a wreath product of the category of pointed sets with the category
of probability distributions, cf.~\cite{Mar19}, Sec.~2. This is the basic example for a more general
construction of probabilistic categories obtained as wreath products of a category $\cC$ (with sum and zero object)
and the category of finite probability distributions $\cF\cP$.

\subsection{Probabilistic Gamma-spaces}\label{ProbGammaSec}

A version of Gamma-spaces based on the category $\Box_*$ of 
cubical sets with connections rather than the category $\Delta_*$
of simplicial sets was introduced in \cite{Mar19}. 
Versions of Gamma-spaces that incorporate probabilistic data, using the category 
$\cC=\cF\cP={\mathbb I}/{\rm FinStoch}$,  were also introduced in \cite{Mar19}.


In the setting of probabilistic Gamma-space of \cite{Mar19} one constructs
functors $\Gamma_{\cP\cC}: \cP\cF_* \to \cP\Box_*$, where $\cP\cC$ is a
probabilistic category of resources (an explicit example is discussed in \S \ref{ProbTransSec}). 
The category $P(X)$
is replaced by the category $P(\Lambda X)$ with $\Lambda X= \sum_i \lambda_i (X_i,x_i)$ an 
object in the probabilistic category $\cP\cF_*$ of pointed sets. This category $P(\Lambda X)$
has objects the subsystems $\Lambda A =\sum_i \lambda_i (A_i, x_i)$ where $A_i\subseteq X_i$
is a pointed subset and morphisms that are the identity on $\Lambda$ and (deterministic)
pointed inclusions on the sets. A summing functor  $\Phi_{\Lambda X}: P(\Lambda X) \to \cP\cC$
has the form $\Phi_{\Lambda X}(\Lambda A)=\sum_i \lambda_i \Phi_{X_i}(A_i)$ where the
$\Phi_{X_i}: P(X_i) \to \cP\cC$ are summing functors. Thus, when we interpret
an object $\Lambda X$ of $\cP\cF_*$ as a probabilistic assignment of
sets $X_i$ (which here we think of as certain systems of neurons), we think of 
a summing functor $\Phi_{\Lambda X}$ as the corresponding probabilistic
assignment of (non-deterministic) transition systems to each $X_i$ and all its 
subsets $A_i$ in a consistent way. The choice of working with cubical rather than
simplicial sets does not alter the homotopy type of the resulting construction.


It is often convenient to work with cubical sets because of the fact that transition systems
and higher dimensional automata have a natural formulation in terms of cubical
sets~\cite{FaRaGou06}. The values in $\cP\Box_*$ of the functor 
$\Gamma_{\cP\cC}: \cP\cF_* \to \cP\Box_*$, simply keep track of the
probabilities $\Lambda=(\lambda_i)$ assigned to the subsystems $X_i$ of an
object $\Lambda X$ of $\cP\cF_*$ by considering the object in
$\cP\Box_*$ given by the same combination of cubical nerves of the
categories of summing functors of the subsystems, 
$\sum_i \lambda_i \cN_{\text{cube}}(\Sigma_{\cP\cC}(X_i,x_i))$.

\subsection{Probabilistic transition systems}\label{ProbTransSec}

As an example of a relevant probabilistic category to consider in this setting, we 
describe more explicitly the probabilistic category of transition systems, where
the probabilistic category $\cP\cC$ can be constructed as in \cite{Mar19}, by taking
a wreath product $\cF\cP\wr \cC$ of the category $\cC$ of transition systems described 
above with a category $\cF\cP$ of finite probabilites. 


The resulting categories has objects that are formal convex
combinations of objects of $\cC$ (finite sets of objects of $\cC$ with a probability distribution)
and morphisms consists of a stochastic matrix that relates the probabilities on the objects,
together with a set of morphisms in $\cC$ with assigned probabilities, with a compatibility
between the probability distribution of this set of morphisms and the stochastic matrix. More
precisely, the resulting category has objects
given by finite combinations $\Lambda \tau :=\sum_k \lambda_k \, (S_k, \iota_k, \cL_k, \cT_k)$
and morphisms given by a stochastic map $S$ with $S\Lambda =\Lambda'$ and morphisms
$F: \Lambda \tau \to \Lambda' \tau'$ with
$F=F_{ab,r}=(\sigma_{ab,r},\lambda_{ab,r})$ with probabilities $\mu^r_{ab}$ with 
$\sum_r \mu^r_{ab}=S_{ab}$.
The objects of this category can be seen as non-deterministic automata with
states set $S=\cup_k S_k$ which are a combination of subsystems $S_k$ that are
activated with probabilities $\lambda_k$. A morphism in this category consists of a
stochastic map affecting the probabilities of the subsystems and non-deterministic maps
$(\sigma_{ab,r},\lambda_{ab,r})$ of the states and labeling systems and transitions, 
applied with probabilities $\mu^r_{ab}$.

\subsection{Persistent Gamma-spaces and persistent spectra}\label{persGamma}

We develop here a new formalism that extends and combines the constructions of \cite{ManMar19}
and \cite{Mar19} of Gamma-spaces enriched with probabilistic data and of persistent topology.


We are interested here in a variant of Segal's construction of Gamma-spaces and
associated spectra, which we will then also combine with probabilistic data as in \cite{Mar19},
and which allows us to also incorporate persistent topology structures, following the point
of view we adopted in \cite{ManMar19}, and the categorical setting for persistence described in 
\cite{BuSco14}.

\subsubsection{Thin categories and persistence diagrams}

A persistence diagram in a category $\cC$ indexed by a thin category $(S,\leq)$ (as in Definition~\ref{thindef}) 
is a functor
\begin{equation}\label{persistentP}
 P: (S,\leq) \to \cC \, .
\end{equation} 
In particular a {\em pointed simplicial persistence diagram} is a functor $P: (S,\leq) \to \Delta_*$
to the category $\Delta_*$ of pointed simplicial sets.  We write 
\begin{equation}\label{persistentcat}
\cC^{(S,\leq)} :={\rm Func}((S,\leq), \cC)
\end{equation}
for the category of persistence diagrams, with objects given by functors as in
\eqref{persistentP} and morphisms given by natural transformations of these functors.
This is the categorical viewpoint on persistent topology developed in \cite{BuSco14} and
used in \cite{ManMar19}.

\subsubsection{Persistent Gamma-spaces}

We can accommodate the notion of persistent topology in the setting of
Gamma-spaces in the following way.

\begin{defn}\label{persGammadef}
Let $\Gamma\cS$ denote the category of Gamma-spaces
\begin{equation}\label{Gammacat}
\Gamma\cS:= {\rm Func}(\cF_*,\Delta_*)\, .
\end{equation}
We define {\em persistent Gamma-spaces} to be persistence diagrams in the category of 
Gamma-spaces
\begin{equation}\label{Gammacat2}
\Gamma\cS^{(S,\leq)}:= {\rm Func}((S,\leq),{\rm Func}(\cF_*,\Delta_*)) \, ,
\end{equation}
which we can equivalently view as functors from the category $\cF_*$ to the category of
pointed simplicial persistence diagrams, or as functors from $\cF_* \times (S,\leq)$ to $\Delta_*$
$$ \Gamma\cS^{(S,\leq)} \simeq {\rm Func}(\cF_*,{\rm Func}((S,\leq),\Delta_*)) \simeq {\rm Func}(\cF_*\times (S,\leq), \Delta_*). $$
\end{defn}


Correspondingly, we introduce a notion of {\em persistent spectra}. Let $\Sigma\bS$ denote the category
of symmetric spectra (see \cite{Schw12}). This category has objects given by sequences $X=\{ X_n \}_{n\in \N}$ 
of pointed simplicial sets with a basepoint-preserving left action of the symmetric group $\Sigma_n$ on $X_n$
and with structure maps $\sigma_n: S^1\wedge X_n \to X_{n+1}$ such that the composition 
$S^k\wedge X_n\to X_{n+k}$ is equivariant with respect to the action of $\Sigma_k \times \Sigma_n$. It has
morphisms $f\in {\rm Mor}_{\Sigma\bS}(X,Y)$ given by a collection $f=\{ f_n \}$ of morphisms
$f: X_n \to Y_n$ in $\Delta_*$ that are $\Sigma_n$-equivariant and compatible with the structure maps
through the commutative diagrams
$$ \xymatrix{ S^1\wedge X_n \dto^{\sigma_n^X} \rto^{id\wedge f_n} & S^1\wedge Y_n \dto^{\sigma_n^Y} \\
X_{n+1} \rto^{f_{n+1}} & Y_{n+1} \, .} $$

\begin{defn}\label{persSpecdef}
Persistent spectra are persistence diagrams in the category of spectra
$$ \Sigma\bS^{(S,\leq)} = {\rm Func}((S,\leq), \Sigma\bS) . $$
\end{defn}

\begin{lem}\label{persGammaSpec}
A persistent Gamma-space determines a persistent spectrum.
\end{lem}

\begin{proof}
  By applying the Segal construction \cite{Segal} pointwise we see that a persistent Gamma-space gives
rise to an associated persistent spectrum, by first upgrading a persistent  Gamma-space $\Gamma: \cF_* 
\to {\rm Func}((S,\leq),\Delta_*)$ to a functor $$\Gamma: \Delta_*\to {\rm Func}((S,\leq),\Delta_*)$$ seen equivalently
as a persistent endofunctor of the category of simplicial sets $$\Gamma: (S,\leq)\to {\rm Func}(\Delta_*,\Delta_*).$$
The associated persistent spectrum is given by the functor determined by assigning $X(s)_n:=F(s)(S^n)$. 
\end{proof}


The Segal construction \cite{Segal} of Gamma-spaces and spectra associated to
categories with coproduct and zero object can be extended to the persistent setting
as follows.


\begin{prop}\label{persSegal}
Let $\cC$ be a category with coproduct and zero object and let $\cC^{(S,\leq)}$
be the category of persistence diagrams in $\cC$ indexed by a thin category $(S,\leq)$.
The category $\cC^{(S,\leq)}$ determines a persistent Gamma-space
$\Gamma_{\cC^{(S,\leq)}}: \cF_*\to \Delta_*^{(S,\leq)}$ and an associated
persistent spectrum $\cS_{\cC^{(S,\leq)}}$ in the category $\Sigma\bS^{(S,\leq)}$.
\end{prop}

\begin{proof}
  If $\cC$ is a category with coproduct $\oplus$ and a zero object $0$, then the category
of persistence diagrams $\cC^{(S,\leq)}={\rm Func}((S,\leq),\cC)$ endowed with the pointwise
coproduct has zero object given by the functor $F_0(s)=0$ for all $s\in S$ and $F_0(s\leq s')={\rm id}_0$.


The nerve construction is given by a functor $\cN: {\rm Cat} \to \Delta$ from the category of small
categories to simplicial sets defined on a small category $\cD$ by $\cN(\cD)_n  := {\rm Obj}({\rm Func}([n], \cD))$ 
where $[n]=\{0<1<\cdots <n \}$ the ordered set seen as a thin category with a unique morphism $i\to j$ for $i\leq j$.


Consider then the categories ${\rm Func}((S,\leq),\cD)$ and ${\rm Func}([n],{\rm Func}((S,\leq),\cD))$,
which we can identify with ${\rm Func}((S,\leq),{\rm Func}([n],\cD))={\rm Func}([n],\cD)^{(S,\leq)}$. 
The set of objects of ${\rm Func}([n],\cD)^{(S,\leq)}$ consists of objects of ${\rm Func}([n],\cD)$
parameterized by elements $s\in S$, hence the nerve 
\mbox{$\cN({\rm Func}((S,\leq),\cD))$} 
determines
a pointed simplicial persistence diagram in $\Delta_*^{(S,\leq)}$.


Given a category $\cC$ with coproduct and zero object, and the associated category
of persistence diagrams $\cC^{(S,\leq)}$, for each finite pointed set $X\in {\rm Obj}(\cF_*)$
we can consider the category $\Sigma_{\cC^{(S,\leq)}}(X)$ of summing functors
$$ \Phi_X: P(X) \to  \cC^{(S,\leq)} $$
from the category of pointed subsets of $X$ with inclusions such that
$\Phi_X(*)=F_0$. The base point of $X$ is sent to the zero object of $\cC^{(S,\leq)}$
and $\Phi_X(A\cup A') =\Phi_X(A) \oplus \Phi_X(A')$, with $\oplus$ the coproduct of $\cC^{(S,\leq)}$,
whenever $A\cap A'=\{ * \}$. Any such summing functor $\Phi_X: P(X) \to  \cC^{(S,\leq)}$
determines a functor $\psi_X: (S,\leq)\to \Sigma_\cC(X)$, where $\Sigma_{\cC}(X)$ is the
category of summing functors $\Phi_X: P(X) \to \cC$, by $\Phi_X(A)(s)=\psi_X(s)(A)$. Thus,
we can identify $\Sigma_{\cC^{(S,\leq)}}(X)$ with 
$\Sigma_\cC(X)^{(S,\leq)}$, and applying
the nerve construction $\cN(\Sigma_\cC(X)^{(S,\leq)})$ as above we obtain a pointed 
simplicial persistence diagram in $\Delta_*^{(S,\leq)}$. Thus, using the Segal construction
of \cite{Segal}, we can associate to a category of persistence diagrams $\cC^{(S,\leq)}$
over a category $\cC$ with coproduct and zero object a persistent Gamma-space
$$ \Gamma_{\cC^{(S,\leq)}}: \cF_* \to \Delta_*^{(S,\leq)} $$
and an associated persistent spectrum $\cS_{\cC^{(S,\leq)}}$.
\end{proof}

\subsection{Probabilistic persistent Gamma-spaces}\label{probpersGamma}

Consider as in \cite{Mar19} the category $\cF\cP={\mathbb I}/{\rm FinStoch}$ of finite
probabilities with stochastic maps as morphisms (see \cite{BaFr14} and Remark~2.2 of \cite{Mar19}). 
Given a category $\cC$ with
categorical sum and zero object, consider the wreath product $\cP\cC:=\cF\cP \wr \cC$,
which has objects given by formal convex linear combinations $\Lambda C:=\sum_i \lambda_i C_i$ 
of objects $C_i\in {\rm Obj}(\cC)$ with $\lambda_i\geq 0$ and $\sum_i \lambda_i =1$,
and morphisms $\phi: \Lambda C \to \Lambda' C'$ given by pairs $\phi=(S,F)$ of a 
stochastic map with $S\Lambda =\Lambda'$ (a morphisms of $\cF\cP$) and a finite collection
$F=\{ F_{ab,r} \}$ of morphisms $F_{ab,r} \in {\rm Mor}_\cC(C_b,C_a')$ with assigned
probabilities $\mu^r_{ab}$ satisfying $\sum_r \mu^r_{ab} = S_{ab}$. 
As shown in \S 2 of \cite{Mar19}
the category $\cP\cC$ constructed in this way has a zero object and a coproduct 
of the form $\Lambda C \oplus \Lambda' C' =\sum_{ij} \lambda_i \lambda'_j C_i \oplus_\cC C'_j$.


The morphisms in the category $\cP\cC$ can be interpreted as a collection of
morphisms $F_{ab,r}$ in $\cC$ that are chosen and applied with probability $\mu_{ab}^r$
where $\sum_r \mu_{ab}^r$ is the stochastic matrix that relates the probabilities
$\Lambda$ and $\Lambda'$ in $\cF\cP$ associated to the objects $\Lambda C$
and $\Lambda' C'$ of $\cP\cC$. Thus, both the objects and the morphisms of $\cC$
are made non-deterministic by passing to $\cP\cC$, in a way that preserves
the property that the category has coproduct and zero object.
Thus, the Segal construction applied to $\cC$ can be extended to the
probabilistic category $\cP\cC$. 


Moreover, it is shown in \cite{Mar19} that the Segal construction itself
can be made probabilistic, by considering a version of Gamma-spaces
based on the category $\Box_*$ of pointed cubical sets with connections 
(which are homotopy equivalent to the usual Gamma-spaces valued in
pointed simplicial sets) and then defining stochastic Gamma-spaces 
as functors
\begin{equation}\label{stochGamma}
\Gamma : \cP\cF_* \to \cP\Box_*
\end{equation}
where $\cP\cF_*$ and $\cP\Box_*$ are the probabilistic
categories associated to pointed sets and to pointed cubical
sets with connections. It is shown in \S 5 of \cite{Mar19}
that to any category $\cC$ with zero object and categorical
sum, one can associate, using the Segal construction, a
probabilistic Gamma-space $\Gamma_{\cP\cC}: \cP\cF_* \to \cP\Box_*$
which is the functor determined by $\Gamma_{\cP\cC}(\Lambda X)=\sum_i \lambda_i 
\cN_{\text{cube}}(\Sigma_{\cP\cC}(X_i))$, seen as an object in $\cP\Box_*$.
The cubical nerve $\cN_{\text{cube}}(\Sigma_{\cP\cC}(X))$ is homotopy equivalent to the 
simplicial nerve $\cN(\Sigma_{\cP\cC}(X))$~\cite{Ant02}, meant as the equivalence of
the respective realizations (see \S 4 of \cite{Ant02}). 


The cubical nerve of a category $\cD$ is obtained by
considering functors ${\rm Func}(\cI^n, \cD)$, where the
objects of $\cI^n$ are the vertices of the $n$-cube (sequences
$(s_1,\ldots,s_n)$ with digits $s_i\in \{0,1\}$) and morphisms
generated by the edges of the cube. The maps 
$\cN_{\text{cube}}(\cD)_n \to \cN_{\text{cube}}(\cD)_m$
are induced by precomposition of the functors $\cI^n \to \cD$
with the morphisms $\cI^m\to \cI^n$ of the cubical category (box category). 


By combining this construction with the construction of persistent Gamma-spaces
we introduced in \S \ref{persGamma}, we obtain the following.

\begin{prop}\label{persprobGammaprop}
Let $\cC$ be a category with zero object and categorical sum, with
$\cP\cC=\cF\cP\wr \cC$ the associated probabilistic category, and let
$(S,\leq)$ be a thin category.
The Segal construction determines a probabilistic persistent spectrum
\begin{equation}\label{persprobGamma}
 \Gamma_{\cP\cC^{(S,\leq)}} : \cP \cF_* \to \cP \Box_*^{(S,\leq)} \, .
\end{equation} 
\end{prop}


The construction described in Proposition~7.2 of \cite{ManMar19} can be seen
as a special case of the construction obtained here above.

\subsection{Modeling constraints and the role of persistence} \label{Persist1Sec}

As observed in \S 3.2 of \cite{Per19}, a diagram in a thin category $(S,\leq)$ is
just a selection of a subset $A \subseteq S$. A cone on $A$ with vertex $x$
is a lower bound $x$ for $A$, since it consists of an arrow from $x$ to each element 
$a\in A$, while a cocone is an upper bound. Limits and colimits then correspond to
greatest lower bounds and least upper bounds for subsets $A\subseteq S$.
Thus, functors that are compatible with limits and colimits can be viewed as
describing constrained optimization settings where certain maximization
or minimization conditions are imposed. 


Consider a thin category $(S,\leq)$ and a category $\cD^{(S,\leq)}={\rm Func}((S,\leq),\cD)$
of persistence diagrams in $\cD$ indexed by $(S,\leq)$. We can interpret the objects
$D(s)$ and morphisms $D(s\leq s'):  D(s) \to D(s')$ in $\cD$ as families of
objects in $\cD$ subject to constraints encoded in $(S,\leq)$.


In our setting here we can consider a theory of resources formulated as in \cite{CoFrSp16}
and \cite{Fr17} (see \S \ref{ResourcesSec} above). In particular, we assume given an 
abelian semigroup with partial ordering $(R,+,\succeq, 0)$ on the set $R$ of isomorphism 
classes of objects in ${\rm Obj}(\cR)$, where $\cR$ is a symmetric monoidal category
describing resources. 


We can use $(R,\succeq)$ as the indexing of persistence diagrams. Here we use the
reverse ordering, since in $(R,+,\succeq, 0)$ the relation $A \succeq B$ means
${\rm Mor}_{\cR}(A,B)\neq \emptyset$ hence resource $A$ is convertible to resource $B$.
Thus, in a category $\cD^{(R,\succeq)}$ we have objects $D_A$, indexed by resources $A\in R$
with a morphism $D_A \to D_B$ whenever $A\succeq B$. The morphism describes the effect
on the object $D_A$ caused by converting the resource $A$ to the resource $B$. The
constraints here are encoded in the convertibility of resources. 


Assuming that we are modeling the possible
concurrent/distributed computing architectures or other resources associated to a population
of neurons via a (probabilistic) Gamma-space $\Gamma_{\cP\cC}: \cP\cF_*\to \cP\Box_*$,
where $\cP\cC$ is the category of (non-deterministic) transition systems (or the probabilistic
version of another category of resources),
we can incorporate in this description the constraints given by the convertibility
properties of computational, metabolic or informational resources by considering an associated
persistent (probabilistic) Gamma-space 
$$ \Gamma_{\cP\cC^{(R,\succeq)}}: \cP\cF_*\to \cP\Box_*^{(R,\succeq)}, $$
where $\cP\cC^{(R,\succeq)}$ is the category of persistence diagrams of (non-deterministic) transition systems
parameterized by the convertibility of resources $(R,\succeq)$ with the reverse ordering, as above.


Thus, for instance, we can view
the category $\cP\cC^{(R,\succeq)}$ as describing families of non-deterministic (computational) resources
(transition systems) $C_A$ associated to the available (metabolic) resources $A\in R$, with maps $C_A\to C_B$
that describe the change to a transition system affected by the conversion of resources $A\succeq B$.
The persistent (probabilistic) Gamma-space carries this information about the dependence on
resources and conversion of resources over into the construction of the resulting objects in
$\cP\Box_*^{(R,\succeq)}$, which we now interpret as a family of (probabilistic) simplicial or cubical 
sets associated to the available resources in $R$ with maps describing the effect of
resource conversion. This provides a description of all the
possible ways of assigning transition systems to a probabilistic set $\Lambda X$ with 
assigned resources $A\in R$.

\subsection{Persistence to model time and scale dependence} \label{Persist2Sec}

There is another use of the persistence structure in these models, which is in line with
the more common use of persistent topology, namely as a way to keep track of the
dependence on time and on scale. 


In the previous subsection we have described how to use persistent (probabilistic) Gamma-spaces
to model the dependence on constraints dictated by resources and convertibility of resources,
where the latter are encoded in the structure of a preordered semigroup $(R,+,\succeq)$.
A more common use of persistent topology is in capturing the dependence of a simplicial
set on either a time variable or a scale factor. In this setting, the thin category of the form
$(\cI,\leq)$ where $\cI$ is a subinterval of the real line $\R$, with its natural ordering $\leq$,
where the variable $s\in \cI$ represents either time or a scale variable. The scale dependence
for example is used in the construction of the Vietoris--Rips simplicial complex associated to
a set of data points. The time dependence is crucial for example in the analysis of the
formation of non-trivial homology in the persistent topology of the simulations of the 
neural cortex in response to stimuli analyzed in \cite{Hess}.


The results of \cite{Hess} show that nontrivial topological structures
arise in the computational architecture of the response of the (simulated) neural cortex to
stimuli. In simulations of the reconstructed microcircuitry, 
following a spatio-temporal stimulus to the network, during correlated activity,
active cliques of increasingly high dimension are detected,  
with a large number of nontrivial homology
generators forming and peaking at around 60--80 ms from the initial stimulus
and then quickly disappearing. While different stimuli give rise to different patterns of
activity, all have this common feature, where  functional relations among increasingly
high-dimensional cliques form and then disintegrate.


This kind of result motivates the introduction of a time scale for
the birth and death of simplices and homology generators in
various dimensions. A dependence on scale may also be similarly needed. 


While one can try to incorporate both the time/scale dependence and
the dependence on resources and their convertibility in a the persistent
structure, it seems more natural to reserve persistence as a way to
capture the time/scale dependence and incorporate the metabolic 
constraints and other resources constraints in a different way. This can
be done by working directly with the symmetric monoidal category of
resources $(\cR,\otimes, \bI)$ instead of using the associated preordered
semigroup $(R,+,\succeq)$.

\subsection{Variants of the nerve construction}\label{VarNerveSec}

We can use Gamma-spaces and their probabilistic and persistent generalizations
to associate in a functorial way to a given population of neurons endowed with 
certain probabilities of activation, described by an object $\Lambda X$ in the category
$\cP\cF_*$, a (probabilistic) simplicial or cubical 
set $\Gamma_{\cP\cC}(\Lambda X)$. The underlying category of summing functors 
describes all the consistent ways of assigning
computational architectures (transition systems) or other kinds of resources, seen as elements
in a category $\cP\cC$, to all subsystems
(subsets with probabilities) of $\Lambda X$. 


The resulting object $\Gamma_{\cP\cC}(\Lambda X)$
can itself be regarded as a computational architecture (a higher dimensional automaton)
attached to $\Lambda X$. It encodes all the possible consistent assignments
of transition systems to $\Lambda X$ and to its constituent parts, and it inputs and
outputs (probabilistic) simplicial data. 


The nerve construction used in the notion of Gamma-spaces (and their probabilistic 
and persistent versions) can be adapted to accommodate other possible categorical models of 
concurrent/distributed computation in the resulting $\Gamma_{\cP\cC}(\Lambda X)$.
Indeed, the usual full and faithful nerve functor $\cN: {\rm Cat} \to {\rm Func}(\Delta^{\operatorname{op}}, {\rm Set})$
provides a way of describing categories through simplicial sets. 
This nerve construction admits generalizations, as shown in \cite{Leister}, \cite{Weber}, obtained
by considering certain classes of monads $T$ on suitable categories (see \cite{Leister} for the
specific conditions on monads and categories), to which it is possible to associate a nerve functor
$$ \cN_T: {\rm Alg}(T) \to {\rm Func}(\Delta_T^{\operatorname{op}}, {\rm Set}) $$
with $\Delta_T^{\operatorname{op}}$ a category of $T$-simplicial sets, and ${\rm Alg}(T)$ the category of algebras
over the monad $T$.


While we do not develop this viewpoint in the present paper, it is worth mentioning the
fact that this can lead to other ways of adapting the formalism of Gamma-spaces to a
probabilistic setting, where probabilities are interpreted from a monad viewpoint,
see \cite{FrPeRe19} (and also \cite{Fr19a}, \cite{FrPe18a}, \cite{FrPe17}, \cite{FrPe18b}, \cite{FrPeRe19})
for recent developments of this approach to probability.

\bigskip
\subsection*{Acknowledgment} The authors thank the anonymous referees for many very detailed comments
and suggestions that greatly improved the exposition and the organization of the material in the paper. 
We also thank the handling editor
for providing extensive feedback and suggestions.


\providecommand{\doi}[1]{}
\renewcommand{\doi}[1]{\href{https://doi.org/\detokenize{#1}}{\detokenize{#1}}}%

\bibliographystyle{plainnat}

\end{document}